%% file: paper.tex
\documentclass[reprint, aps, prd, letterpaper, showpacs, amsmath, %
amssymb, amsfonts, nofootinbib, floatfix, superscriptaddress, %
twoside]{revtex4-1}

\pdfoutput=1

\usepackage{graphicx}
\usepackage{subfigure}
\usepackage{amssymb, amsmath}
\usepackage{mathtools}
\usepackage[usenames]{color}
\usepackage[dvipsnames]{xcolor}
\usepackage{mathrsfs} 
\usepackage{xspace}
\usepackage[colorlinks, plainpages=false, hyperfigures=true]{hyperref}
\definecolor{CiteColor}{rgb}{0.18039, 0.18824, 0.57255}
\definecolor{UrlColor} {rgb}{0.741, 0.173, 0.000}
\definecolor{LinkColor}{rgb}{0.25098, 0.47843, 0.04706}
\hypersetup{linkcolor=LinkColor}
\hypersetup{citecolor=CiteColor}
\hypersetup{urlcolor=UrlColor}

\input{Macros}

\begin{document}

\graphicspath{%
  {figs/}%
}

\title{Comparing Gravitational Waveform Extrapolation\\ to 
Cauchy-Characteristic Extraction in Binary Black Hole Simulations}

\newcommand{\Caltech}{\affiliation{Theoretical Astrophysics 350-17,
    California Institute of Technology, Pasadena, California 91125,
    USA}} %
\newcommand{\Cornell}{\affiliation{Center for Radiophysics and Space
    Research, Cornell University, Ithaca, New York 14853, USA}}%
\newcommand{\CITA}{\affiliation{Canadian Institute for Theoretical
    Astrophysics, University of Toronto, 60 St.~George Street,
    Toronto, Ontario M5S 3H8, Canada}} %


\author{Nicholas W.~Taylor} \Caltech 
\author{Michael Boyle} \Cornell 
\author{Christian Reisswig} \thanks{Einstein Fellow}\Caltech 
\author{Mark A.~Scheel} \Caltech 
\author{Tony Chu} \CITA 
\author{Lawrence E.~Kidder} \Cornell 
\author{B\'{e}la Szil\'{a}gyi} \Caltech 

\date{\today}

\begin{abstract}
  We extract gravitational waveforms from numerical simulations of
  black hole binaries computed using the Spectral Einstein Code.
  We compare two extraction methods: direct construction of
  the Newman-Penrose (NP) scalar $\Psi_4$ 
  at a finite distance from the source
  and Cauchy-characteristic extraction (CCE).  
  The direct NP approach
  is simpler 
  than CCE, but NP waveforms can be
  contaminated by near-zone effects---unless the waves 
  are extracted at several distances from the source and
  extrapolated to infinity.  Even then, the resulting
  waveforms can in principle
  be contaminated by gauge effects.  In contrast, 
  CCE directly provides, by construction, gauge-invariant 
    waveforms at future null infinity.
  We verify the gauge invariance of CCE by running the same
  physical simulation using two different gauge conditions.  We find
  that these two gauge conditions
  produce the same CCE waveforms
    but show differences in extrapolated-$\Psi_4$ waveforms.
  We examine data from several different binary 
    configurations and measure the dominant sources of error in
  the extrapolated-$\Psi_4$ and CCE waveforms.
  In some cases, we find that
  NP waveforms extrapolated to infinity agree
  with the corresponding CCE waveforms to within the estimated
  error bars.
  However, we find that in other cases 
  extrapolated and CCE waveforms disagree, most notably
  for $m=0$ ``memory'' modes.
\end{abstract}


\maketitle


\section{Introduction}
\label{section:introduction}
In the next few years, the second generation of ground-based
gravitational-wave interferometers is expected to make the first
direct detection of gravitational waves (GWs) from the inspiral and
coalescence of compact binaries, 
marking the beginning of the era of gravitational wave
astronomy~\cite{Harry2010,Somiya:2012,aVirgo,Grote:2010zz}.
Because of the very low compact binary coalescence rate~\cite{Abadie:2010cfa}, 
observable GW events 
are expected to originate from sources at the
edge of the detectable range, 
with signal to noise ratios of order
unity. Detecting these exceptionally weak GW signals requires the use
of matched filtering, in which the noisy data are
 compared with a
template bank of expected waveforms (see, e.g., Ref.~\cite{Owen:1998dk} and
references therein). For black hole binaries, 
these expected waveforms
can be accurately computed only 
by using full numerical solutions of Einstein's equations. 
However, because these simulations are computationally expensive,
analytical or phenomenological models of GW emission are required
in order to densely cover the parameter space.  Because these models
must be calibrated using results from numerical 
simulations~\cite{Buonanno99,Ajith-Babak-Chen-etal:2007b,Ajith:2011ec,
Buonanno:2009qa,PanEtAl:2011,Taracchini:2012},
it is essential that accurate waveforms from numerical
simulations are available. Moreover, 
it is crucial that the uncertainties
in these numerical waveforms are well understood.

There are several sources of uncertainty in numerical waveforms.
Perhaps the
most straightforward to understand and measure
is the numerical truncation error in the binary black hole 
simulation itself, which we refer to as the
``Cauchy error''.  Numerical relativity codes for black hole binaries
solve the full nonlinear Einstein equations.  
These are formulated as an initial value (Cauchy) problem, 
in which initial data (satisfying the Einstein
constraints) are 
provided on some spacelike surface labeled by
coordinate time $t$.  
The Einstein evolution equations are then used to determine
data at subsequent times.  
The Cauchy error is the error made in solving these evolution equations
numerically. It depends on the truncation error of the employed 
numerical scheme and the coarseness of the computational grid.

Another source of uncertainty in numerical simulations 
is the error
associated with the use of a finite outer boundary.  
In principle the solution of Einstein's equations
is needed for the entire spacetime, but
most simulations solve the equations only on a finite spatial domain.
For example, simulations 
performed using the
Spectral Einstein Code (\texttt{SpEC}) typically have
outer boundaries located at about $500\, M$ 
(where $M$ is the total mass
of the system), while the total
simulation time may be thousands of 
$M$~\cite{SpECwebsite,Szilagyi:2009qz,Buchman:2012dw,Hemberger:2012jz}. The
effects of a finite outer boundary can be mitigated by choosing 
constraint-preserving
boundary conditions (see, e.g., Ref.~\cite{Rinne2007}). 
However, such boundary conditions are not exact and cannot
account for physical effects such as the backscatter of GWs off the
spacetime curvature from regions outside the boundary.  Previous
studies have shown that this outer boundary
error is typically comparable to or smaller than the Cauchy 
error~\cite{Scheel2009,Buchman:2012dw}.

Yet another source of uncertainty is the error
associated with waveform extraction from finite-radius numerical data 
to future null infinity ($\scri^+$).
A waveform at $\scri^+$ represents what
would be measured by an Earth-based GW observatory that detects an
astrophysical source.  The simplest approach to waveform extraction is to 
compute the Newman-Penrose scalar $\Psi_4$ (see 
Section~\ref{section:psi4} for details) at a large but finite distance
from the source~\cite{Newman1962}, 
and to use this as an approximation to the waveform at $\scri^+$.
This can be inaccurate, because the 
quantity $\Psi_4$ represents measurable outgoing 
gravitational radiation only in the limit of infinite distance
from the source 
(see, e.g., Refs.~\cite{Stewart:1990uf,Penrose1986V2,Penrose1965})
and in the Bondi gauge~\cite{Stewart:1990uf}
(rather than the gauge of the simulation).

A better approximation is a popular refinement of this
single-extraction-radius method: $\Psi_4$ is extracted as
before, but at several different radii 
instead of at a single radius, and this information 
is used to extrapolate $\Psi_4$ to $\scri^+$ 
(see Section~\ref{section:extrapolation} for details).  
This extrapolation
procedure can remove near-zone effects and some 
gauge effects from the resulting waveform.  
However, as we
show below, 
extrapolation does not always succeed in a convergent way, and
even when it does, 
it is possible for 
some near-zone and gauge effects to remain.
Estimating the magnitude of these remaining effects 
is difficult; 
it currently requires
either repeating simulations using multiple gauge conditions
or comparing with an independent wave-extraction method.  
Most of the currently
available numerical-relativity waveforms, either published or in use
by groups working on calibration of analytical methods, employ
(low-order) extrapolation of $\Psi_4$ or simply $\Psi_4$ extracted at
a finite 
radius~\cite{Hinder:2013oqa, Ajith:2012az, AjithEtAl:2009, Healy:2013jza}.

A more robust method of waveform extraction is Cauchy-characteristic
extraction (CCE).   This procedure uses
a characteristic 
evolution code to solve
Einstein's equations on a foliation
of outgoing null hypersurfaces rather than on
spacelike
hypersurfaces~\cite{Bishop97b,Winicour05,Gomez:2007cj,Reisswig:2012ka}.
Radial compactification enables the use of
null hypersurfaces 
that extend all the way to future null infinity,
so a waveform at $\scri^+$ can be directly computed.
Furthermore, the waveform at $\scri^+$
can be computed in a
gauge-invariant way~\cite{Bishop97b}.  
In practice, the strong-field region near the source is evolved using a 
Cauchy code, while the asymptotic region is evolved with a 
characteristic code.  The Cauchy evolution supplies data on a timelike, 
finite-radius worldtube, which serves as the inner boundary for the 
characteristic evolution (see Fig.~\ref{fig:ccecartoon}).
This technique has been 
used in Refs.~\cite{Reisswig:2009us, Babiuc:2010ze, Pollney:2010hs} for 
simulations of 
binary black hole mergers and in 
Refs.~\cite{Reisswig:2010cd, Reisswig:2013sqa, Reisswig:2012nc, Ott2011a} for
simulations of stellar collapse, binary neutron star mergers, and black 
hole formation. 
The primary disadvantages of CCE are its computational expense and its 
complexity (because it requires two separate methods of solving the 
Einstein equations).
Our binary black hole simulations
typically require weeks of walltime, and performing CCE 
can add several additional days of computation time.  
By comparison, the extrapolation procedure requires only 
minutes.

Other methods of waveform extraction have been considered in the
literature.  In addition to the methods discussed above, the most
widely used is the Regge-Wheeler-Zerilli-Moncrief
method~\cite{ReggeWheeler1957,Zerilli1970b,Zerilli:1971wd,Moncrief},
in which the far-field solution is treated as a perturbation about a
fixed background (typically Schwarzschild or Minkowski), and the
perturbed solution is constructed by reading off gauge-invariant
perturbation coefficients from the numerical solution on 
a finite extraction sphere. See Ref.~\cite{NagarRezzolla2005} for a
review.  
Related methods for finding the asymptotic form of the waves
  from the finite-radius behavior were considered by Abrahams and
  Evans~\cite{Abrahams1988, AbrahamsEvans:1990}, Lousto
  \etal~\cite{LoustoEtAl:2010}, and have recently been generalized by
  Benedict \etal~\cite{BenedictEtAl:2013}.  However, these analyses
  rely on certain assumptions about gauge that we do not make.
In Ref.~\cite{Reisswig:2010cd}, a 
comparison between CCE, $\Psi_4$, and Regge-Wheeler-Zerilli-Moncrief 
extraction was performed in the context of stellar collapse.
In this paper we consider only two extraction methods:
$\Psi_4$-extrapolation and CCE.

The goal of this paper is to compare extrapolated-$\Psi_4$
and CCE waveforms for binary black hole
simulations performed using \texttt{SpEC}.
We estimate the uncertainties in the waveforms associated
with each extraction method, and we
examine the differences between the waveforms relative to 
these estimated errors. 
In particular, by comparing 
extrapolated-$\Psi_4$ and CCE waveforms, 
we can estimate the unknown gauge error that 
may be present in the former.
One important question we wish to address is whether
it suffices to use the (simpler and 
less computationally expensive) 
extrapolation
method, or whether the gauge-invariance of CCE is necessary,
given the current accuracy of our simulations.

Some previous comparisons of CCE and extrapolation have 
been done using binary black hole simulations performed with the 
finite-difference code
\texttt{Llama}~\cite{Pollney:2009yz}.  
In Refs.~\cite{Reisswig:2009us, Reisswig:2009rx}, it was found that 
differences between extrapolated
and CCE quantities were on the order of the
discretization error of the Cauchy simulation.  Additionally, the differences
were found to be non-convergent, suggesting that the waveform extraction
error could become dominant for high-accuracy simulations.
These previous studies focused on short simulations of
equal-mass and spin-aligned binaries.
Here, we also consider longer 
unequal mass and generic 
precessing configurations, 
we compare multiple $Y_{\ell m}$ modes, and we use 
more sophisticated
extrapolation and waveform-alignment methods.

This paper is organized as follows.  In
Section~\ref{section:gwextraction} we review different methods of
waveform extraction. 
We discuss direct construction of $\Psi_4$ 
on finite-radius extraction spheres, 
extrapolation of $\Psi_4$ to infinity,
and waveform extraction using CCE.
In Section~\ref{sec:binary-black-hole} we describe 
the black hole binary simulations that we 
use, briefly discussing
the initial data, gauge conditions, and
evolution algorithms. In
Section~\ref{sec:EstimatingErrors} we discuss how to  
estimate the various sources of error
in the gravitational waveforms, including
errors in the Cauchy
evolution as well as 
in the waveform extraction methods.  
In Section~\ref{sec:results} we verify that CCE is 
indeed gauge-invariant
by comparing waveforms from two simulations with identical 
physics but with different gauge conditions.  
We compare the relative magnitudes of the
various errors, and we show 
that the error associated with the location of the CCE inner boundary
(which we attribute to mismatch of characteristic and Cauchy
initial data) 
is typically greater
than the numerical error in the characteristic evolution. 
We also show that, except for
modes with $m=0$, 
extrapolated-$\Psi_4$ and CCE waveforms 
agree to within 
the estimated error bars. We summarize in Section~\ref{sec:discussion}.
Note that we will refer to extrapolated-$\Psi_4$ waveforms
simply as ``extrapolated waveforms'', 
and we 
will use the terms uncertainty and error interchangeably
when discussing error estimates.


\section{Gravitational wave extraction}
\label{section:gwextraction}
In this section, we review some of the mathematical preliminaries 
as well as the GW extraction methodology.  
We discuss how gravitational radiation content is extracted from the 
finite-radius numerical simulation, and we review the 
extrapolation and CCE methods.

\subsection{Direct extraction of Newman-Penrose $\Psi_4$}
\label{section:psi4}

\subsubsection{The Newman-Penrose scalar $\Psi_4$}
\label{sec:newm-penr-scal}
The GW content of a spacetime can be defined in terms of a particular
component of the Weyl tensor using the Newman-Penrose (NP) 
formalism~\cite{Newman1962}.  This formalism is based on a complex tetrad of 
null vectors
$\{l^\mu, n^\mu, m^\mu, \bar{m}^\mu\}$ 
that satisfy $l^\mu n_\mu=-m^\mu\bar{m}_\mu=1$.  Here, a bar denotes complex
conjugation.  The Weyl tensor
$C_{\alpha\beta\gamma\delta}$ can be uniquely represented via five
complex scalars by contracting
with elements of the null tetrad:
\begin{subequations}
  \begin{align}
    \Psi_0 &\define l^\alpha m^\beta l^\gamma m^\delta
    C_{\alpha\beta\gamma\delta}\,, \\
    \Psi_1 &\define l^\alpha n^\beta l^\gamma m^\delta
    C_{\alpha\beta\gamma\delta}\,, \\
    \Psi_2 &\define l^\alpha m^\beta \bar{m}^\gamma n^\delta
    C_{\alpha\beta\gamma\delta}\,, \\
    \Psi_3 &\define l^\alpha n^\beta \bar{m}^\gamma n^\delta
    C_{\alpha\beta\gamma\delta}\,, \\
    \label{eq:Psi4FromWeyl}
    \Psi_4 &\define n^\alpha \bar{m}^\beta n^\gamma \bar{m}^\delta
    C_{\alpha\beta\gamma\delta}\,.
  \end{align}
\end{subequations}

In asymptotically flat spacetimes, by virtue of the peeling theorem,
the Weyl tensor obeys
\begin{equation}
  \label{eq:Peeling}
  C_{\alpha\beta\gamma\delta} \sim \frac{[N]}{r} + \frac{[III]}{r^2} +
  \frac{[II]}{r^3} + \frac{[I]}{r^4}+\OfOrder(r^{-5})\,,
\end{equation}
where letters in brackets denote Petrov types
(see, e.g., Refs.~\cite{Stewart:1990uf,Penrose1986V2,Penrose1965}).  
As the distance from the source tends toward infinity, 
the spacetime approaches type $N$.
Petrov type $N$ spacetimes are
outgoing plane-wave solutions, with $\Psi_4$ the only non-zero
component of the Weyl tensor for a suitable choice of null tetrad.  
Consequently, in the
limit of infinite distance from the source, $\Psi_4$ is identified as
containing purely outgoing gravitational radiation.  Assuming Bondi
gauge~\cite{Stewart:1990uf}, $\Psi_4$ can be directly related to the 
measurable plus
and cross polarization modes of the strain $h$ via two time integrals,
\begin{equation} \label{eq:h_from_psi4} h_+-ih_\times =
  \lim_{r\rightarrow\infty}\int_{-\infty}^t dt'\int_{-\infty}^{t'}dt''
  \Psi_4 |_{S^2}\,,
\end{equation}
on a spherical surface $S^2$ at $\scri^+$.

\subsubsection{$\Psi_4$ extraction at finite distance}
\label{sec:psi_4-extraction-at}
To extract $\Psi_4$ from a numerical simulation, one chooses a tetrad
$\{l^\mu, n^\mu, m^\mu, \bar{m}^\mu\}$, computes the Weyl tensor 
by differentiating the metric, and 
then constructs $\Psi_4$ via
Eq.~\eqref{eq:Psi4FromWeyl}. 
Since the computational domain is of finite size, 
it is not possible to compute $\Psi_4$ at an
infinite distance from the source.  Instead,
we typically compute
$\Psi_4$ on finite-radius coordinate spheres. On each of these
spheres, we expand $\Psi_4$ in spin-weighted spherical harmonics,
\begin{equation}
  \label{eq:SWSHDecomposition}
  \Psi_{4}(t, r, \vartheta, \varphi) = \sum_{\ell,m}
  \Psi_{4}^{\ell,m}(t,r)\, \mTwoYlm{\ell, m} (\vartheta, \varphi),
\end{equation}
where $(\vartheta, \varphi)$ are the usual polar coordinates on the
sphere, in the coordinate system used by the simulation.  In \texttt{SpEC}, 
we choose a coordinate tetrad that is only asymptotically null
and orthonormal, 
in anticipation of extrapolation to infinity (see
Section~\ref{section:extrapolation}).  Details of the 
$\Psi_4$ extraction method 
used by \texttt{SpEC} are described in
Refs.~\cite{Pfeiffer-Brown-etal:2007,Boyle2007,Scheel2009}.

This procedure has three drawbacks. First, it computes
$\Psi_4$ at a finite radius where the spacetime is not necessarily of
Petrov type $N$.  This means that even if in the proper gauge,
$\Psi_4$ may not be the only non-zero component in
Eq.~(\ref{eq:Peeling}), and furthermore $\Psi_4$ does not necessarily
correspond only to purely outgoing gravitational radiation.  Second,
we choose a coordinate-based tetrad $\{l^\mu, n^\mu, m^\mu,
\bar{m}^\mu\}$, which only asymptotically has the properties that lead
to the peeling theorem, Eq.~(\ref{eq:Peeling}).  Third, we do not
impose Bondi gauge, but instead we use whatever gauge is used by the
code that evolves Einstein's equations.  This may lead to mixing of
the $\Psi_n$, and hence it can invalidate Eq.~(\ref{eq:h_from_psi4}),
which relates $\Psi_4$ to the GW strain $h$ even in the limit of
infinite distance from the source.

The first two of these drawbacks can be reduced by
extracting $\Psi_4$ on
multiple coordinate spheres with different radii, and then
extrapolating these results to $r\to\infty$, as described in
Section~\ref{section:extrapolation} below.  This extrapolation
procedure not only handles the problem of finite extraction radius,
but it also corrects error terms introduced by the choice of a
coordinate-based tetrad, since these error terms scale like higher
powers of $1/r$. 
Extrapolation can also correct some gauge errors,
provided that they 
fall off faster than $1/r$.  However,
it is possible that some gauge choices may produce 
effects that
persist even after
extrapolation, and some gauge choices may prevent accurate
extrapolation altogether. This could occur, for example, if 
the gauge-induced leading-order falloff of the
extracted $\Psi_4$ were slower than $1/r$.  
We will show an
example of the latter case in Section~\ref{sec:errors-from-diff}.

\subsubsection{Extrapolation}
\label{section:extrapolation}
To extrapolate $\Psi_4$ to infinite radius, using data extracted on a
series of finite spheres of different radii, we follow the procedure
of Ref.~\cite{BoyleMroue:2009}.  In this section we summarize the
technique, including certain minor improvements.

We measure the coefficients $\Psi_4^{\ell,m}$ of
Eq.~(\ref{eq:SWSHDecomposition}) at a set of coordinate times
$\{t_{i}\}$ on a set of coordinate spheres of radii $\{R_{j}\}$, using
the procedure described in Section~\ref{sec:psi_4-extraction-at}.  At
each time, we also compute the areal radius $\ra$ of each sphere by
integrating over the sphere using the full spatial metric, and we
compute the average value of the metric component $g^{tt}$
over each sphere. From the initial data we compute the
Arnowitt--Deser--Misner (ADM) mass~\cite{ArnowittEtAl:1962} $\ADMMass$
of the spacetime.

We then construct a retarded time that slightly generalizes the
  usual Schwarzschild definition to account for simple
  time dependence of the lapse and the radial coordinate.  We define
  the retarded time as
\begin{subequations}
  \label{eq:RetardedTime}
  \begin{gather}
    \tr \define \tcorr - \rt\,, %
    \intertext{where} %
    \rt \define \ra + 2\, \ADMMass\, \ln\left(\frac{\ra}{2\ADMMass} -
      1 \right)\,, %
    \intertext{and} %
    \tcorr \define \int_{0}^{t}\, \sqrt{\frac{-1/g^{tt}}{1 -
        2\ADMMass/\ra}} \, dt'\,.
  \end{gather}
\end{subequations}
Here, $\rt$ is the standard tortoise coordinate of the Schwarzschild
metric, with the Schwarzschild radial coordinate replaced by the areal
radius $\ra$, and the Schwarzschild mass parameter replaced by the ADM
mass $\ADMMass$ of the initial data for simplicity.  The
corrected time $\tcorr$ is constructed so that if the metric in the
given coordinates has the standard Schwarzschild form except for the
lapse, then $\tr$ will be precisely a null coordinate.  This does
not account for other
departures of the metric from Schwarzschild.

The quantities $\tr$ and $\ra$ defined above may not be the most
optimal choices of coordinates; for instance, there may be other
choices that make $\tr$ more nearly a null coordinate.  The final
result of extrapolation, however, will not be affected by imperfect
choices of $\tr$ and $\ra$ as long as two conditions are satisfied:
(1) our choices differ from the optimal choices by factors of at most
$1+\OfOrder(1/\ra)$, and (2) the extrapolated quantities can be
expanded in convergent power series in $1/\ra$.  In most cases it
appears that these conditions are satisfied.  However, in
Sec.~\ref{sec:errors-from-diff} we show an example where at least one
of them fails.

Having measured $\Psi_{4}^{\ell,m}(t,R)$, which is a function of
coordinate time and coordinate radius, we can instead express $\Psi_4$
as a function of retarded time and areal radius:
$\Psi_{4}^{\ell,m}(\tr,\ra)$.  The straightforward way to extrapolate
to infinity is to fit $\Psi_{4}^{\ell,m}(\tr,\ra)$ to a polynomial in
$1/\ra$ at a fixed value of $\tr$, and then evaluate the polynomial in
the limit $1/\ra \to 0$, thus obtaining
$\Psi_{4}^{\ell,m}(\tr,\infty)$.  

Because $\Psi_{4}^{\ell,m}(\tr,\ra)$ may be rapidly oscillating in
$\tr$, however, errors made in computing $\tr$ can lead to large
errors in $\Psi_{4}^{\ell,m}(\tr,\ra)$ and subsequently in the
extrapolated value $\Psi_{4}^{\ell,m}(\tr,\infty)$.  For this reason,
it would be better to extrapolate a function that is slowly varying in
$\tr$.  For most modes of nonprecessing systems, a slowly
varying representation is obtained by decomposing the complex quantity
into amplitude and phase as
\begin{equation}
  \label{eq:ModulusAndArgument}
  \ra\, M\, \Psi_{4}^{\ell,m}(\tr,\ra) \define
  A^{\ell,m}(\tr,\ra)\, e^{i\, \phi^{\ell,m}(\tr,\ra)}\,,
\end{equation}
where $M$ is the sum of the initial Christodoulou masses of the two
holes.  We include the factor of $M$ to make the amplitude
dimensionless. (The use of Christodoulou mass is simply a conventional
choice; we could have also used the ADM mass here.)  For these modes,
we extrapolate $A^{\ell,m}$ and $\phi^{\ell,m}$ rather than the real
and imaginary components of $\Psi_{4}^{\ell,m}$, and then we
reconstruct the extrapolated $\Psi_{4}^{\ell,m}(\tr,\infty)$ from
$A^{\ell,m}(\tr,\infty)$ and $\phi^{\ell,m}(\tr,\infty)$.  For other
modes, such as modes in which $\Psi_{4}^{\ell,m}$ is purely real
($m=0$ modes in certain cases), or modes in which the amplitude
$A^{\ell,m}$ passes through zero, the phase $\phi^{\ell,m}$ can be
discontinuous, ill-defined, or numerically difficult to determine.  In
these cases, the real and imaginary parts of $\Psi_{4}^{\ell,m}$ are
extrapolated directly.  

Similarly, it is possible to decompose
  the modes in a corotating frame~\cite{Boyle:2013a}, so that the
  modes show very little time dependence---and in particular,
  essentially no oscillations.  Because they are slowly varying, the
  real and imaginary parts are extrapolated directly for all modes.
  This is the preferred method for precessing systems (although
  it can be applied to nonprecessing systems as well). 

To find the form of the extrapolating functions, we consider standard
expressions for the general formal radiative solution of the Einstein
vacuum equations~\cite{Thorne:1980, BlanchetDamour:1986}.  It turns
out~\cite{BoyleThesis:2008} that we can expect solutions to have
finite-radius behavior in the form of expansions in
  $\lambdabar/\ra$, where $\lambdabar
  = \lambda / 2\pi$ is
 the typical (reduced) wavelength of a given mode.
Because $\lambdabar$ may be several hundred times the mass of the
system, fitting to polynomials in $1/r$ would be numerically
problematic---the fit coefficients for high-order terms would quickly
become very large.  Therefore we fit to polynomials in
$\lambdabar/\ra$, measuring $\lambdabar$ from the frequency of the
$(\ell,m)=(2,2)$ mode.  Note that the purpose of this correction is
only to improve numerical behavior; fitting to $1/\ra$ should produce
the same answer modulo numerical issues.

To reiterate, our extrapolation of nonprecessing systems involves
the following steps.  First, the extracted waveform
$\Psi_{4}^{\ell,m}(t,R)$ is re-expressed as
$\Psi_{4}^{\ell,m}(\tr,\ra)$.  A set of retarded times $\{\tri\}$ is
then constructed---the times at which we want the final
extrapolated waveform.  Next, for each time $\tri$, the waveforms at
each radius are interpolated in retarded time to produce
$\Psi_{4}^{\ell,m}(\tr,\ra)$.  At each time $\tri$, the reduced
wavelength of the $(2,2)$ mode is read off as
$\lambdabar^{2,2} = 1/ \dot{\phi}^{2,2}$, as measured on the
outermost extraction sphere.  The set of finite-radius waveforms is
then fit to a polynomial in $\ra$ using
\begin{subequations}
  \label{eq:ExtrapolatingPolynomials}
  \begin{gather}
    \label{eq:ExtrapolatingPolynomial_A}
    A^{\ell,m}(\tri,\ra) \approx \sum_{k=0}^{N}\,
    A_{(k)}^{\ell,m}(\tri)\,
    \left(\frac{2\, \lambdabar^{2,2}} {m\, \ra} \right)^{k}\,, \\
    \label{eq:ExtrapolatingPolynomial_phi}
    \phi^{\ell,m}(\tri,\ra) \approx \sum_{k=0}^{N}\,
    \phi_{(k)}^{\ell,m}(\tri)\,
    \left(\frac{2\, \lambdabar^{2,2}} {m\, \ra} \right)^{k}\,,
  \end{gather}
\end{subequations}
for oscillatory modes ($m\neq 0$), or
\begin{equation}
  \label{eq:ExtrapolatingPolynomial_ReIm}
  \ra\, M\, \Psi_{4}^{\ell,0}(\tri,\ra) \approx \sum_{k=0}^{N}\,
  \psi_{(k)}^{\ell,0}(\tri)\, \left(\frac{1} {\ra} \right)^{k}\,,
\end{equation}
for non-oscillatory modes ($m=0$), where the $\psi$ are complex
  fitting coefficients.  The time-dependent $k=0$ coefficients are
then used as the amplitude and phase (or for $m=0$, the real and
imaginary parts) of the extrapolated waveform.

Our extrapolation of the precessing system follows the same
  steps, except that the finite-radius data are transformed to the
  corotating frame~\cite{Boyle:2013a} of the outermost extracted data,
  and modes with $m \neq 0$ are fit to polynomials of the form
\begin{equation}
  \label{eq:ExtrapolatingPolynomial_Precession}
  \ra\, M\, \Psi_{4}^{\ell,m}(\tri,\ra) \approx \sum_{k=0}^{N}\,
  \psi_{(k)}^{\ell,m}(\tri)\, \left(\frac{1} {m\, \Omega\, \ra}
  \right)^{k}\,,
\end{equation}
where $\Omega$ is the angular velocity of the
waveform~\cite{Boyle:2013a} as measured on the outermost extraction
sphere.  Modes with $m=0$ are again extrapolated using
Eq.~\eqref{eq:ExtrapolatingPolynomial_ReIm}.  The final result is
then transformed back to the inertial frame.

In all cases, the choice of order of the extrapolating polynomial $N$
is somewhat arbitrary.  Early in the simulation, during the slow
inspiral, $\lambdabar/\ra$ is typically relatively large, so
higher-order terms may still be important.  This means that the
polynomial approximation will converge slowly, suggesting that higher
$N$ may be necessary.  On the other hand, during the merger and
ringdown, $\lambdabar/\ra$ will typically be quite small.  In this
case, we generally find that small $N$ is sufficient; using large $N$
simply over-fits the data.  In practice, the extrapolation procedure
is never strictly convergent, because we have data at a finite number
of extraction radii (typically about $20$), and because these data
inevitably contain some amount of truncation-level noise.  This leads
to extrapolating polynomials that converge for the first few orders,
but eventually begin to diverge because of over-fitting.  The effect
of the choice of $N$ is discussed further in Sec.~\ref{sec:extrap}.

When the wavelength of a given mode is comparable to or larger than
the extraction radii, it is possible that the convergence of
extrapolation will be adversely affected.  In particular, the
convergence for non-oscillatory modes ($m=0$) tends to be slow because
of their large wavelength (except possibly during merger and
ringdown).  Even though we expand such modes in powers of $1/\ra$ in
Eq.~\eqref{eq:ExtrapolatingPolynomial_ReIm} (rather than in powers of
$\lambdabar/\ra$), the coefficients in the expansion will accordingly
be large.  As previously mentioned, this can be numerically
problematic and can limit the accuracy of the extrapolating fits.
Indeed, we will see below that the quality of extrapolation is poor
when $m=0$.

\subsection{Cauchy-characteristic GW extraction}
\label{section:cce}

\begin{figure}
  \includegraphics[width=0.6\linewidth]{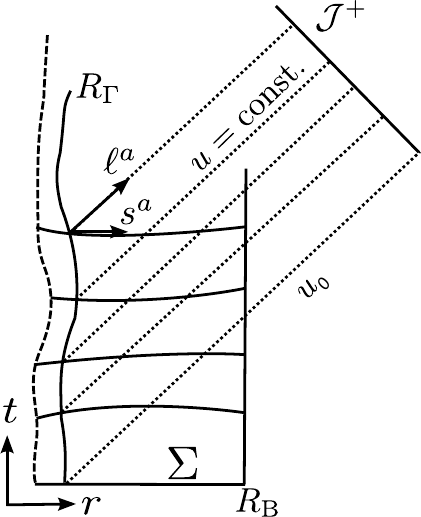}
  \caption{Spacetime diagram illustrating CCE, with two spatial
    dimensions suppressed.  The Cauchy evolution code advances its
    solution of Einstein's equations on successive spatial
    hypersurfaces $\Sigma$ bounded by the outer boundary $R_B$. 
    The wavy, dashed line on the left represents the small-radius
      part of the Cauchy simulation, whose details are not important here.
    The characteristic code advances its solution of Einstein's equations
    on successive null hypersurfaces (labeled $u=$ constant).  It uses
    data from the Cauchy code on the inner boundary (the worldtube
    labeled by $R_\Gamma$) to produce 
    a solution that is valid all the way to $\scri^+$, 
    where gravitational radiation is well defined.
    The characteristic code requires initial data
    on the null surface $u_0$.  
  }
  \label{fig:ccecartoon}
\end{figure}

Cauchy-characteristic extraction (CCE) is a method of computing
gravitational radiation unambiguously and gauge-invariantly at future
null infinity~\cite{Winicour05, Bishop97b, Bishop1998}. This method is by
construction immune to uncertainties associated with
finite-radius and gauge effects.  
The essential idea is to couple a
Cauchy evolution used to evolve the strong field region containing the
black holes to a characteristic evolution evolving the far
gravitational field (see Fig.~\ref{fig:ccecartoon}).  
As opposed to the spatial hypersurface foliation
in Cauchy evolutions, characteristic evolutions are based on null
hypersurface foliations of spacetime.  Without loss of accuracy, this
allows one to apply a compactification of the radial coordinate
to include infinity on the computational grid.
Note that in CCE, the interface between Cauchy and characteristic foliation
is only a one-way boundary. Metric data is propagated from the Cauchy domain onto the characteristic
domain, but not vice versa. The full two-way coupling is achieved by 
Cauchy-characteristic \textit{matching}~\cite{Winicour05, Bishop1998}, 
which has been implemented in the linearized limit
in Ref.~\cite{Szilagyi:2002kv}.

\subsubsection{Characteristic evolutions}

We use the \texttt{PITTNull} characteristic code~\cite{Bishop97b,Winicour05} 
to evolve the gravitational far-field out to future null infinity.
This code uses the framework established by Bondi and 
Sachs~\cite{Bondi1962,Sachs1962,Bishop97b,Winicour05}.
In this framework, the metric is written in the form
\begin{eqnarray}
    ds^2 &=& -\Big(\e^{2\beta}(1+r\hat{W})-r^2h_{AB}U^AU^B\Big)\,du^2
    - 2\e^{2\beta}\,du\,dr \nonumber \\
    & &\qquad{}
    -2r^2h_{AB}U^B\,du\,dy^A+r^2h_{AB}\,dy^A\,dy^B\,,{}
    \label{eq:Bondi-metric}
\end{eqnarray}
where $u=r-t$ is a retarded time coordinate, $r$ is an areal radial 
coordinate, and $y^A$ with $A=2,3$ are angular coordinates.
The variables $\beta$, $\hat{W}$, $U^A$, and $h_{AB}$ are free metric 
coefficients that must satisfy the Einstein equations.
In addition, $h_{AB}$ satisfies $h^{AB}h_{BC}={\delta^A}_C$ 
and $\det(h_{AB})=\det(q_{AB})$, where 
$q_{AB}$ is the unit sphere metric. 
In the \texttt{PITTNull} code, angular components $A$ are represented by 
means of complex spin-weighted scalars:
\begin{equation}
J\equiv q^Aq^B h_{AB}\,,\qquad U\equiv q_A U^A\,, 
\end{equation}
where $q^A$ is a complex dyad satisfying $q^{A} = q^{AB}\,q_{B}$,
$q^{A}\,q_{B}=0$, and $q^{A}\bar{q}_{A}=2$.

Recasting the Einstein equations in terms of the line element above results 
in a set of hypersurface equations, evolution equations, and constraint 
equations.
The hypersurface and evolution equations are solved to determine the 
metric variables $\beta$, $U$, $\hat{W}$ and $J$ between a 
worldtube $\Gamma$ at a radius $R_\Gamma$
and future null infinity $\scri^+$.
To place future null infinity on the computational grid, a compactified 
radial coordinate $x(r)= r/(R_{\Gamma}+r)$ is introduced.

On the worldtube $\Gamma$,
inner boundary data in the form of the metric 
coefficients $\beta$, $U$, $\hat{W}$ and $J$ must be supplied.
Following the prescription of Ref.~\cite{Bishop1998}, 
these quantities are
obtained via a transformation of 
metric data produced by the 
Cauchy evolution (see below).
In addition, the
metric variable $J$ is required on 
the initial null hypersurface. Currently, there exists no solution 
for binary black hole initial data
for the characteristic system. 
Instead, we impose 
a reasonable approximation: 
we use the value of $J$ 
obtained from metric data on the initial Cauchy hypersurface at the worldtube,
and we smoothly blend $J$ to zero on the initial null hypersurface 
so that $J|_{\scri^+}=0$~\cite{Babiuc:2010ze}.
Note that for conformally flat Cauchy initial data, this corresponds to $J=0$ 
everywhere on the initial null hypersurface (see Ref.~\cite{Bishop:2011iu} 
for a discussion).  We have also tried
setting $J=0$ on the initial null hypersurface for a case with 
non-conformally
flat Cauchy initial data (case 4 in Table~\ref{tab:models}, described below).
We find that, at least in this case, it makes no significant 
difference to any of the results whether $J=0$ or $J$ is smoothly blended
to $J|_{\scri^+}=0$.

This choice of characteristic initial data will in general be 
inconsistent with the Cauchy initial data: 
the time evolution of Cauchy initial data 
in the region $R>R_\Gamma$ yields a 
solution on the outgoing initial null hypersurface 
(see Fig.~\ref{fig:ccecartoon}), and this solution
does not generally agree with the
supplied characteristic initial data there.
In the error analysis in 
Sec.~\ref{sec:cceFiniteRadius}, we refer to 
the associated waveform uncertainty as 
the ``CCE initial-data error''.

The characteristic equations are solved on a finite difference grid 
consisting of $N_x$ radial points that discretize the compactified radial 
direction.
For each radial point, the angular discretization of $S^2$ consists of two 
overlapping stereographic patches. 
Each patch contains $N_{\rm ang}$ 
points per angular 
direction.
The two angular patches use circular boundaries to 
eliminate noise from patch corners~\cite{BabiucEtAl2008}.

As detailed in Ref.~\cite{Babiuc:2010ze}, 
the radial and time directions are evolved
using second-order finite difference derivatives together with a 
second-order null-parallelogram integration algorithm 
(see Ref.~\cite{Reisswig:2012ka}
for a new full fourth-order algorithm with spectral angular derivatives).
The angular derivative operators are discretized by fourth-order finite 
differences. Interpatch boundary data are
 obtained via fourth-order 
interpolation.

\subsubsection{Worldtube boundary data}
\label{sec:cce-worldtube}

We obtain boundary data from 3+1 Cauchy
metric data as described analytically 
in Ref.~\cite{Bishop1998} and as implemented 
in Refs.~\cite{Reisswig:2009rx, Babiuc:2010ze}.
We define the worldtube $\Gamma$ as a time succession of spheres of constant 
coordinate radius $R_\Gamma=\sqrt{x^2+y^2+z^2}$, with surface 
normal $s^\alpha$ (see Fig.~\ref{fig:ccecartoon}). 
On $\Gamma$, we construct outgoing null rays $\ell^\alpha$ that induce the 
null foliation.
As detailed in Refs.~\cite{Bishop1998, Reisswig:2009rx, Babiuc:2010ze}, 
the transformation from Cauchy to 
characteristic metric data requires two steps.
The first step involves transformation of the 4-metric from a Cartesian 
to an affine null coordinate system. The second step involves transformation 
of the affine 4-metric
to the characteristic Bondi coordinate system $(u,r,y^A)$. 
The intermediate transformation step to the affine coordinate system is 
necessary since the areal radius of the Bondi coordinates can only be 
computed once angular metric components are known.

The characteristic code requires Cauchy
metric data in the form of the 
spherical harmonic modes of 
the 3-metric $g_{ij}$, lapse $\alpha$, shift $\beta^i$, 
and their radial and time derivatives.
In the evolutions we have performed using \texttt{SpEC}
(see Sec.~\ref{sec:binary-black-hole} below), we decompose the 
required quantities into modes up to $\ell=16$.

\subsubsection{Wave extraction in Bondi gauge}
\label{sec:cce-wave-extract}

We extract waveforms at $\scri^+$ using the methods described 
in Refs.~\cite{Bishop97b,BabiucEtAl2008}.
The original wave extraction method of Ref.~\cite{Bishop97b} 
computes the gravitational news function $\mathcal{N}$, which in Bondi gauge
is related to the metric component $J$
by $\mathcal{N}=J_{,ur}$. 
An alternative and independent method computes the Weyl 
scalar $\Psi_4$ \cite{BabiucEtAl2008}, which
is related to the news function by $\Psi_4=\mathcal{N}_{,u}$. 
Note that this last relation is not used in the characteristic
code; the two quantities $\Psi_4$ and $\mathcal{N}$
are computed independently.
It is also possible to directly extract the 
strain $h$ at $\scri^+$ as well, and this
could potentially remove the problems associated with time 
integration of $\Psi_4$ or $\mathcal{N}$~\cite{Reisswig:2010di}.
An algorithm to accomplish this has recently been implemented~\cite{Bishop:2013}, but
is not used here.

During a simulation, the gauge at $\scri^+$ is induced by the boundary 
data at the worldtube, and the assumption of Bondi gauge generally does 
not hold.
As detailed in Refs.~\cite{Bishop97b,BabiucEtAl2008}, it is necessary to 
apply a transformation from the induced gauge, denoted by 
coordinates $(u,r,y^A)$, to
Bondi gauge, denoted by coordinates $(u_B, r_B, y^A_B)$.
The code presented in Refs.~\cite{Bishop97b,BabiucEtAl2008} applies 
the relevant 
transformation to spatial Bondi gauge $(r_B,y^A_B)$ and computes the Bondi 
time $u_B(u,y^A_B)$ 
as a function of coordinate time $u$ and angular coordinates $y^A_B$. 
In a final step,
it is necessary to make the transformation $u\rightarrow u_B$ to constant 
Bondi time $u_B(y^A_B)=\rm{const.}$ by means of time interpolation 
at
each point on the sphere at $\scri^+$ \cite{Reisswig:2009rx}. 

\subsubsection{Convergence order}
\label{sec:cce-convergence-order}

The characteristic evolution algorithm of the \texttt{PITTNull}
code is expected to exhibit at least 
second-order convergence (see, e.g., Ref.~\cite{Reisswig:2006nt} for tests 
with linearized solutions).
In combination with the algorithm for obtaining worldtube boundary data 
from a Cauchy evolution (Sec.~\ref{sec:cce-worldtube}), 
however, we observe first-order convergence in certain 
quantities~\cite{Reisswig:2009rx,Babiuc:2010ze}.
This may be due to a term at the worldtube which is only known to first order. 
In addition, the numerical algorithm for evaluating $\Psi_4$ 
at $\scri^+$ (Sec.~\ref{sec:cce-wave-extract}) involves a large number of 
terms, some of
them including one-sided finite difference derivatives. As noted 
in Ref.~\cite{Babiuc:2010ze}, 
the convergence order may be negatively affected by this, 
particularly for 
quantities measured at $\scri^+$.


\section{Binary black hole simulations}
\label{sec:binary-black-hole}

\begin{table*}
  \caption{Parameters of BBH runs. Columns indicate mass ratio $q$, 
    dimensionless spins
    $\chi_1,\chi_2$, type of initial data, gauge conditions, 
    number of orbits before merger, 
    initial orbital eccentricity,
    and the initial gravitational-wave frequency $M\omega_{\text{ini}}$ 
    of the $(2,2)$ mode.
  }
  \label{tab:models}
  \begin{ruledtabular}
    \begin{tabular}{ccccccccc}
      case & $q$ & $\chi_1$ & $\chi_2$ & ID & gauge & orbits & ecc
      & $M\omega_{\rm ini}$ \\
      \hline
      $1$ & $1$ & $0$ & $0$ & CF & F$\rightarrow$W  & 16 & $5\times10^{-5}$
      & $0.034$ \\
      $2$ & $1$ & $0$ & $0$ & CF & F$\rightarrow$H$\rightarrow$DH 
      & 16  & $5\times10^{-5}$ & $0.034$ \\
      $3$ & $6$ & $0$ & $0$ & CF & F$\rightarrow$DH & 22 & $4\times10^{-5}$ 
      & $0.038$ \\
      $4$ & $3$ & $(0.7,0,0.7)/\sqrt{2}$ 
      & $(-0.3,0.3,0)/\sqrt{2}$ & SKS & F$\rightarrow$DH 
      & $26$  & $1\times10^{-3}$ & $0.032$ \\
    \end{tabular}
  \end{ruledtabular}
\end{table*}

In this section we describe the binary black hole (BBH) simulations
that we use for comparing wave extraction techniques.  All simulations
were performed using the Spectral Einstein Code
(\texttt{SpEC})~\cite{SpECwebsite} described in Refs.~\cite{Szilagyi:2009qz,
  Lovelace:2010ne,Buchman:2012dw,%
  Hemberger:2012jz,Ossokine:2013zga} and references therein.
This code evolves a first-order representation~\cite{Lindblom2006} of
the generalized harmonic
system~\cite{Friedrich1985,Garfinkle2002,Pretorius2005c} with
constraint damping~\cite{Gundlach2005,Pretorius2005c,Lindblom2006}.
Outgoing-wave boundary
conditions~\cite{Lindblom2006,Rinne2006,Rinne2007} designed to
preserve the
constraints~\cite{Stewart1998,FriedrichNagy1999,Bardeen2002,Szilagyi2002,%
  Calabrese2003,Szilagyi2003,Kidder2005} are imposed at the outer
boundary.  Interdomain boundary conditions are enforced with a penalty
method~\cite{Gottlieb2001,Hesthaven2000}.

We consider four simulations, which are listed in
Table~\ref{tab:models}.  The first two are equal-mass non-spinning
binary simulations that have identical initial data but different
gauge conditions; these are used to test the gauge dependence of the
two wave-extraction methods
in Section~\ref{sec:errors-from-diff}.  
Case 1 is described in
Ref.~\cite{Scheel2009}, and Case 2 is the $q=1$ run discussed in
Refs.~\cite{Buchman:2012dw,PanEtAl:2011}. Case 3 is a BBH with no spin
but with a mass ratio of 6, and is the $q=6$ run discussed in
Refs.~\cite{Buchman:2012dw,PanEtAl:2011}. Case 4 is a generic,
precessing BBH with a mass ratio of 3, and spins on both holes in
generic directions; this simulation is new and has not been presented
elsewhere.

In the generalized harmonic system, the gauge is chosen by freely
specifying four gauge source functions $H_a$.  The simulations in
Table~\ref{tab:models} utilize several different gauge choices.  For
Case 1, $H_a$ is fixed (F) in the corotating frame during inspiral
and smoothly transitions to a solution of an auxiliary wave equation
(W) of the form $\nabla^c \nabla_c H_a=\ldots$ during the plunge and
ringdown.  The gauge used in Case 1 is described in detail in
Ref.~\cite{Scheel2009}.  Case 2 begins with the same fixed gauge as
Case 1, but transitions smoothly to harmonic (H) gauge $H_a=0$ very
quickly (after about $t\sim 40M$) and remains in harmonic gauge
throughout the inspiral.  It then transitions to the damped harmonic
(DH) gauge~\cite{Lindblom2009c,Choptuik:2009ww,Szilagyi:2009qz} of
Ref.~\cite{Szilagyi:2009qz} before merger, and maintains the DH gauge
through merger and ringdown.  Case 3 uses the same fixed gauge as Case
1 during the inspiral, and transitions to damped harmonic gauge about
1.5 orbits before merger.  Case 4 uses fixed gauge for only the first
$t\sim 40M$ of the inspiral, and transitions directly to damped
harmonic gauge for the remainder of the simulation.

The simulations in Table~\ref{tab:models}
employ two different methods of
constructing initial data.  For the non-spinning cases, we use
conformally flat data, as described in Ref.~\cite{Boyle2007}.  For the
spinning, precessing case we use superposed Kerr-Schild
data~\cite{Lovelace2008}.  Both of these methods 
can produce
astrophysically relevant initial data, but the superposed Kerr-Schild
method is more flexible and (for example) allows construction of
initial data with higher
spins~\cite{Lovelace2008,Lovelace:2010ne,Lovelace:2011nu}.

For all cases in Table~\ref{tab:models}, the initial orbital parameters
are adjusted via the iterative method of
Refs.~\cite{Pfeiffer-Brown-etal:2007,Buonanno:2010yk} so as to reduce
the orbital eccentricity of the binary.  In addition, all of the
simulations were done at multiple numerical resolutions 
in order to
provide a means of estimating Cauchy error.


\section{Estimating errors in waveforms}
\label{sec:EstimatingErrors}

A main goal of this paper is to estimate the gauge-related error in 
extrapolated
waveforms by comparing to CCE waveforms, which are gauge invariant.
In order for this comparison to be meaningful, 
we must first estimate the other sources
of error in the numerical waveforms.

We first consider 
the numerical truncation error of the Cauchy 
simulation (``Cauchy error''); this contributes to both 
extrapolated and CCE
waveforms.  For waveforms extrapolated to infinity, we also
estimate the uncertainty introduced by the extrapolation procedure.
For CCE waveforms, we estimate two sources of error in addition to
Cauchy error: the numerical truncation error of the characteristic
evolution and the error associated with the location of the
CCE extraction
worldtube.  The latter error
is due to incompatibility of 
the Cauchy solution and
the data chosen on the initial
null hypersurface of the characteristic code.

We do not estimate the error associated with imperfect
outer-boundary conditions in the Cauchy simulation.  This error has
previously been estimated~\cite{Scheel2009,Buchman:2012dw} by comparing 
otherwise-identical Cauchy simulations with the outer boundary placed
at different locations; this outer-boundary error was found to be 
comparable to or smaller than the Cauchy error. 

For most of this section, we concentrate on errors in the amplitude
and phase of the waveform, as these are the errors most often quoted
by the numerical relativity community.  However, in some cases the
phase of a waveform can become ill-defined.  Therefore, in
Section~\ref{sec:altern-error-meas} we consider alternative error
measures.

\subsection{Waveform Alignment}
\label{sec:waveform-alignment}

Our error estimates for a given (complex) waveform $\psi(t)$ are
obtained by computing the difference between two versions of that
waveform, $\psi_A(t)$ and $\psi_B(t)$, that are generated by slightly
different methods (for instance extrapolation vs. CCE, 
or two different numerical resolutions).  
In matched filtering, the procedure for comparing a
signal waveform against a template 
waveform 
includes a global time and phase shift of the template in order
to best match the signal.  These shifts effectively account for the
arrival time of the signal and the orbital phase at that time.
Therefore, when computing the difference between two waveforms
$\psi_A(t)$ and $\psi_B(t)$ that might ultimately be used as 
templates, it is
appropriate to 
likewise introduce a global time and phase shift between
$\psi_A(t)$ and $\psi_B(t)$, which are 
chosen to minimize some measure
of the difference between the waveforms.  This procedure is 
referred to as waveform alignment.

Waveform alignment in matched filtering is done implicitly by Fourier
transforming and working in the frequency domain. The measure of
comparison is typically an overlap integral that includes the noise
spectrum of the detector \cite{Damour98, Flanagan1998a}.  
The integral and alignment may be
done simultaneously by inverse Fourier transforming the integrand,
taking the absolute value, and finding the maximum value as a function
of time.  
In this paper we choose instead to work in the time domain,
and we do not include noise from a specific detector.

For nonprecessing systems, we use an alignment procedure described in
Ref.~\cite{Boyle:2008}, in which $\psi_A(t)$ is given a time shift
$\Delta t$ and a phase shift $\Delta\Phi$ that are chosen to minimize
\begin{equation}
  \label{eq:AlignmentMinimization}
  \Xi(\Delta t, \Delta \Phi) \coloneqq \int_{t_{1}}^{t_{2}}\, \left(
    \phi_{A}(t) - \phi_{B}(t+\Delta t) - 2 \Delta \Phi \right)^{2}\, d
  t\,,
\end{equation}
where the waveform phases are defined according to
Eq.~\eqref{eq:ModulusAndArgument}.  We choose the range $[t_1,t_2]$
to be early in the inspiral, but late enough to avoid the junk
radiation, and wide enough to average over numerical noise
($t_2-t_1>700M$, where $M$ is the sum of the Christodoulou masses of
the two holes).  We determine the phase and time offsets $\Delta\Phi$
and $\Delta t$ by matching only the $(\ell,m)=(2,2)$ spin-weighted
harmonic modes of 
$\psi_A(t)$ and $\psi_B(t)$; we then use the same
$\Delta t$ and $\Delta\Phi$ (the latter scaled by $m$ for each mode)
to shift all other spin-weighted harmonic modes $(\ell,m)$.

This method is a special case of the more general one needed for
precessing systems.  For precessing systems, the alignment must apply
an arbitrary rotation rather than the simple one shown
above~\cite{SchmidtEtAl:2011, OShaughnessyEtAl:2011, Boyle:2011gg,
  OchsnerOShaughnessy:2012, SchmidtEtAl:2012, Boyle:2013a}.  The
waveform modes transform just as ordinary spherical harmonics do under
rotations, by application of the Wigner $\mathfrak{D}$
matrices.\footnote{In the case of nonprecessing systems, the symmetry
  allows us to pick out a preferred direction: the axis of rotation,
  which we choose to coincide with the $z$ axis.  For a rotation about
  the $z$ axis through an angle $\gamma$, the Wigner matrices simplify
  to $\mathfrak{D}^{(\ell)}_{m',m} = \exp[i\, m\, \gamma]\,
  \delta_{m',m}$, which is why we simply multiply the modes by
  $\exp[i\, m\, \Delta \phi/2]$ in the nonprecessing case.}
Reference~\cite{Boyle:2013a} describes the method we use for achieving
this alignment in the precessing case.  Essentially, the corotating
frame of each waveform is found.  Because these frames are physically
and geometrically meaningful measures of the waveform, it is
meaningful to compare them.  We can define a phase difference between
the two frames using the logarithms of their orientations, which are
represented as unit quaternions $\mathbf{R}_{A}$ and $\mathbf{R}_{B}$.
This phase difference is inserted into an expression that is the
appropriate generalization of Eq.~\eqref{eq:AlignmentMinimization} to
full three-dimensional rotations:
\begin{equation}
  \label{eq:AlignmentMinimizationFour}
  \Xi(\Delta t, \mathbf{R}_{\Delta}) \coloneqq \int_{t_{1}}^{t_{2}}\,
  4 \left \lvert \log \left[ \mathbf{R}_{A}(t)\, \bar{\mathbf{R}}_{B}
      (t + \Delta t) \bar{\mathbf{R}}_{\Delta} \right]
  \right\rvert^{2}\, d t\,,
\end{equation}
which is then minimized over $\Delta t$ and all three degrees of
freedom in the unit quaternion $\mathbf{R}_{\Delta}$.  
Once the optimum rotation is found, it is applied to the
waveform $\psi_{A}(t)$.

For some purposes, alignment need not be done at all when estimating
errors.  For example, when estimating extrapolation error by
subtracting waveforms of different extrapolation orders, 
alignment
is not strictly necessary because all finite-radius waveforms have
already been shifted by $\rt$ when expressing them as functions of
retarded time.  However, our goal is to compare extrapolated and
CCE waveforms, and these 
cannot be compared directly without alignment.
This is because the
extrapolated waveforms are
shifted (in retarded time) by some
$\rt$, whereas the CCE waveforms are 
shifted by
a different offset that depends on the radius of the CCE worldtube.
Therefore, for consistency we estimate every source of error using the
same waveform alignment procedure that is used to compare CCE with
extrapolated waveforms.

Because small time shifts can lead to large accumulated phase differences,
especially for the relatively long waveforms that we consider, it is 
important that the alignment procedure be robust.
For example, we find that aligning waveforms at peak amplitude
is sensitive to small amounts of noise in the waveforms.
For the procedure we use, we have verified
that small changes in the alignment window $[t_1,t_2]$ 
do not affect the results.  
Furthermore, we have repeated all of the analysis in
this paper with an alignment window 
near merger, $[t_\text{merger}-450M, t_\text{merger}+50M]$,
instead of in the early inspiral.  
We find that although this
changes the shapes of error-versus-time plots,
the main results of this paper (namely, the relative magnitudes of
different sources of error) 
are not affected.

\subsection{Cauchy error}
\label{sec:cauchy-error}

To estimate the waveform uncertainties
associated with 
numerical truncation error in the
Cauchy simulations, we use waveforms  
computed at different numerical resolutions. 
Each case in Table~\ref{tab:models} was evolved at three different 
resolutions (not necessarily the same in different cases), which we refer to 
as \emph{low}, \emph{medium}, and \emph{high} resolution.

For a sufficiently fast convergence rate (we expect exponential
convergence for spectral simulations of smooth problems), the
difference between the waveforms at low and medium resolution is a good 
estimate for
the low-resolution Cauchy error, while the difference between 
the medium- and high-resolution
waveforms is a good estimate for the medium-resolution Cauchy error.  
We prefer to err on the side of caution, so we use the difference 
between the medium-
and high-resolution waveforms as an estimate for the high-resolution Cauchy 
error.

\begin{figure}
  \includegraphics[width=1.\linewidth]{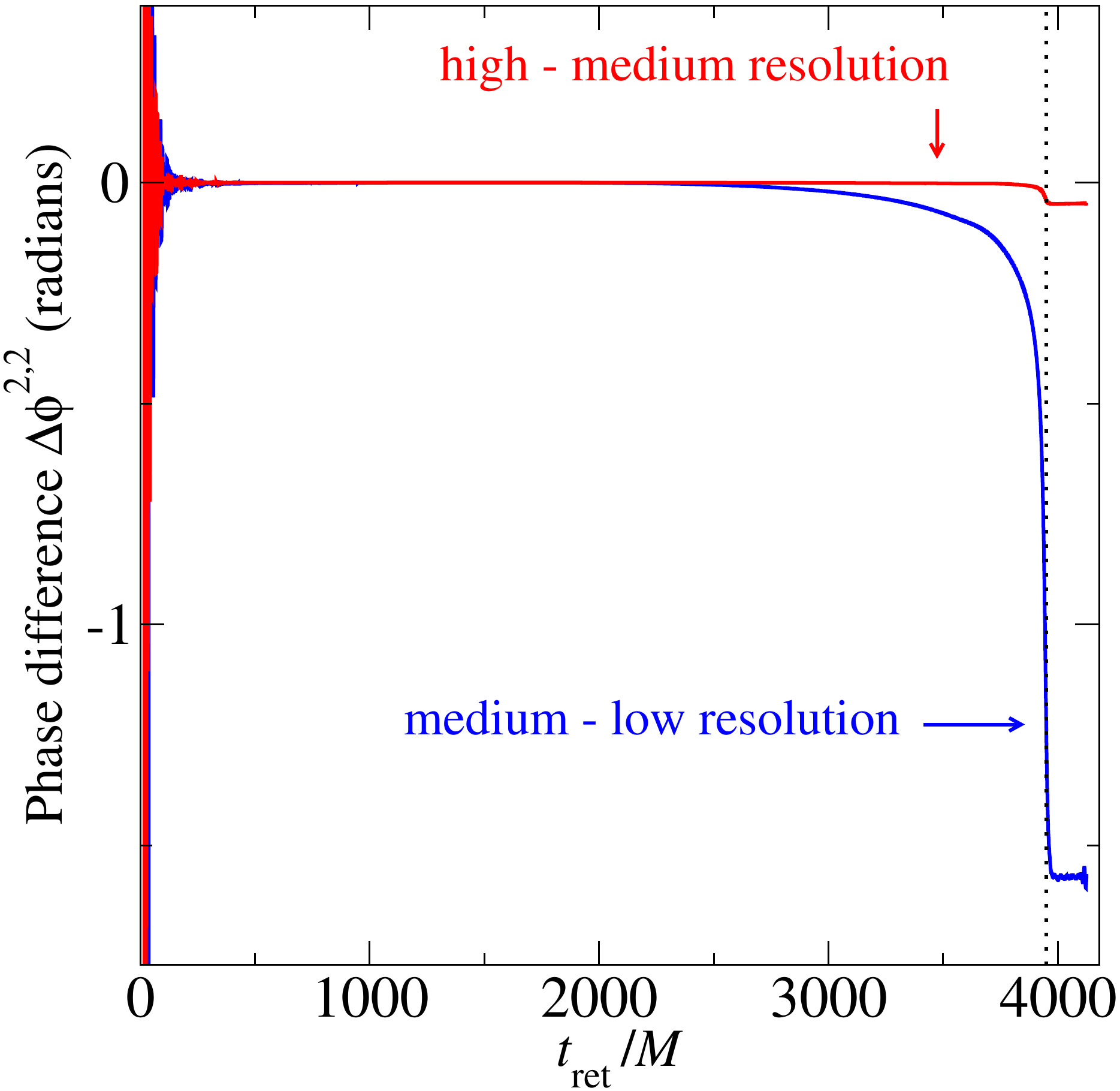}
  \caption{Phase differences in the extrapolated $\Psi_4^{2,2}$ 
    mode between
    successive Cauchy resolutions for simulation
    2 of Table~\ref{tab:models}, using extrapolation 
    order $N=5$.  The
    waveforms at different resolutions have been aligned 
    over the interval
    $[1000M,2000M]$.  The maximum amplitude occurs at $\tr \approx
    3952\,M$, shown here as the dotted vertical line.
  }
  \label{fig:extrapCauchy-phase-1000-2000}
\end{figure}

Figure~\ref{fig:extrapCauchy-phase-1000-2000} shows phase differences
between the $\Psi_4^{2,2}$ modes 
from Case~2 of
Table~\ref{tab:char-grid} at different Cauchy resolutions. 
For each resolution, $\Psi_4^{2,2}$ has been
extrapolated to infinity using $N=5$ in
Eq.~(\ref{eq:ExtrapolatingPolynomial_phi}).  The waveforms for
different resolutions are aligned early in the inspiral before taking
differences.  These phase differences represent the 
estimated Cauchy error in the medium- and high-resolution extrapolated 
$(2,2)$ modes.
Relative amplitude differences between Cauchy resolutions 
show similar convergence.
We compute the Cauchy error for each $(\ell,m)$ mode in the
extrapolated and CCE waveforms in an analogous way.

\subsection{Extrapolation fit error}
\label{sec:extrap}

The extrapolation fit error 
is the uncertainty in the extrapolated
waveform $\Psi_4^{\ell,m}(\tr)$ computed by the procedure of
Section~\ref{section:extrapolation}, given $\Psi_4^{\ell,m}(t,R)$ on
extraction spheres of several radii $R$.  Recall that this procedure
involves fitting the modulus and argument of $\ra\, M\,
\Psi_4^{\ell,m}(\tr,\ra)$ to $N$\textsuperscript{th}-order polynomials
in $\lambdabar/\ra$ (where $\lambdabar$ is the reduced wavelength),
and that the extrapolated result is the coefficient of the constant
term in the polynomial.

\begin{figure*}
  \includegraphics[width=1.\columnwidth]{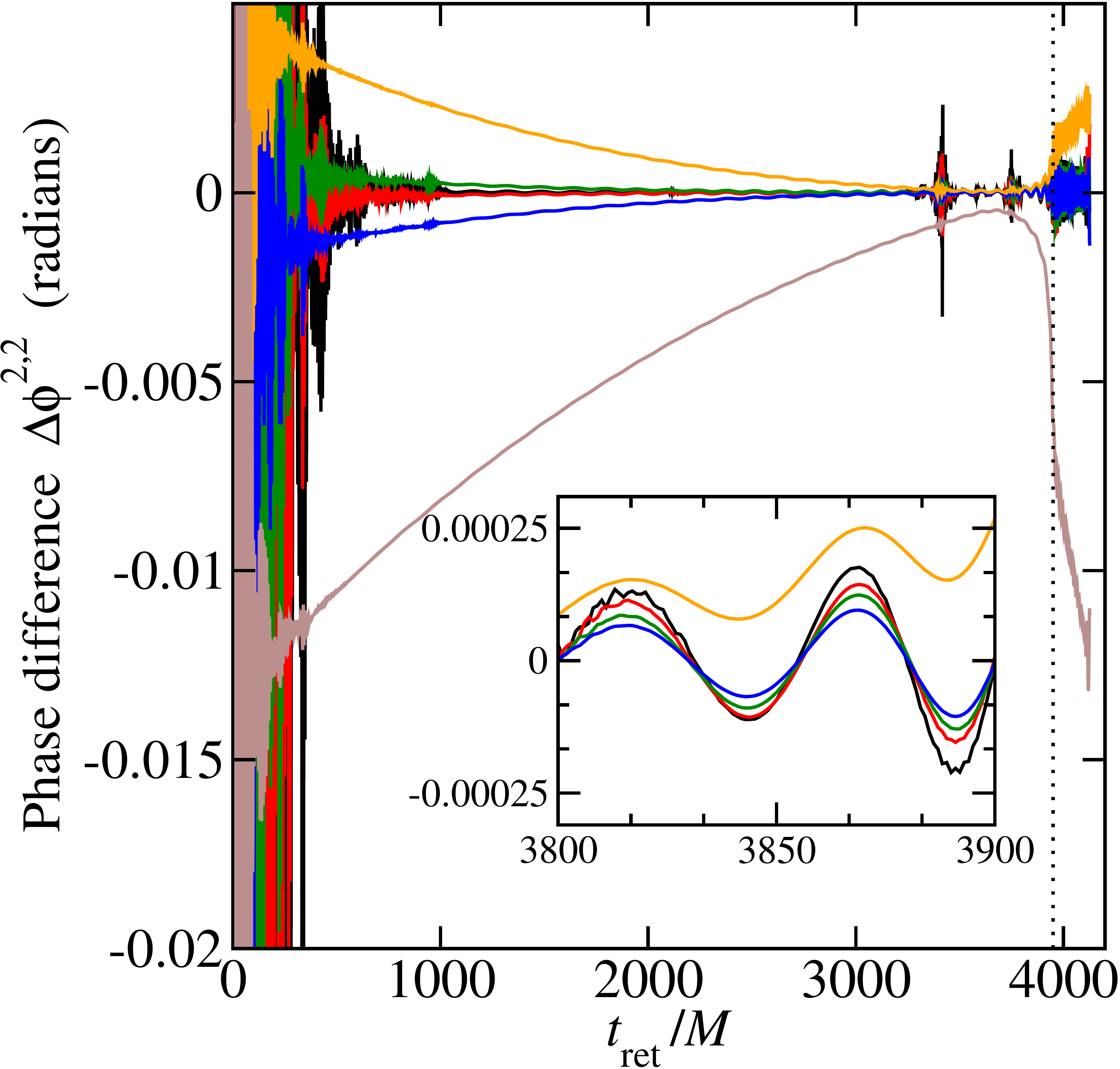}
  \hfill
  \includegraphics[width=1.\columnwidth]{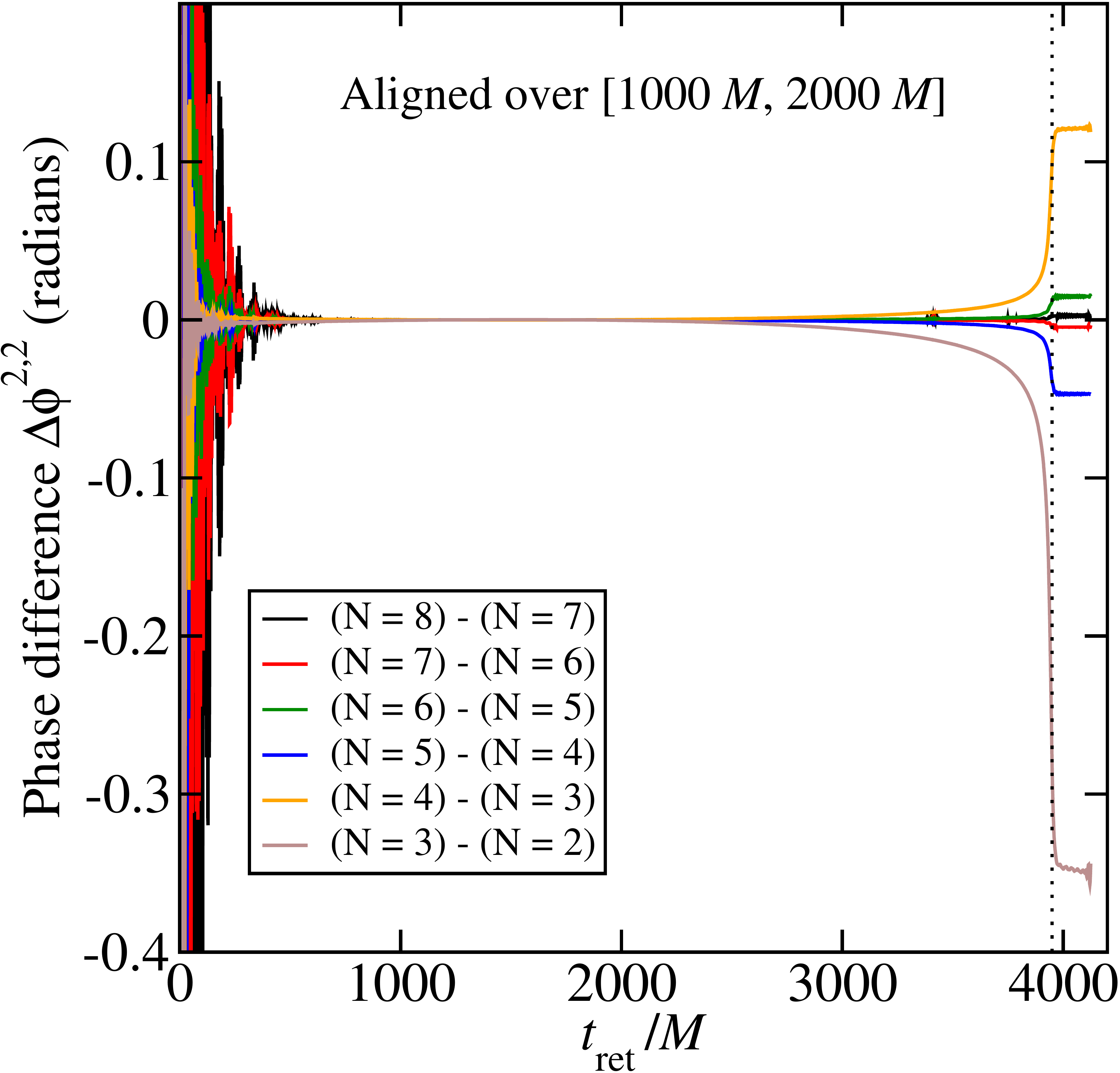}
  \caption{Phase differences in the high-resolution, extrapolated
    $\Psi_4^{2,2}$ waveform between different extrapolation orders
    $N$, for simulation 2 of Table~\ref{tab:models}.  In the left
    panel, no alignment has been done; in the right, each pair of
    waveforms has been aligned over $[1000M,2000M]$.  Each curve is
    the phase difference between a waveform with order $N+1$ and an
    otherwise identical waveform with order $N$.  The maximum
    amplitude occurs at $\tr \simeq 3952M$, shown in each plot as a
    vertical dotted line.  Note the difference in vertical scales.
      Because the frequency is greatest near merger, 
      small time shifts in the alignment window
    produce large phase differences 
    in the plot on the right.
    The two noisy regions in the left panel between
      $\tr \approx 3400-3700$ correspond
      to gauge or grid changes in the Cauchy simulation.
  }
  \label{fig:extrapconv-phase-1000-2000}
\end{figure*}

There are several ways one might seek to estimate this error, 
although we are not aware of a method that
can provide a rigorous estimate.  
One approach is to compare waveforms
extrapolated using different polynomial orders $N$.  
Phase differences from such a comparison
are shown 
in Fig.~\ref{fig:extrapconv-phase-1000-2000}, where
\begin{equation}
  \label{eq:PhaseDifference}
  \Delta \phi^{2,2} \coloneqq \phi^{2,2}_{N} - \phi^{2,2}_{N-1}
\end{equation}
is the error estimate for extrapolation of order $N$.  
The left
panel of this figure shows the comparisons without alignment; the
right panel shows the comparisons after aligning the waveforms
in the early inspiral.
Although the phase errors generally decrease with $N$, the amount of
noise in the extrapolated waveforms increases with $N$.  The noise is
largest in the first few hundred $M$, during the junk-radiation phase,
and near merger at times corresponding to grid or gauge changes in the
Cauchy simulation.

In the non-aligned 
case (left panel), 
the phase differences are well
described as constant multiples of $\lambdabar^{N}$ during the
inspiral. This is presumably due to near-field
effects~\cite{BoyleThesis:2008} 
and is the reason for our choice of 
$\lambdabar/r$ as the extrapolation variable.  
There is very little
work for extrapolation to do near merger, when $\lambdabar$ becomes
comparatively small.  
In fact, as shown in the inset, the differences grow
slightly with increasing order of extrapolation.  Presumably, this is
because the higher-order polynomial coefficients are fitting to
noise in the data when there are no significant physical
features present.

Note the very different vertical scales in the two panels.  
The large phase differences near merger in the right panel are
a result of aligning waveforms
in the early inspiral.  
Alignment introduces
time offsets between waveforms, which 
are necessary to
make the phase and frequency agree as much as possible in the alignment
window.  But even small time offsets in the early inspiral can 
result in large phase differences
near merger, because the frequency is large there.

It may seem that alignment unfairly inflates 
the estimate of extrapolation fit error, but 
the relevant error for many applications is the one that includes 
alignment.  For
example, if we were to attach
the numerical waveform to an analytic
waveform for hybridization, 
we would have to do so at a time when both waveforms are
valid---presumably during the early inspiral.  
The relevant uncertainty in the numerical waveform for this situation is 
the one computed with alignment in the hybridization region.

As an alternative measure of the uncertainty, one might consider
the variance
$\sigma_{N}$ of extrapolation 
at order $N$, as inferred from the
least-squares fit to the data.   In a classical model, with
unbiased and uncorrelated errors, the variance gives the standard
error in the fit coefficients.  
But here we have no reason to assume that the errors
are unbiased and uncorrelated.  If we
simply assign equal, arbitrary
errors to the input waveforms
then even 
in the best case, this leaves the
overall scale of $\sigma_N$ arbitrary
(although it would at least provide some
relative measure of goodness of fit).  

Yet another approach is to obtain an error estimate by
Richardson extrapolation instead of simply comparing neighboring
values of $N$.  The idea is to first
estimate the waveform that one would obtain with $N\to\infty$, and
then construct the error for order $N$ by subtracting the order-$N$
waveform from the order-$\infty$ waveform.  This approach, 
and to a lesser extent the
approach used in Fig.~\ref{fig:extrapconv-phase-1000-2000}, assumes
that the extrapolated waveform converges as $N\to\infty$. 
However, the extrapolation series usually begins to
diverge at some order (which is time-dependent), 
as shown in the inset of the left panel of
Fig.~\ref{fig:extrapconv-phase-1000-2000}.  
We can take the difference
between two orders as some kind of measure, but we cannot justify a
rigorous error bound because of the lack of convergence with
extrapolation order $N$.

The above considerations indicate a need for further investigation
into the complicated issue of extrapolation fit error.  For now, we defer
these issues to a future work, and we henceforth choose the simplest
approach of estimating extrapolation fit errors: taking the difference
between two waveforms of successive extrapolation orders.

Finally, there is 
the question of how to choose the value of
$N$ when constructing the nominal extrapolated waveform.  
One must balance the desire for small error in smooth 
regions (such as in the left panel of 
Fig.~\ref{fig:extrapconv-phase-1000-2000}) 
with the desire for low
noise.  For concreteness in this paper we choose $N=5$, 
and we estimate the error as the difference between the 
$N=5$ and $N=4$ waveforms.  
An alternative method would be to vary the extrapolation order $N$ as the
simulation progresses, choosing a large value of $N$ during the smooth
inspiral, and a smaller value of $N$ to reduce the noise in the merger
and ringdown.  We do not consider this refinement here.

\subsection{CCE truncation error}
\label{sec:cce-truncation-error}

The waveform uncertainty associated with numerical truncation error
on the characteristic grid can be estimated by considering
a sequence of
CCE resolutions, which we label $r0,r1,r2$.  
The actual timestep size
$\Delta u$, the number of radial points $N_x$, and the number of
angular points $N_{\rm ang}$ for each of these resolutions 
are listed in Table~\ref{tab:char-grid}.

\begin{table}
  \centering
  \caption{Resolution of the characteristic grid.}
  \label{tab:char-grid}
  \begin{ruledtabular}
    \begin{tabular}{lccc}
      Resolution & $\Delta u$ [M] & $N_x$ & $N_{\rm ang}$ \\
      \hline
      $r0$  & $0.37500$  & $101$ & $41$  \\
      $r1$  & $0.25000$  & $151$ & $61$  \\
      $r2$  & $0.18750$  & $201$ & $81$  \\
    \end{tabular}
  \end{ruledtabular}
\end{table}

Let $\phi_k(\tr)$ denote the phase of a CCE waveform computed with
resolution $k$, and let $\Delta \phi_{k,k+1} \define |\phi_k - \phi_{k+1}|$ be
the phase difference between the waveforms from different resolutions.
If we measure the convergence rate of $\Delta \phi_{k,k+1}$ with increasing
$k$ and find a consistent convergence order, then we can use 
Richardson extrapolation 
to estimate the error in the highest resolution 
(see, e.g., Ref.~\cite{alcubierreBook}).

\begin{figure}
  \includegraphics[width=\columnwidth]{%
    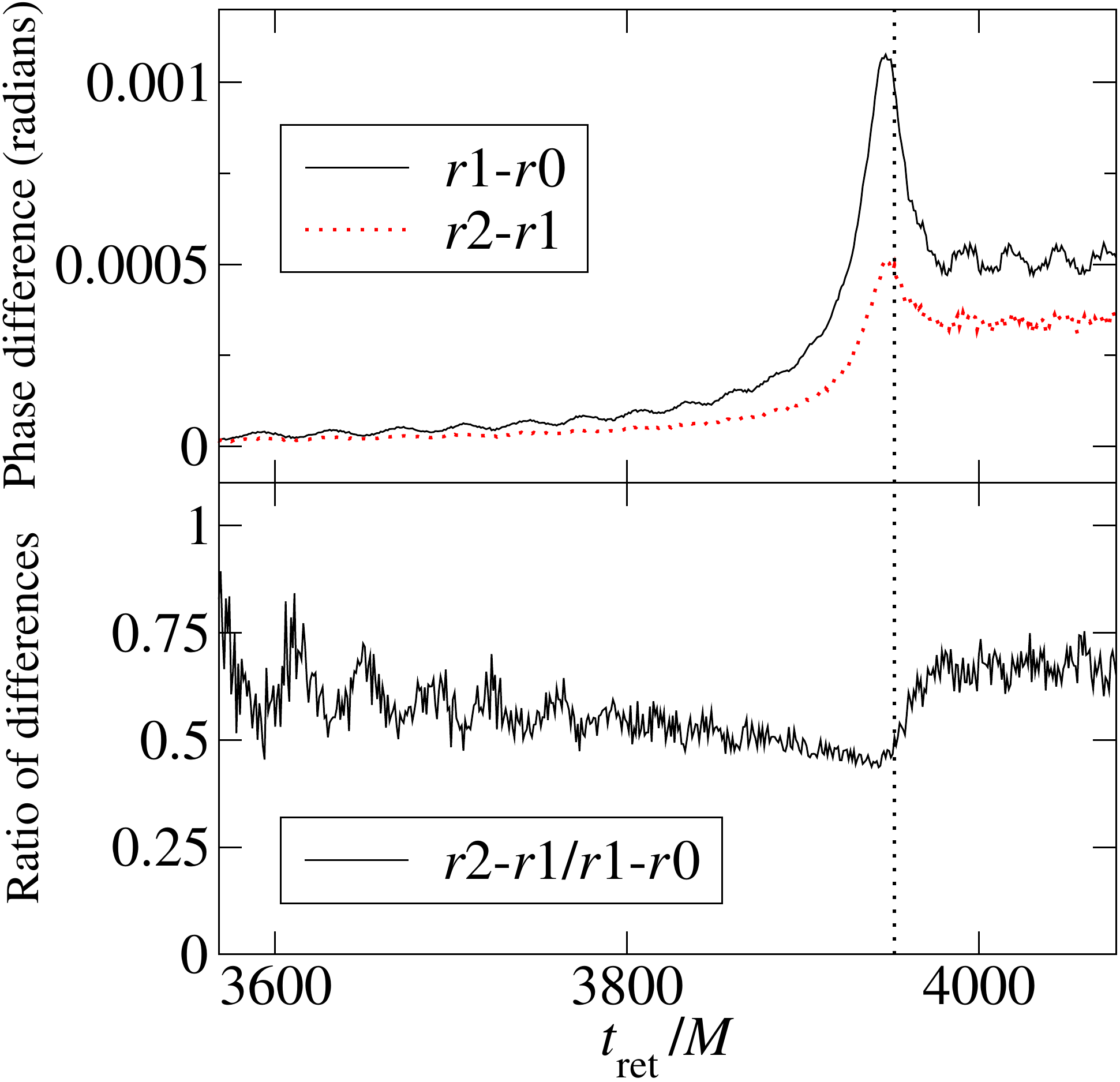}
  \caption{Top: 
    Phase differences 
    between $\Psi_4^{2,2}$ 
    near merger, computed using different CCE resolutions
    (as labeled in Table~\ref{tab:char-grid}), for
    simulation 2 in Table~\ref{tab:models}.
    Bottom: Ratio of the phase differences from the top panel. We find
    roughly first-order convergence, i.e. a ratio of about 0.5.
    All waveforms use the same high-resolution
    Cauchy data and a
    worldtube radius of $R=385M$.  Waveforms are aligned in the
    interval $[1000M,2000M]$.      
    The maximum amplitude 
        occurs at $\tr \simeq 3952M$,
      denoted here by the vertical dotted line.
    }
  \label{fig:cceConv-phase-1400-2400}
\end{figure}

The top panel of Fig.~\ref{fig:cceConv-phase-1400-2400} shows the 
phase differences
$\Delta \phi_{k,k+1}$ for $\Psi_4^{2,2}$ CCE waveforms from simulation
2 in Table~\ref{tab:models}.  To estimate the convergence order, we 
assume that the phase obeys
\begin{equation}
  \phi(h) = \phi(0) + \mathcal{O}(h^n),
\end{equation}
where $h$ represents the grid spacing and $n$ the convergence order.
Note that because time, radial, and angular resolutions are all refined
by the same factor between successive resolutions, we can use a single
measure $h$ here.  We then compute
\begin{equation}
  \frac{\Delta \phi_{12}}{\Delta \phi_{01}} = \frac{h_1^n -
    h_2^n}{h_0^n - h_1^n} + \mathcal{O}(h_0^{n+1}), \label{eq:cceconv}
\end{equation}
where $h0$, $h1$, and $h2$ represent the grid spacings in resolutions
$r0$, $r1$, and $r2$, respectively.
For the values shown in Table~\ref{tab:char-grid}, we expect $\Delta
\phi_{12}/\Delta \phi_{01} = 0.5$ for first-order convergence, 
and $\Delta \phi_{12}/\Delta \phi_{01} = 0.35$ for second-order convergence.

As shown in the bottom panel of
Fig.~\ref{fig:cceConv-phase-1400-2400}, the ratio
of phase differences is roughly consistent with first-order convergence.
We note here that without any alignment of the waveforms, 
the phase convergence of the CCE
data is very cleanly first order.  Aligning early in the inspiral renders the
phase convergence somewhat less uniform.
Doing the same for the amplitude error, we find good second-order
convergence (independent of alignment). 
These measured convergence orders are consistent with the 
theoretically expected convergence discussed in 
Sec.~\ref{sec:cce-convergence-order}.

Assuming first-order
convergence in phase, we estimate the phase error in the $r2$
waveform using
Richardson extrapolation to be 
$3\cdot\Delta \phi_{12}$. Similarly, assuming second-order convergence,
we estimate the (relative)
amplitude error to be 
$9/7 \cdot \Delta A_{12}$.  
Henceforth, we use resolution r2 as the nominal CCE waveform.

\subsection {CCE initial-data error}
\label{sec:cceFiniteRadius}

Waveforms evolved using CCE may depend on the location of the 
characteristic worldtube (the surface labeled $R_\Gamma$ in 
Fig.~\ref{fig:ccecartoon}) for two reasons.
Most significantly, 
the characteristic evolution requires
data on an initial null hypersurface (the surface
labeled $u_0$ in Fig.~\ref{fig:ccecartoon}).
In the simulations we consider, these
data are chosen to be blended to conformally flat
as described in Sec.~\ref{section:cce}.
However, this does not necessarily 
agree with the Cauchy evolution, which may contain
physical backscattered radiation, 
junk radiation,
and ingoing radiation from imperfect outer boundary conditions.
This incompatibility 
is a source of uncertainty in the CCE waveforms.
As the radius of the worldtube is increased, this mismatch 
and the resulting error should decrease.

Another reason one might expect a CCE waveform to depend
on worldtube location is that the length scale of
dynamical features in the spacetime decreases as the worldtube
is moved closer to the source.
Unless there is a corresponding increase in the resolution of the 
characteristic code, one would therefore expect a smaller worldtube radius to
result in larger truncation errors.  However, we find this contribution
to the overall error to be insignificant; 
the estimated error is essentially independent of the 
characteristic-code resolution.

Because of these observations, 
we refer to this error as the ``CCE initial-data error'', even though
we measure it by varying the finite-radius worldtube location.
One method for estimating this error 
is simply to
take the difference
between waveforms computed using two different worldtube radii. This
approach is inadequate because it depends too heavily on which radii are 
chosen.  If the two radii are very near to each other, then 
this would result
in an arbitrarily small estimate.  On the other hand, if the two radii were
very far apart, this method might
yield an incorrectly large estimate of the error.

\begin{figure}
  \includegraphics[width=1.\linewidth]{%
    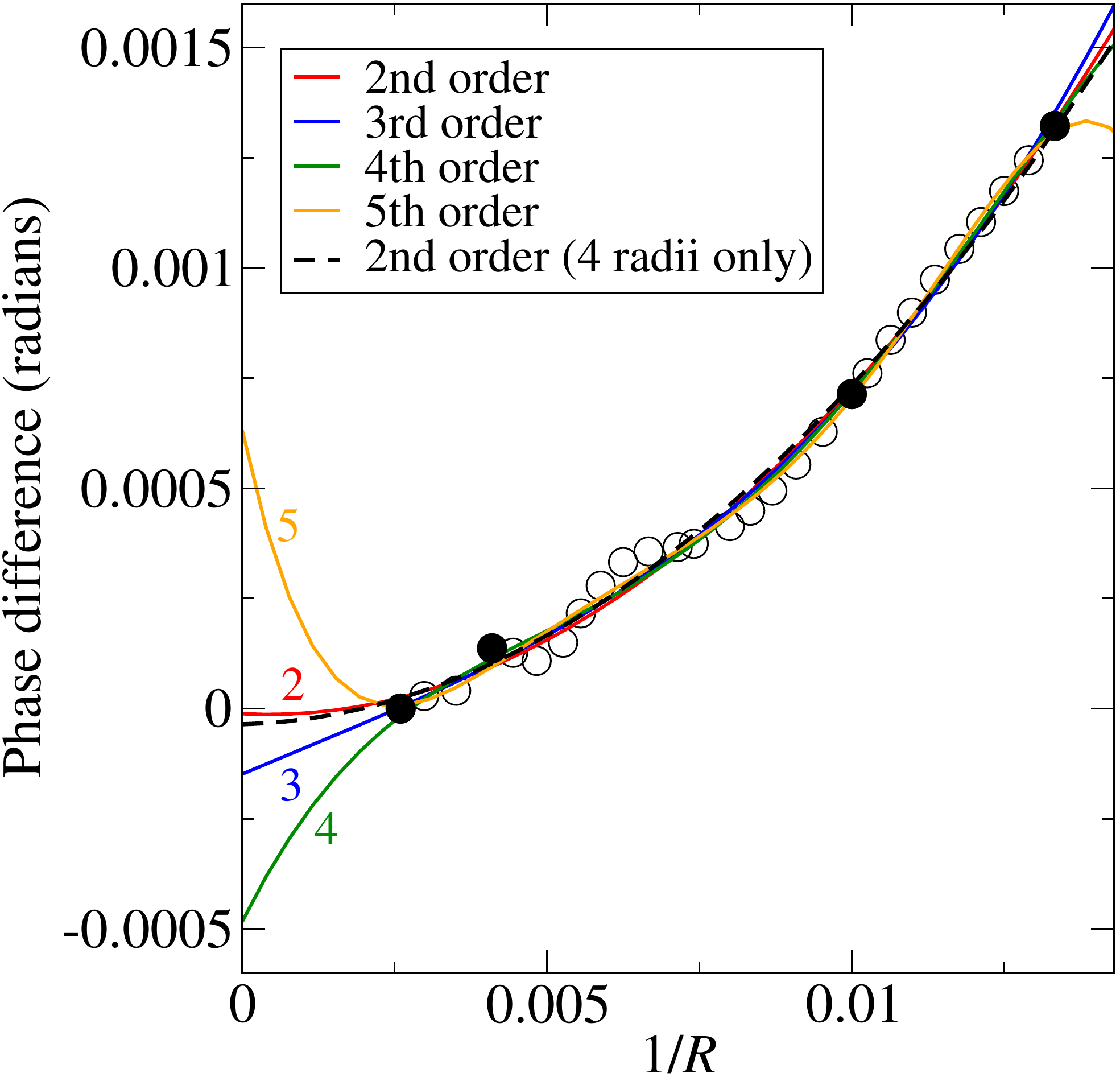}
  \caption{
    Phase differences as a function of inverse worldtube
      radius $R$.
    Each of the 28
    circles (both open and closed) is the 
    phase difference, evaluated at $\tr \simeq 2600M$,
    between the CCE waveform
    $\Psi_4^{2,2}$ computed with worldtube radius $385M$ 
    and the
    same waveform computed with worldtube 
    radius $R$.
    Solid curves are polynomial fits of different orders 
    in powers of $1/R$.
    The four closed circles represent the
    typical worldtube radii used in most of the 
    simulations, and the dashed
    curve shows the second-order fit to just these four 
    points. 
    All waveforms are from 
        the high-resolution run of simulation 2 
        in Table~\ref{tab:models}, using CCE resolution $r2$.  
    Waveform alignment is 
    done using the interval $[1000M,2000M]$.
  }
  \label{fig:cceRadial-phase-times-new-3000}
\end{figure}

For the high-resolution run of simulation 2 in Table~\ref{tab:models},
we have computed CCE waveforms
from 28 different worldtube radii
(ranging from $R=77.5M$ to $R=385M$).
We calculate the phase difference between the waveform 
from each radius and the waveform from the outermost radius, where
the two waveforms are aligned over $[1000 M, 2000M]$.
Figure~\ref{fig:cceRadial-phase-times-new-3000} shows these phase differences
at a particular time ($\tr \simeq 2600M$), plotted against the inverse 
worldtube radius $1/R$.  Note that the outermost
worldtube radius has a phase difference of zero in this plot by definition.
It is immediately evident that
the phase differences decrease predominantly like
$1/R^2$ as $R$ increases.  
We find this same feature at other times
and for relative amplitude differences 
as well as phase differences.

We can estimate the CCE initial-data error at 
each time by fitting such phase
differences to a polynomial in $1/R$ and then extrapolating 
$1/R \rightarrow 0$.  
The
solid curves in Figure~\ref{fig:cceRadial-phase-times-new-3000} show
these fits for polynomials of different orders.  
We see that this extrapolation diverges as the
polynomial order is increased.  This is presumably the same issue 
(overfitting to noisy data) that arises
in waveform extrapolation, as discussed in
Sec.~\ref{sec:extrap}.   Since 
we are interested here only in an
estimate of the CCE finite-radius error and not in 
extrapolating the CCE waveforms
to infinite worldtube radius, we simply choose a quadratic fit.

In the above procedure for estimating CCE 
initial-data error, we
extrapolate CCE phase differences (such as those shown in
Figure~\ref{fig:cceRadial-phase-times-new-3000}) to infinity.
One may ask why we do not instead extrapolate these phase differences
to the outer boundary of the Cauchy simulation.
After all, 
placing the worldtube at the outer boundary
would seem to eliminate any mismatch between characteristic and Cauchy 
initial data.
But imagine a perfect Cauchy code with infinite
resolution, and with perfect
outer boundary conditions so that even with a finite outer boundary,
it exactly reproduces the true solution
of Einstein's equations including effects such as backscatter
and tails.  If the CCE
worldtube were placed at the outer
boundary of this perfect Cauchy domain, 
then there would still be a mismatch between the
(blended to conformally flat)
characteristic initial data and the true solution.
Extrapolating phase differences to infinity estimates the error induced
by this mismatch.

It is important to verify that the procedure for estimating
CCE initial-data error works well when using fewer worldtube
radii, because most of the runs we  
consider have CCE data from only four 
radii.  The dashed black line in
Fig.~\ref{fig:cceRadial-phase-times-new-3000} shows the second-order
fit using only the four radii $R=75, 100, 244, 385 M$ (the four solid
black dots in the figure). 
As can be seen in the figure, this fit 
is quite consistent with the fit 
using all 28 radii.
We find this to be the case
at other times
(not only at the time shown in the figure) 
and
for (relative) amplitude 
differences as well. 
We therefore use
this quadratic extrapolation procedure to estimate the 
CCE initial-data
error, and we 
use the waveform computed from the outermost worldtube 
as the nominal CCE waveform.

\subsection{Alternatives to measuring phase error}
\label{sec:altern-error-meas}

In the previous sections we estimated errors by computing phase and
amplitude differences between otherwise-identical waveforms 
computed using different resolutions, worldtube radii, 
or extrapolation orders.
However, phase errors 
are not always well-defined.  
In this section we illustrate some of
the ways in which phase errors can become difficult to measure, and in
Sec.~\ref{sec:complexmeasure} we describe another error measure that
obviates this difficulty.

\subsubsection {Problems with phase differences}
\label{sec:problems-with-phase}

The phase of a $\Psi_4^{\ell,m}$ mode may become ill-defined
because the amplitude momentarily vanishes, or it
may simply
vary rapidly as the amplitude passes near zero~\cite{Boyle:2011gg}.
The imaginary part of the waveform may
be zero analytically, but at truncation level numerically.
This can cause the phase to change
randomly and discontinuously between
$\epsilon$ and $2\pi-\epsilon$, depending on 
the numerical errors.
Issues like these
can cause trouble even when the waveform at 
$\scri^+$ has a
well-defined phase, 
because the waveforms computed from 
(some of) the worldtube
or extraction radii may 
exhibit such problems.

For example, in the precessing case (simulation 4 in 
Table~\ref{tab:models}), we find that phase differences between 
otherwise-identical CCE waveforms
computed from different worldtube radii sometimes jump by $2\pi$
(similar examples can be found in most cases).
Such jumps enter into
the CCE initial-data error estimate, 
as described in
Sec.~\ref{sec:cceFiniteRadius}, where they can lead to 
estimated phase uncertainties of $\mathcal{O}(\pi)$.
This renders the error estimate much less meaningful 
(although it is at least consistent).

\begin{figure}
  \includegraphics[width=1.\linewidth]{%
    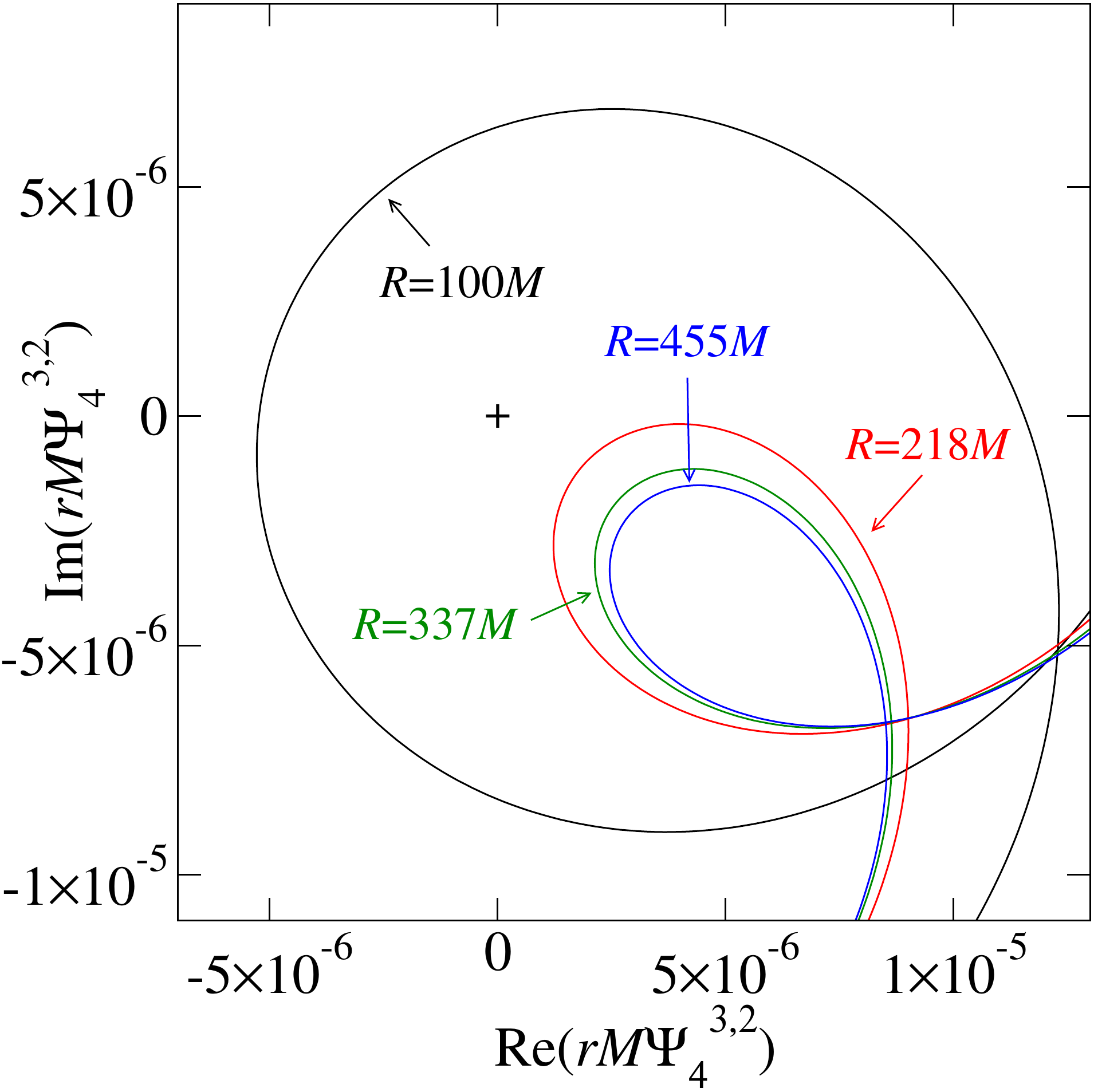}
  \caption{The CCE $rM\Psi_4^{3,2}$ modes
    computed from four different
    worldtube radii, plotted in the complex plane for times in
    the approximate interval
    $[5495M,5565M]$, for case 4 in Table~\ref{tab:models}.
    The trajectory of
        each waveform is traversed in a clockwise
      direction with time, entering on the right and leaving on the bottom.
      The maximum amplitude (of the $(2,2)$ mode) 
      occurs at $\tr \simeq 6722M$.
    The waveform from each radius has been aligned 
    over $[1000M,2000M]$
    with the waveform from the largest radius
    ($455M$).  A jump of $2\pi$ in
    the phase differences 
    between the waveforms
    occurs because only the waveform from $R=100M$ encircles
    the origin (shown as + in
    the figure). 
  }
  \label{fig:q3cceReImL03Mp02}
\end{figure}

It is interesting to examine the real and imaginary
parts of the waveform from 
different 
worldtube radii 
near the time of such a jump. 
Figure~\ref{fig:q3cceReImL03Mp02} shows an example of this.  We
plot $\Psi_4^{3,2}$ in the complex plane 
for times corresponding to an observed $2\pi$
jump in phase differences.
Evidently, the 
jump corresponds to 
the trajectories 
of the waveforms from different radii encircling 
the origin a different number of times.
This occurs even though the trajectories
shown in Fig.~\ref{fig:q3cceReImL03Mp02} are clearly converging 
to a nonvanishing amplitude as the worldtube radius is increased.

Similar problems can occur for extrapolation, but they can be even
worse because the nearly discontinuous jumps in phase end up not only
in the error estimate (as in the case of CCE) but in the extrapolated
waveform itself.  Even in mildly precessing cases, one may thus
encounter non-extrapolatable waveforms---at least with the naive
extrapolation algorithm, which extrapolates phase and amplitude
separately according to Eqs.~\eqref{eq:ExtrapolatingPolynomial_A} and
\eqref{eq:ExtrapolatingPolynomial_phi}.  We solve this problem
  for the precessing system by first transforming the data at all
  radii to a common corotating frame (the
corotating frame of the outermost extraction radius) before
extrapolation~\cite{Boyle:2013a}, as described in 
Section~\ref{section:extrapolation}.

The corotating frame method
  also gives rise to another way to measure phase error,
  because the phase information is almost entirely recorded in the
  orientation of the corotating frame.  The phase difference between
  the two frames is described completely\footnote{See Eq.~(19) and
    surrounding discussion in Ref.~\cite{Boyle:2013a}.} by the
  logarithm of the ratio of the two orientations, as in the integrand
  of Eq.~\eqref{eq:AlignmentMinimizationFour}.  This difference is not
  subject to the sudden phase jumps seen above, and is invariant under
  overall rotations of the physical system or the coordinate system.
  This provides a robust and uniform method that can be used in
  nonprecessing and precessing systems alike.  
  However, this definition of phase error applies to an entire
  waveform including all $(\ell,m)$ modes.  We prefer to use an error
  quantity that can be defined separately for each mode, as described
  below.

\subsubsection{Error measure in the complex plane}
\label{sec:complexmeasure}

Motivated by the difficulty of defining phase errors in some generic
BBH simulations, we employ an alternative error measure that is an
$L^2$-norm of the difference between two (complex) waveforms,
integrated over all positions on the sky:
\begin{equation}
  \centering \left\|\Psi_4^A - \Psi_4^B \right\|^2 = \int\limits_{S^2}
  \left|\Psi_4^A - \Psi_4^B\right|^2 \,d\Omega.
  \label{eq:ComplexErrorDef}
\end{equation}
Expanding each waveform in spin-weighted spherical harmonics using
Eq~(\ref{eq:SWSHDecomposition}), and using orthonormality relations,
one obtains
\begin{equation}
  \label{eq:ComplexErrorTotal}
  \left\| \Psi_4^A - \Psi_4^B \right\|^2 = \sum\limits_{\ell,m}
  \left|\Psi_4^{\ell,m}{}^A - \Psi_4^{\ell,m}{}^B \right|^2.
\end{equation}
This quantity could be normalized by 
a norm of the individual
waveforms (computed using the same measure), 
such as $\|\Psi_4^A\| + \|\Psi_4^B\|$.  
However, if one is
interested in comparing errors in a particular spin-weighted
harmonic mode, then the normalization
(which is the same for
each mode) can be neglected. 
In this case,
amplitude and phase errors are combined into
a single measure,
\begin{equation}
  \label{eq:ComplexErrorMeasure}
  \begin{split}
    \Delta_{\ell m}^2 &= 
    |\Psi_4^{\ell,m}{}^A - \Psi_4^{\ell,m}{}^B|^2 \\
    &=     (\Delta A^{\ell,m})^2 
+ 2 A^{\ell,m}{}^AA^{\ell,m}{}^B(1-\cos \Delta
    \phi^{\ell,m}), 
  \end{split}
\end{equation}
which has the advantage of being 
immune to ill-defined phase
errors, as well as properly ignoring phase differences when
amplitudes are small.  It also provides 
the option of combining all errors
for a mode-independent measure.  The sum in
Eq.~\eqref{eq:ComplexErrorTotal} is invariant under overall
rotations of both waveforms, making this a particularly useful
measure in precessing systems (this is true even when
considering a single value of $\ell$).  
Using this measure, we can estimate the various sources of error in the
same way as we did above for phase and amplitude errors.  

\subsection {Combination of errors}
\label{sec:combination}

In the preceding discussion, we concentrated on computing various
error quantities: Cauchy error, extrapolation fit error, CCE truncation
error, and CCE initial-data error.  
In this section we discuss how to
combine these quantities into a single error bar.  Here we still
consider each $(\ell,m)$ mode separately.

In addition to constructing a combined total error
bar for a waveform, we would also like to compare the relative 
magnitudes of the different sources of error.
The above error measures are all time-dependent, so 
we must either
compare them at each 
value of $t$, or we must construct a
time-averaged error measure.  
We choose the latter, and average the
absolute value of each error over an 
interval $[t_1,t_2]$, where $t_1$
represents a time after junk radiation (usually about $500M$), and
$t_2$ represents the time after merger when the amplitude of the
waveform has decayed to truncation level.  
The early-time and
late-time cutoffs avoid portions of the waveform where the phase is
ill-conditioned and difficult to measure, or where the
waveform is unphysical.  The relative 
magnitudes of
these time-averaged errors then allow
us to see at 
a glance how the different sources of error 
compare.

\subsubsection{Error bar for an individual waveform}
\label{sec:combined-error-an}

To determine the uncertainty in an individual waveform, 
we combine the various sources of error using 
an $L_1$ norm---i.e., we add the absolute values of
each source of error.  For independent, normally-distributed
random errors it would be more appropriate to sum the errors in 
quadrature (see, e.g., Ref.~\cite{Taylor1997introduction}).
In the present case, however, we do not know how the 
errors are distributed, and we have no reason to 
expect them to be independent or normally distributed.  
So, we assume the worst case and combine errors by adding magnitudes.

For the uncertainty in a CCE waveform, we combine Cauchy error 
(measured using CCE waveforms),
CCE truncation error, and CCE initial-data error.
Similarly, for the error in an extrapolated waveform, we combine
Cauchy error (measured using extrapolated waveforms)
and extrapolation fit
error.  This error bar is incomplete
for extrapolated waveforms, as it does not include any contribution from gauge error;
we estimate the magnitude of the gauge error in Sec.~\ref{sec:results}
below by comparing extrapolated waveforms with CCE.

\subsubsection{Error bar for difference between CCE and extrapolated
waveforms}
\label{sec:comb-error-diff}

We wish to determine whether a CCE waveform and an extrapolated waveform
agree to within some error bar.  
If they do, then we
can regard the gauge error in extrapolated waveforms as small, and
we can use the extrapolation procedure instead of the more
complicated and computationally expensive CCE procedure 
to obtain waveforms at $\scri^+$.
The estimated error bar for the difference between CCE and extrapolated
waveforms is 
constructed as 
the $L_1$ norm of the  CCE truncation error, 
CCE initial-data error, extrapolation fit error, and Cauchy error. 

Because the CCE and extrapolated waveforms
each have their own Cauchy error, it is not immediately clear
which Cauchy error should enter into the error bar.
Let 
$C^C$ and $C^E$ denote the Cauchy error
determined using CCE and 
extrapolated waveforms, respectively.
Both $C^C$ and $C^E$ 
arise from the same
source: truncation error in the Cauchy simulation.  To define the
Cauchy error for the difference between a CCE and an extrapolated
waveform, we take the average of $C^C$ and $C^E$.

It is not obvious that averaging $C^C$ and $C^E$ is the correct
procedure: the issue is whether they
are correlated.  To
pursue this further, note that there are two contributions to
both $C^C$ and $C^E$.  
The first contribution corresponds to the error made
in determining the motion of the black holes; this 
affects
$C^C$ and $C^E$ in an identical way. The second contribution
corresponds to the error made in propagating waves through the grid;
this affects $C^C$ and $C^E$ differently, because the
extraction radii and the quantities read from the Cauchy code are
different for extrapolated waveforms than for CCE waveforms.  If the
first contribution is dominant, than $C^C$ and $C^E$ are correlated,
so it would be appropriate to use their average. But if the second
contribution is dominant, then $C^C$ and $C^E$ are uncorrelated, so
it would be appropriate to use their sum.

\begin{figure}
  \includegraphics[width=1.\linewidth]{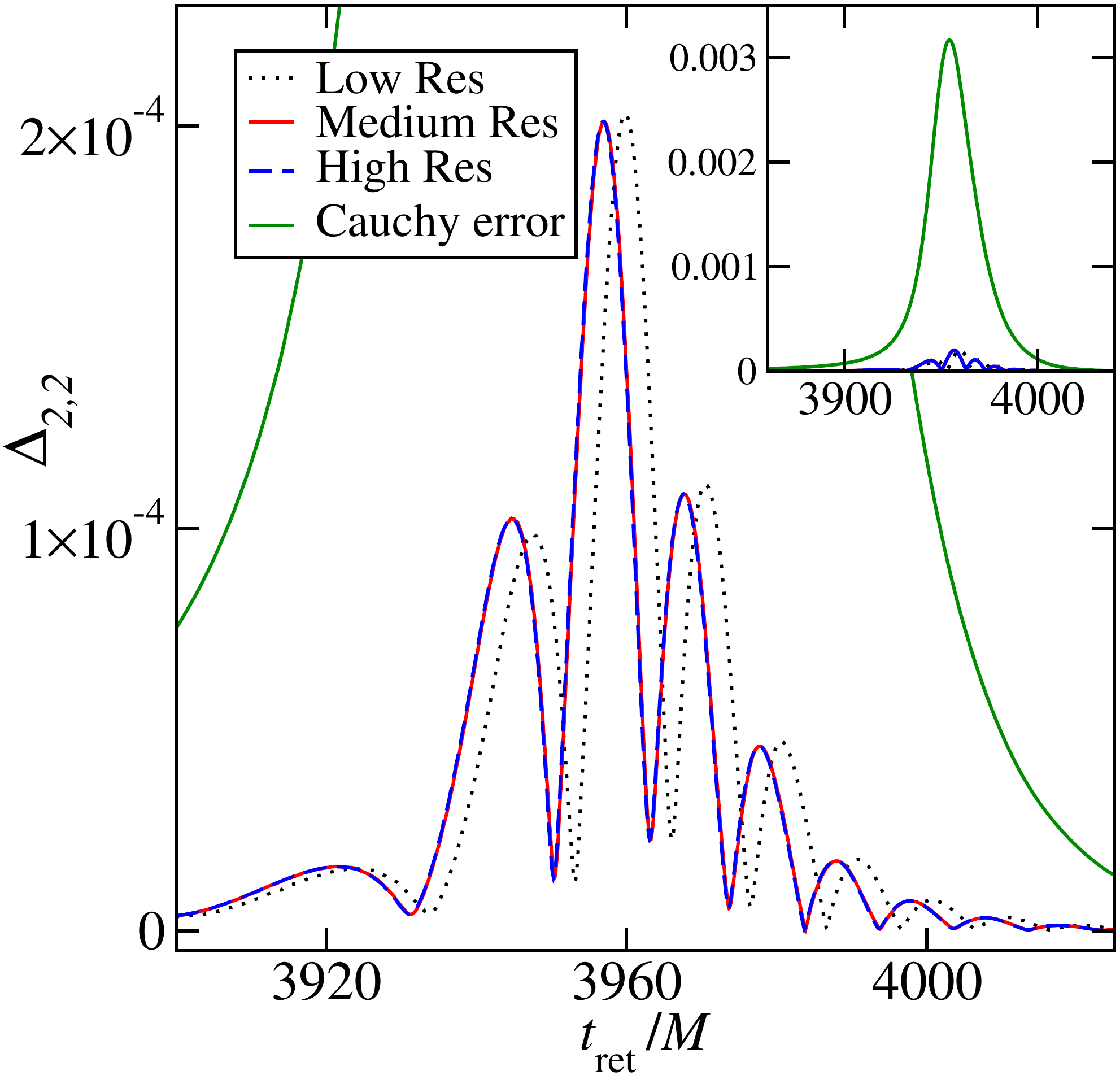}
  \caption{Cauchy-resolution dependence of $\Delta_{\ell,m}$ 
    (cf. Eq.~(\ref{eq:ComplexErrorMeasure}))
    between CCE and extrapolated $\Psi_4^{2,2}$,
    shown near peak amplitude at $\simeq 3952M$.
    for simulation 2 in Table~\ref{tab:models}.
    At each lower Cauchy resolution, the extrapolated waveform is aligned
    with the high-resolution extrapolated waveform.  Then the CCE
    waveform for each Cauchy resolution is aligned with the
    corresponding extrapolated waveform and $\Delta_{\ell,m}$
    is computed. The differences are nearly independent of
    resolution. Also shown (labeled ``Cauchy error'')
    is the difference between the extrapolated $\Psi_4^{2,2}$ 
    waveforms from the high and medium resolutions.
  }
  \label{fig:q01Extrap-CceConv}
\end{figure}

We can determine which part of the Cauchy error is dominant by
plotting the difference between CCE and extrapolated waveforms
taken from a low-resolution simulation, and comparing with the
difference between CCE and extrapolated waveforms taken from a
high-resolution simulation.  Such a plot is shown in
Fig.~\ref{fig:q01Extrap-CceConv}. 
We find that the difference
between CCE and extrapolated waveforms is largely independent of
resolution, indicating that the dominant effect of Cauchy error is to
change the trajectories of the black holes, and that $C^C$ and $C^E$
are highly correlated rather than independent. 
Therefore, 
we are justified in computing
the combined Cauchy error
as the average of $C^C$ and $C^E$, rather than their sum.

Additionally, we find that the difference between CCE and 
extrapolated waveforms
is significantly smaller than the estimated
Cauchy error, as shown in the figure---at least for the $(2,2)$ mode.  
Accordingly, the measures $C^C$ and $C^E$ are not merely
correlated, but are also nearly identical to each other.  This 
continues to hold even for
subdominant modes, for which the Cauchy error can be comparable to the
difference between CCE and extrapolated quantities 
(cf. Fig.~\ref{fig:CauchyErrHist} and the discussion below).

\section{Results}
\label{sec:results}

In this section we compare the relative magnitudes of the various
error quantities for both extrapolated and CCE waveforms.  
We verify the gauge-dependence of
extrapolated waveforms and the gauge-invariance of CCE by
examining waveforms from two physically 
identical simulations performed using 
different gauge conditions.  
For each BBH configuration in Table~\ref{tab:models}, 
we evaluate the quality of extrapolated waveforms by
comparing with CCE waveforms. This allows us to 
determine whether the gauge error in extrapolated waveforms is smaller
than the other sources of error, and hence whether we can justify 
using the extrapolation
method instead of CCE.

\subsection{Is waveform extraction to $\scri^+$ necessary?}
\label{sec:comparefiniterad}
\begin{figure}
  \includegraphics[width=1.\linewidth]{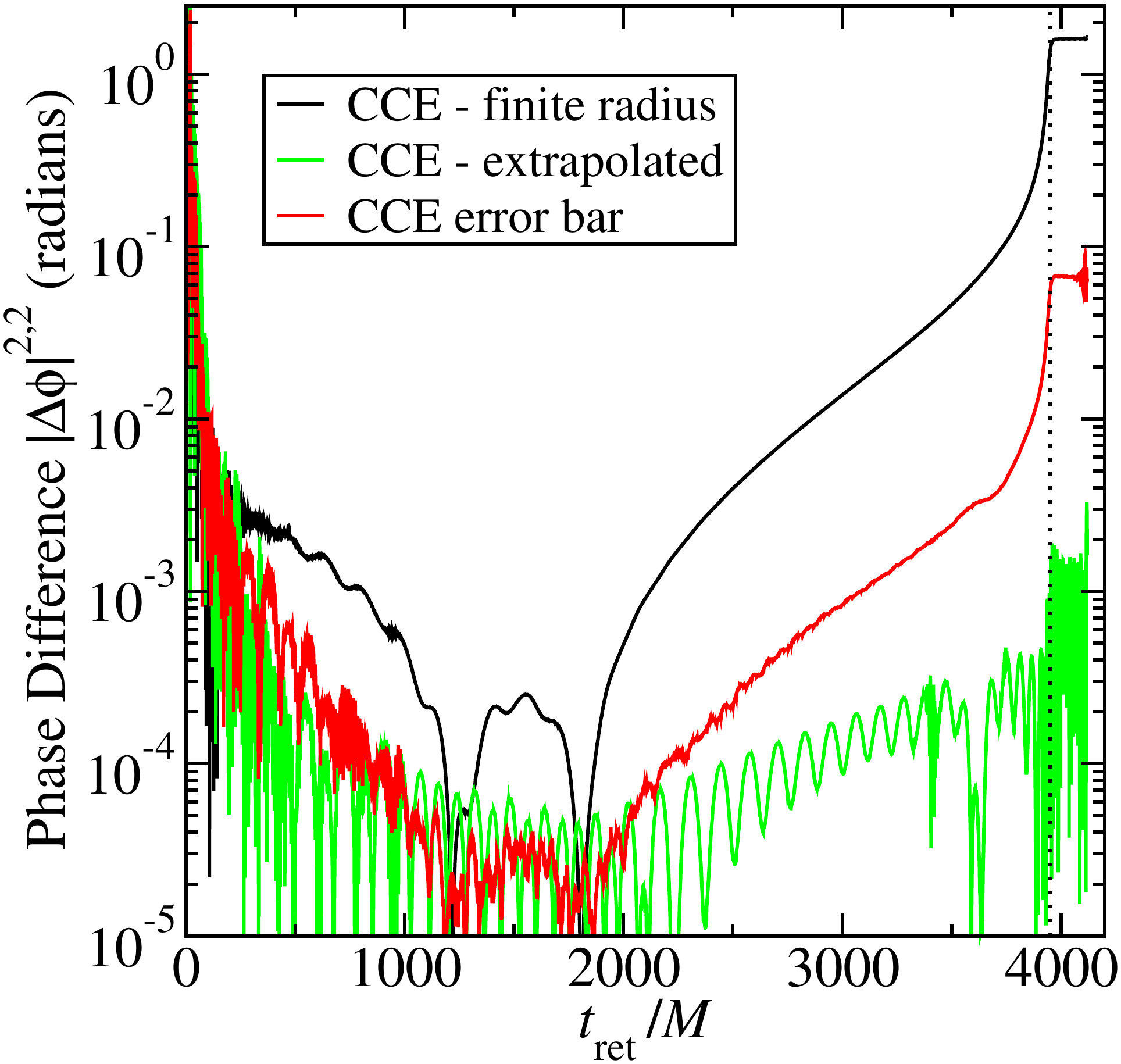}
  \caption{
    Magnitude of phase difference between the CCE
    and the outermost finite-radius ($R=385M$) $\Psi_4^{2,2}$ waveforms,
    for case 2 in Table~\ref{tab:models}.  
    The error bar for phase of the CCE waveform and the phase difference
    between extrapolated and CCE waveforms are also shown.
    The error bar includes Cauchy error (measured using CCE waveforms),
    CCE truncation error, and CCE initial-data error.
    Waveforms are aligned over $[1000M,2000M]$.  The maximum amplitude
    occurs at $\tr \sim 3952M$, indicated by the dotted vertical line.
  }
  \label{fig:q01CceFiniteRad22}
\end{figure}

We first address the question of whether waveform extraction to $\scri^+$
is even necessary, or whether finite-radius waveforms are sufficient,
given the accuracy of our simulations.
Consider the finite-radius $\Psi_4^{2,2}$ mode, 
for case 2 in Table~\ref{tab:models}, computed from the outermost
extraction radius ($R=385M$).
Figure~\ref{fig:q01CceFiniteRad22} shows the phase difference between
this finite-radius waveform and the corresponding CCE waveform.  Also shown
are the difference between the CCE and extrapolated $\Psi_4^{2,2}$ waveforms
and the estimated error bar for the phase of the CCE waveform.
The phase of the finite-radius waveform falls far outside of the estimated
error bar, while in this case the extrapolated and CCE waveforms agree very 
well.  This indicates that the finite-radius waveform is a 
poor proxy for the waveform at $\scri^+$, and that some form of waveform
extraction (either extrapolation of CCE) is required.

\subsection{Comparing different sources of uncertainty}
\label{sec:average-errors}
\begin{figure}
  \includegraphics[width=1.\linewidth]{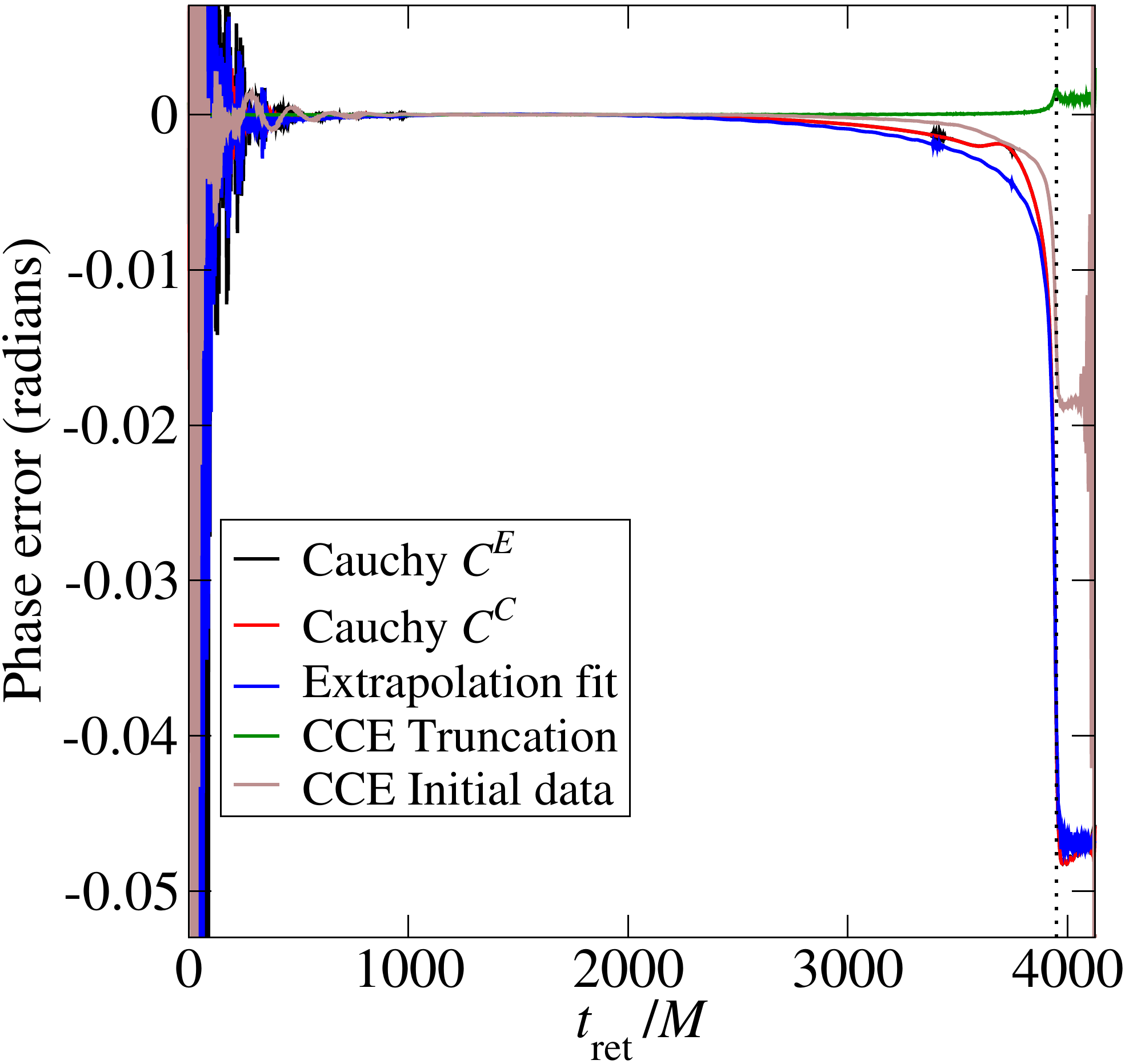}
  \caption{Phase errors in the $\Psi_4^{2,2}$ waveform at $\scri^+$
    from various sources, for simulation 2 in Table~\ref{tab:models}.
    Cauchy errors determined from both extrapolated ($C^E$) and
      CCE ($C^C$) waveforms are shown (see Sec.~\ref{sec:comb-error-diff}).
    The outermost extraction radius is $R=385M$, and all 
    waveforms are aligned over $[1000M,2000M]$.
    The maximum amplitude occurs at $\tr \sim 3952M$, shown here
      as the dotted vertical line.
  }
  \label{fig:q01phase22}
\end{figure}

Here we examine the average magnitudes of errors from different
sources for 
both extrapolated and CCE waveforms. To
illustrate the typical sizes of these errors,
Fig.~\ref{fig:q01phase22} shows the estimated phase 
errors in $\Psi_4^{2,2}$ for
the equal-mass, non-spinning simulation 
(case 2 in Table~\ref{tab:models}).  
All uncertainties are computed using the procedures described in 
Sec.~\ref{sec:EstimatingErrors}.
The errors shown in Fig~\ref{fig:q01phase22} include
the Cauchy error measured using 
extrapolated and CCE waveforms, as
well as the extrapolation fit
error, CCE truncation error (on the null grid), 
and the CCE initial-data error.

We find that the Cauchy error measured from CCE waveforms is
essentially indistinguishable from 
that measured from extrapolated
waveforms, on the scale of Fig.~\ref{fig:q01phase22}.  This 
is consistent with the discussion in 
Section~\ref{sec:comb-error-diff}, i.e.~that these Cauchy errors
are highly correlated.  For extrapolated waveforms, we find
that the Cauchy and extrapolation fit errors are 
about equal.  
For CCE waveforms, the Cauchy error dominates, followed by the CCE 
initial-data error, and finally by the very 
small CCE truncation error.

\begin{figure}
  \includegraphics[width=1.\linewidth]{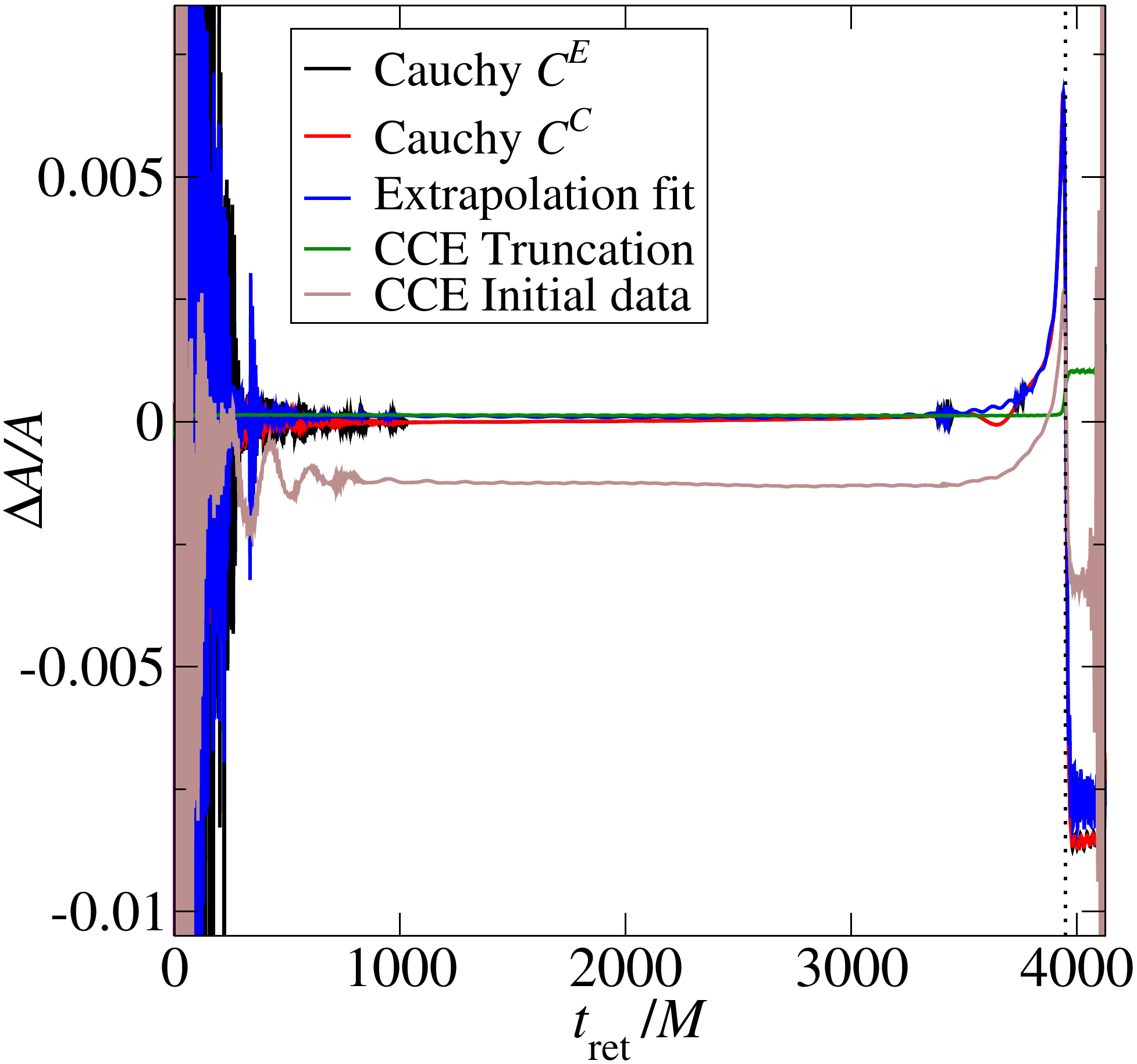}
  \caption{Same as Figure~\ref{fig:q01phase22} except showing relative
    amplitude errors instead of phase errors.
  }
  \label{fig:q01relamp22}
\end{figure}

Figure~\ref{fig:q01relamp22} shows relative amplitude errors for the
same simulation as Fig.~\ref{fig:q01phase22}.  
During merger and
ringdown, the relative contributions of each error source are the same
as for phase error, with Cauchy error being the largest and 
CCE truncation
error the smallest.  Interestingly, we find that the CCE 
initial-data error is the dominant 
source of amplitude error during the inspiral,
although in absolute terms is it still a small error at
$\mathcal{O}(10^{-3})$.  Near merger and during the ringdown, the
relative amplitude errors are small compared with the phase errors
shown in Fig.~\ref{fig:q01phase22}.  Hence, during this portion of the
waveform, the error measure $\Delta_{\ell,m}$ given by
Eq.~(\ref{eq:ComplexErrorMeasure}) will be essentially the same as the
phase error.

\begin{figure*}
  \centering \subfigure[\ Simulation 1: q=1, nonspinning, gauge 1] {
    \includegraphics[width=\columnwidth]{%
      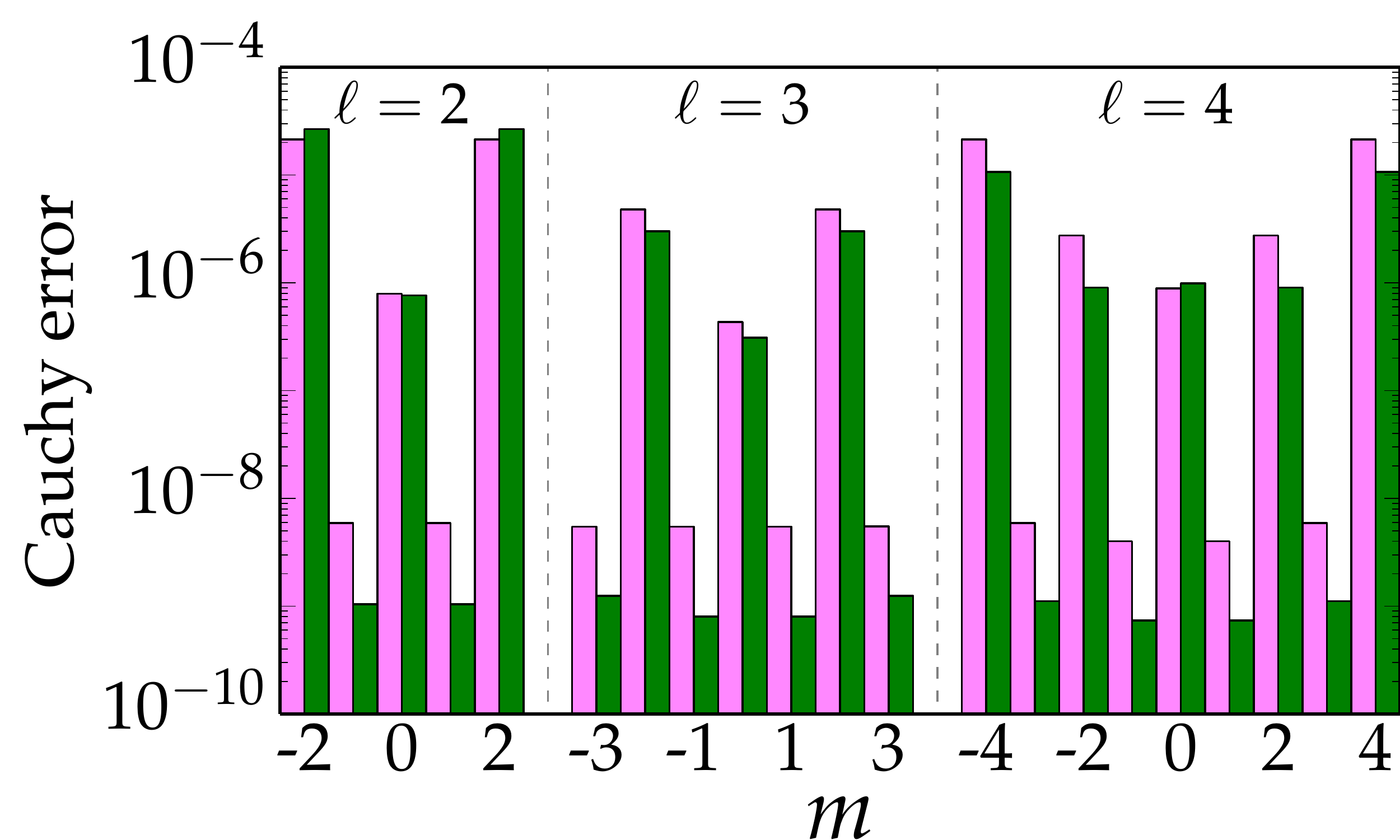}
  } \subfigure[\ Simulation 2: q=1, nonspinning, gauge 2] {
    \label{fig:CauchyErrHistCase2}
    \includegraphics[width=\columnwidth]{%
      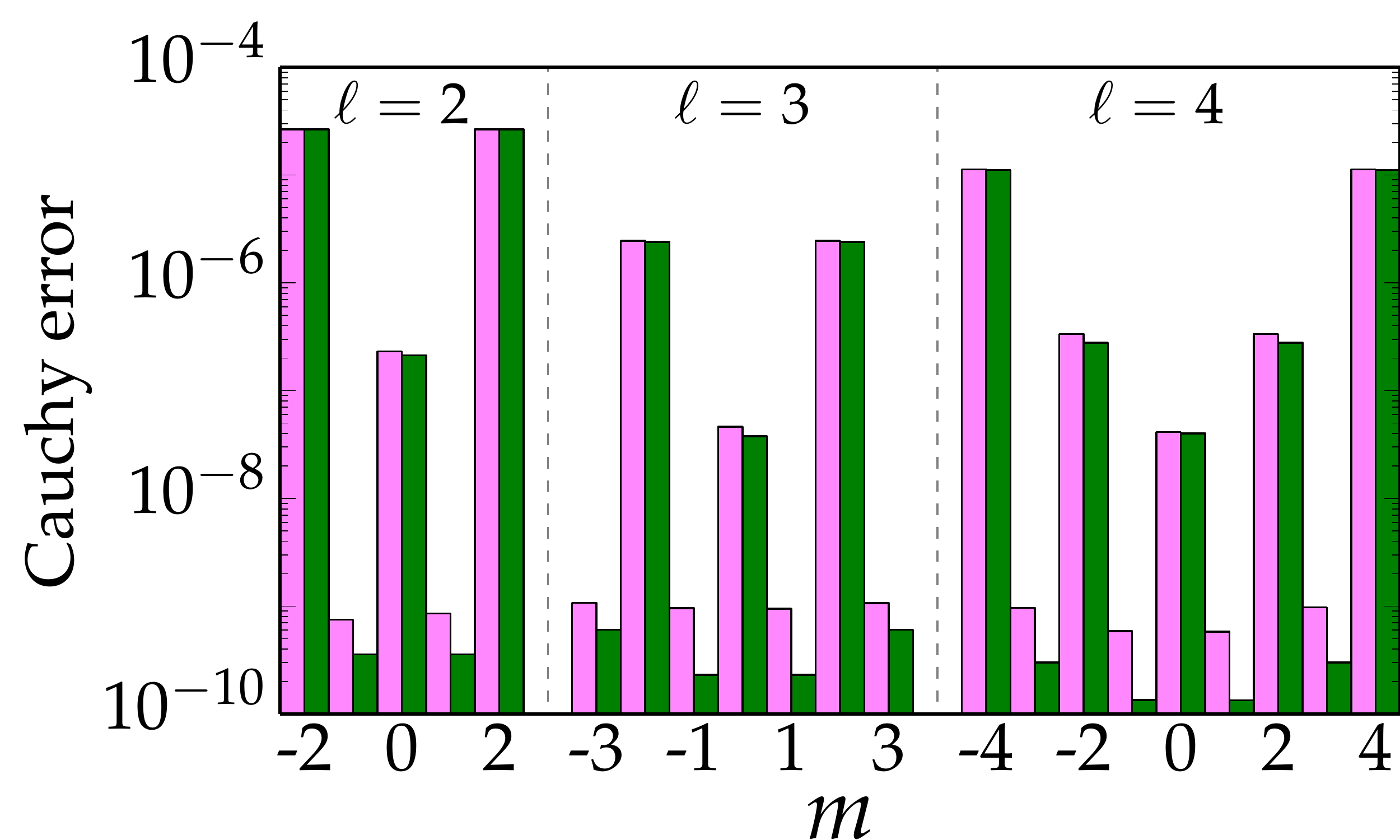}
  } \subfigure[\ Simulation 3: q=6, nonspinning] {
    \label{fig:CauchyErrHistq06}
    \includegraphics[width=\columnwidth]{%
      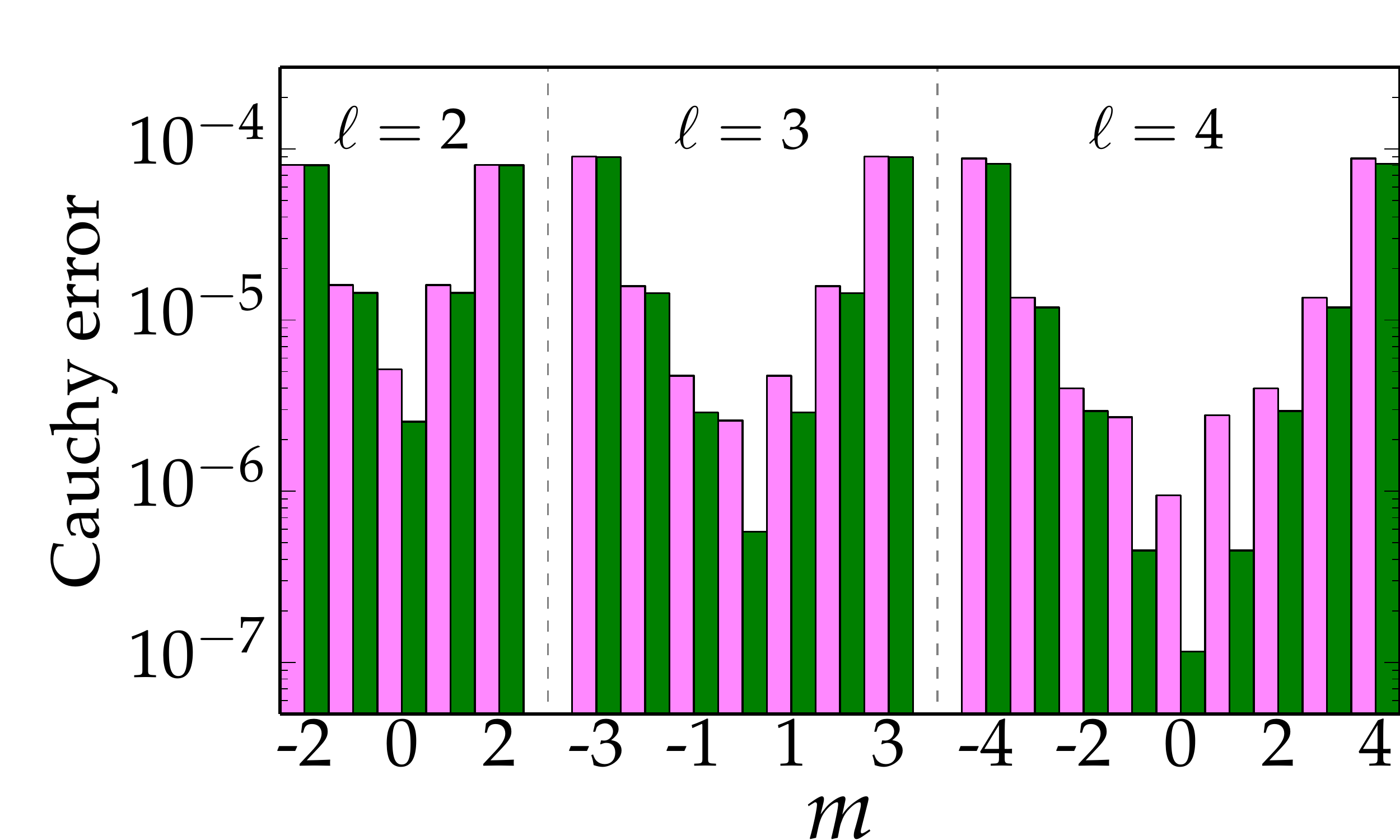}
  } \subfigure[\ Simulation 4: q=3, precessing] {
    \includegraphics[width=\columnwidth]{%
      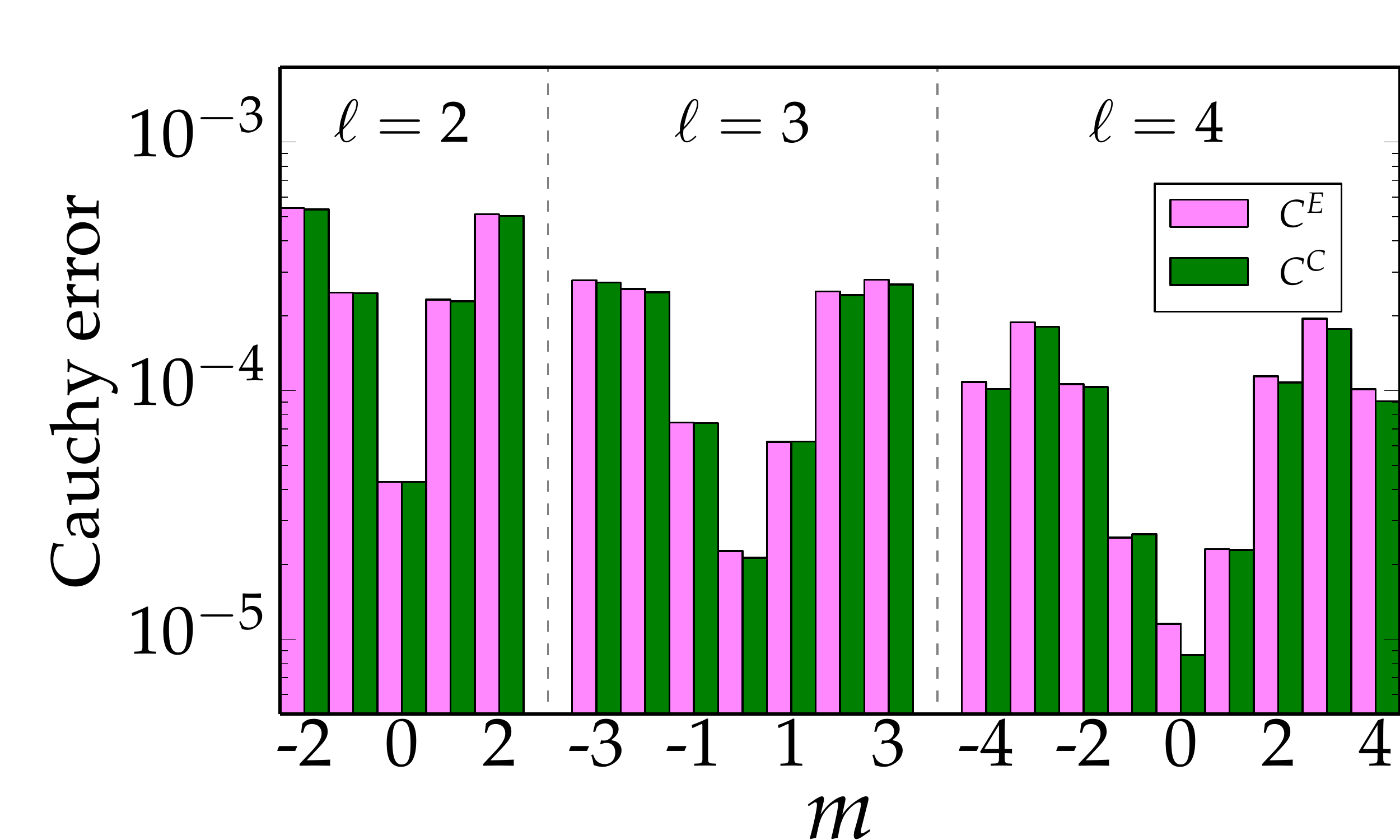}
  }
  \caption{\label{fig:CauchyErrHist} Cauchy errors $C^C$ and $C^E$ (see
    Section \ref{sec:comb-error-diff}) as a function of
    $(\ell,m)$ spherical harmonic mode for different simulations in
    Table~\ref{tab:models}.  The vertical axis is the error measure 
    $\Delta_{\ell,m}$ of
    Eq.~(\ref{eq:ComplexErrorMeasure}), time-averaged so that each
    source of error is described by a single number for each
    $(\ell,m)$ mode.  The horizontal axis represents the spherical
    harmonic $m$ index; vertical dashed lines separate $\ell=2$,
    $\ell=3$, and $\ell=4$ modes, and for each $\ell$, every other
    value of $m$ is labeled on the horizontal axis. 
    The pink bars represent the
    Cauchy error $C^E$
    in the extrapolated waveforms, and the dark green bars 
    represent the Cauchy error $C^C$ in the CCE waveforms.  
  }
\end{figure*}

Having investigated the error in $\Psi_4^{2,2}$ for simulation 2 of
Table~\ref{tab:models}, we now consider the errors 
for the other simulations and for other spin-weighted spherical
harmonic modes.  
To condense information from many modes and
several simulations into a smaller number of figures, we compute
time-averaged errors as described in
Section~\ref{sec:combination}, and we use
the error measure $\Delta_{\ell,m}$
of Eq.~(\ref{eq:ComplexErrorMeasure}) 
 instead of measuring phase and amplitude
errors separately.  This reduces each
error measure for a
given $(\ell,m)$ mode to a single number.

Figure~\ref{fig:CauchyErrHist} shows the time-averaged Cauchy errors
in extrapolated and CCE waveforms for all
$\Psi_4^{\ell,m}$ up to $\ell=4$ and for 
all 
simulations in Table~\ref{tab:models}.  
Although only modes up to $\ell=4$ have been
included in this figure, the qualitative features are unchanged if
modes up to $\ell=8$ (the maximum mode we have computed) are included.

There are a few general features evident in the figure. First, the
Cauchy errors in CCE and extrapolated waveforms have similar
magnitudes.  In addition, the modes with
$|m|=\ell$ have the largest errors. This is to be expected, because
these are the modes with the greatest amplitudes.  Along the same
lines, we see that in the $q=1$ cases, the average error is very small
for the modes with odd $m$, because by symmetry 
(rotation through $\pi$) these modes should
have vanishing amplitude.

\begin{figure*}
  \centering \subfigure[\ Simulation 1: q=1, nonspinning, gauge 1] {
    \includegraphics[width=\columnwidth]{%
      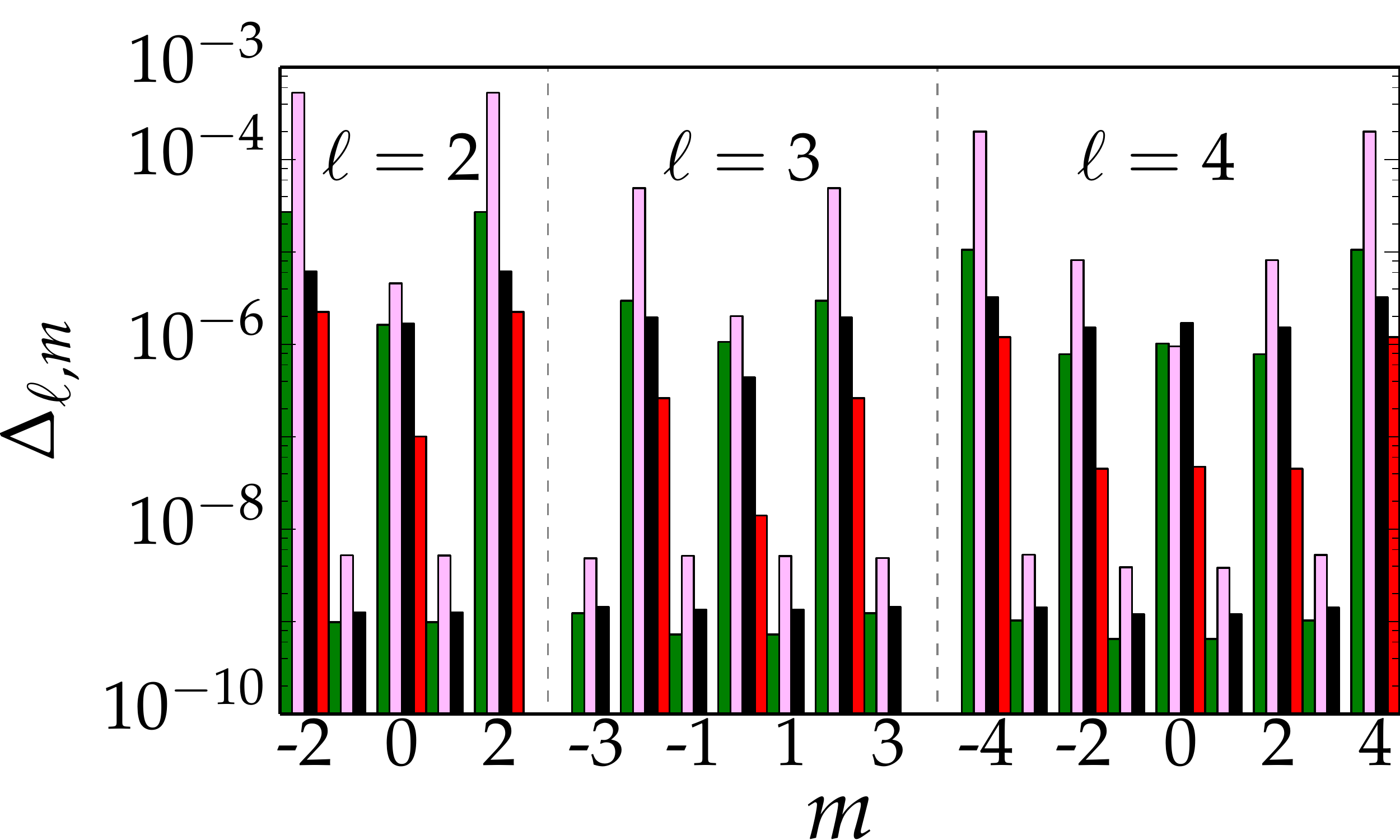}
  } \subfigure[\ Simulation 2: q=1, nonspinning, gauge 2] {
    \includegraphics[width=\columnwidth]{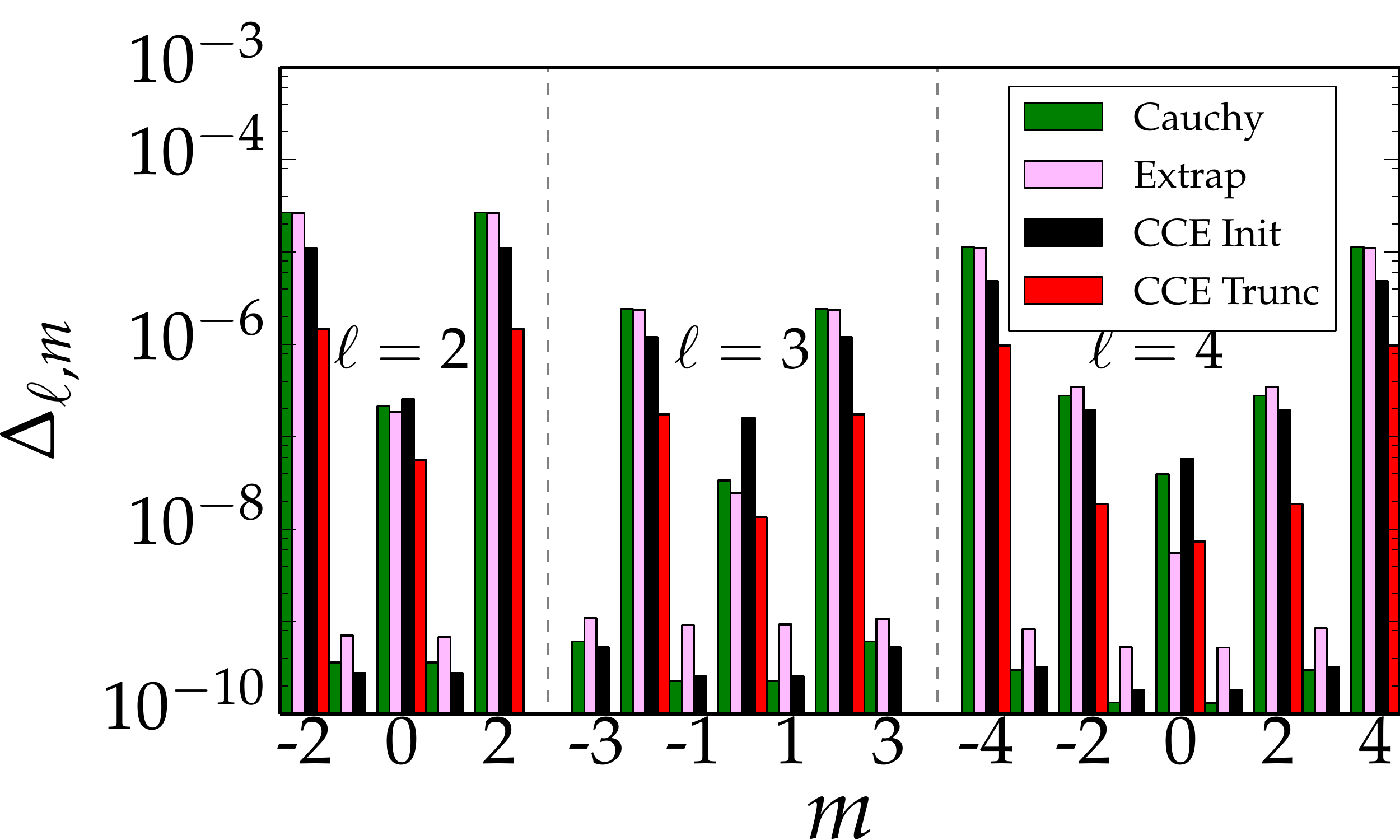}
  } \subfigure[\ Simulation 3: q=6, nonspinning] {
    \label{fig:ComplexErrHistq06}
    \includegraphics[width=\columnwidth]{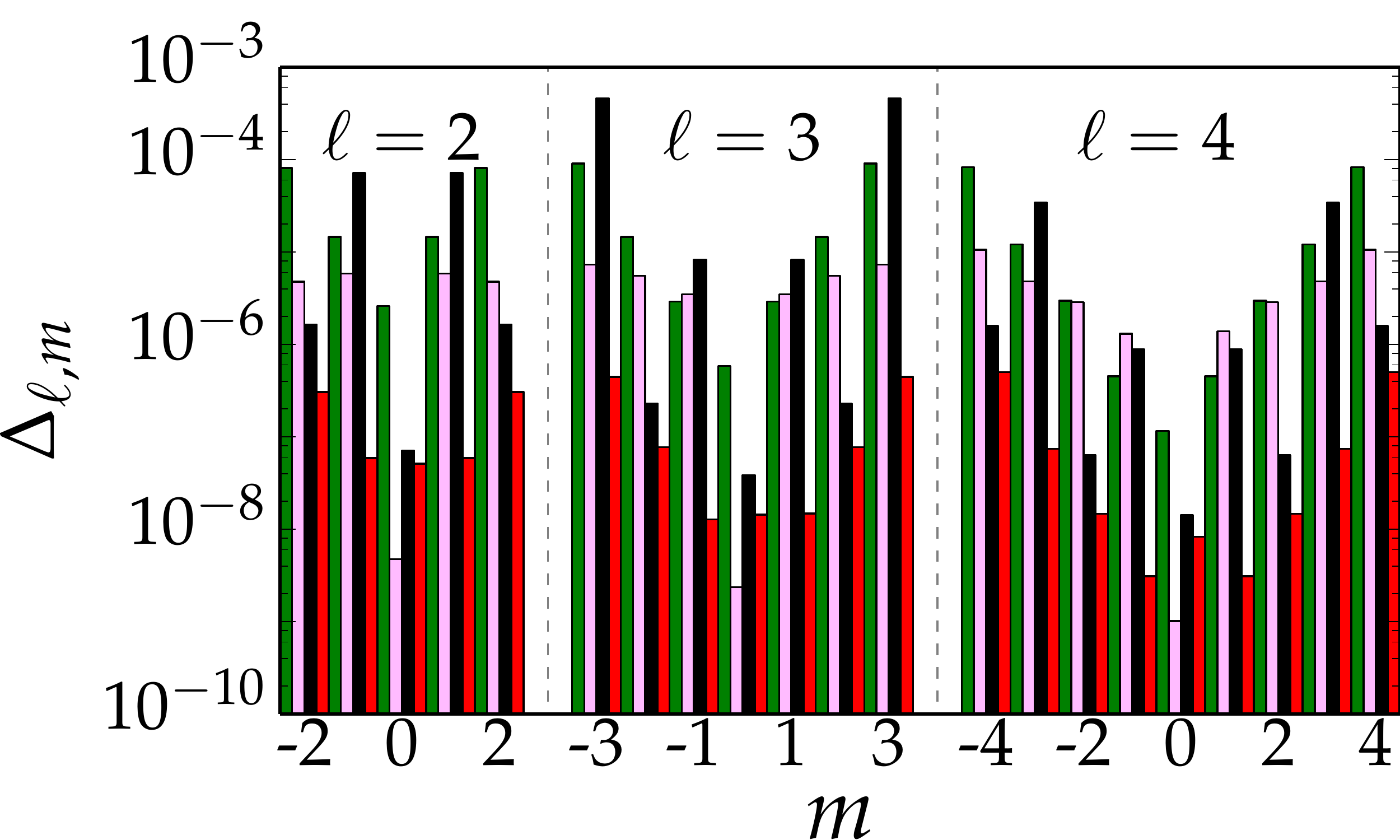}
  } \subfigure[\ Simulation 4: q=3, precessing] {
    \label{fig:ComplexErrHistq03}
    \includegraphics[width=\columnwidth]{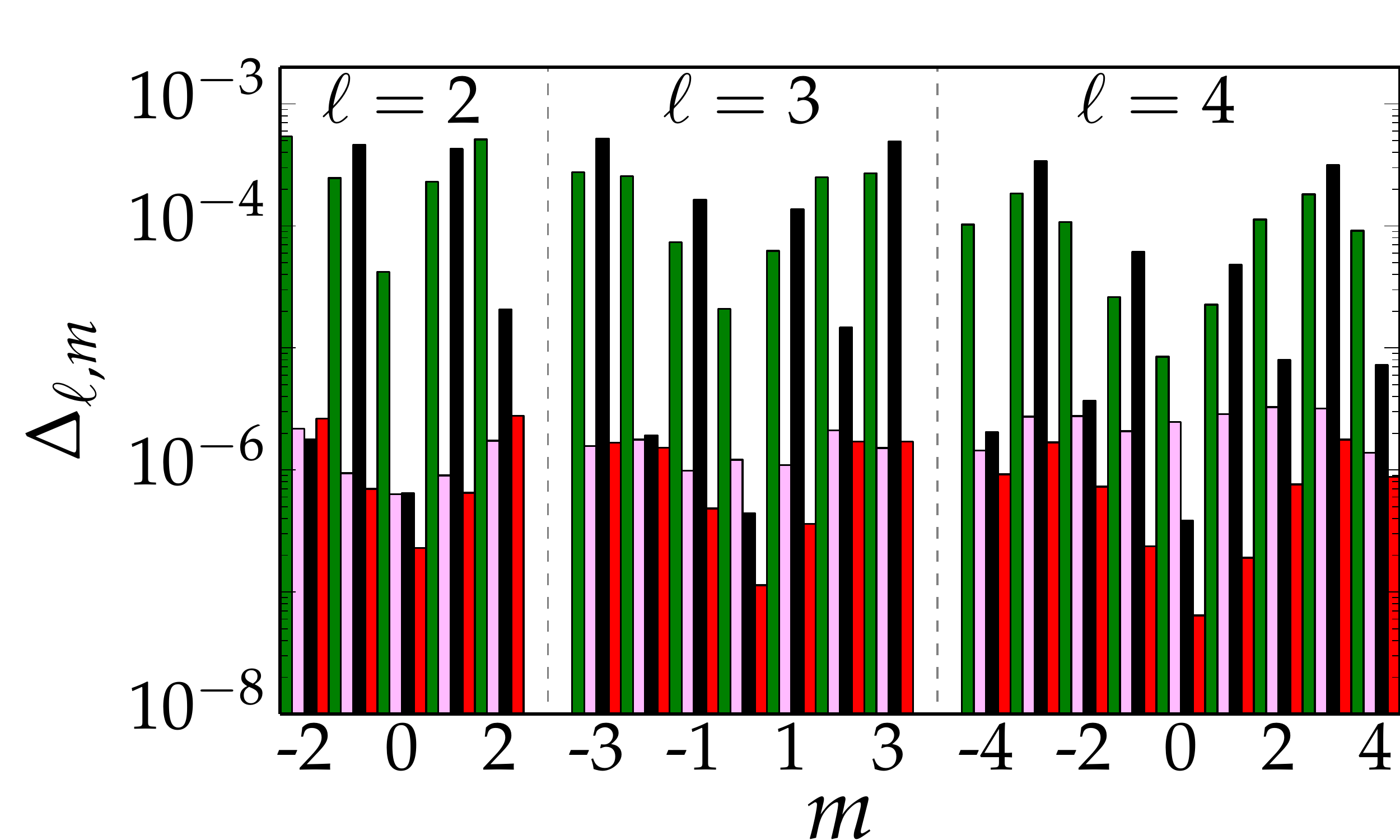}
  }
  \caption{\label{fig:ComplexErrHist} Same as
    Fig.~\ref{fig:CauchyErrHist}, but showing multiple sources of
    error. The Cauchy error shown here is the average of those
      computed using CCE and extrapolated waveforms, for each $(\ell,m)$.
  }
\end{figure*}

Figure~\ref{fig:ComplexErrHist} shows the time-averaged Cauchy error,
extrapolation fit 
error, CCE truncation error, and CCE initial-data error
in $\Psi_4^{\ell,m}$ for all $(\ell,m)$ up to $\ell=4$.  
The Cauchy error shown here 
is the average of 
those computed
from the CCE and extrapolated waveforms.  
As was the case for Fig.~\ref{fig:CauchyErrHist}, 
we only show results up to
$\ell \le 4$, 
but the
qualitative features are the same for all modes we have examined 
(up to $\ell=8$).

The truncation error on the CCE null grid is by far the smallest
source of error in each case.  The largest source of error 
varies, depending
both on the simulation and on the mode.  For most cases, the
CCE initial-data error and the Cauchy error are the largest, except in the
$q=1$ cases where the extrapolation fit error dominates.

\subsection{Gauge dependence}
\label{sec:errors-from-diff}

In principle, extrapolated waveforms may be
contaminated by gauge effects, whereas 
CCE waveforms should be gauge
invariant.  Here we directly investigate the gauge dependence of both
extraction methods by comparing two equal-mass, zero-spin BBH
simulations (the first two cases in Table~\ref{tab:models})
with identical initial data but with different gauge
conditions.
The first simulation is the one described in 
Ref.~\cite{Scheel2009}. It uses a gauge
in which the gauge-source function obeys a wave equation, and the
source terms of this wave equation are fine-tuned by hand.  We have
found previously that this gauge does not work well for black-hole
binaries with unequal masses or large
spins~\cite{Lindblom2007,Lindblom2009c,Szilagyi:2009qz}, so current
BBH simulations using \texttt{SpEC} employ a damped harmonic gauge
condition~\cite{Lindblom2009c,Choptuik:2009ww,Szilagyi:2009qz}, which
is the gauge used in simulation 2 of Table~\ref{tab:models}.

\begin{figure}
  \includegraphics[width=1.\linewidth]{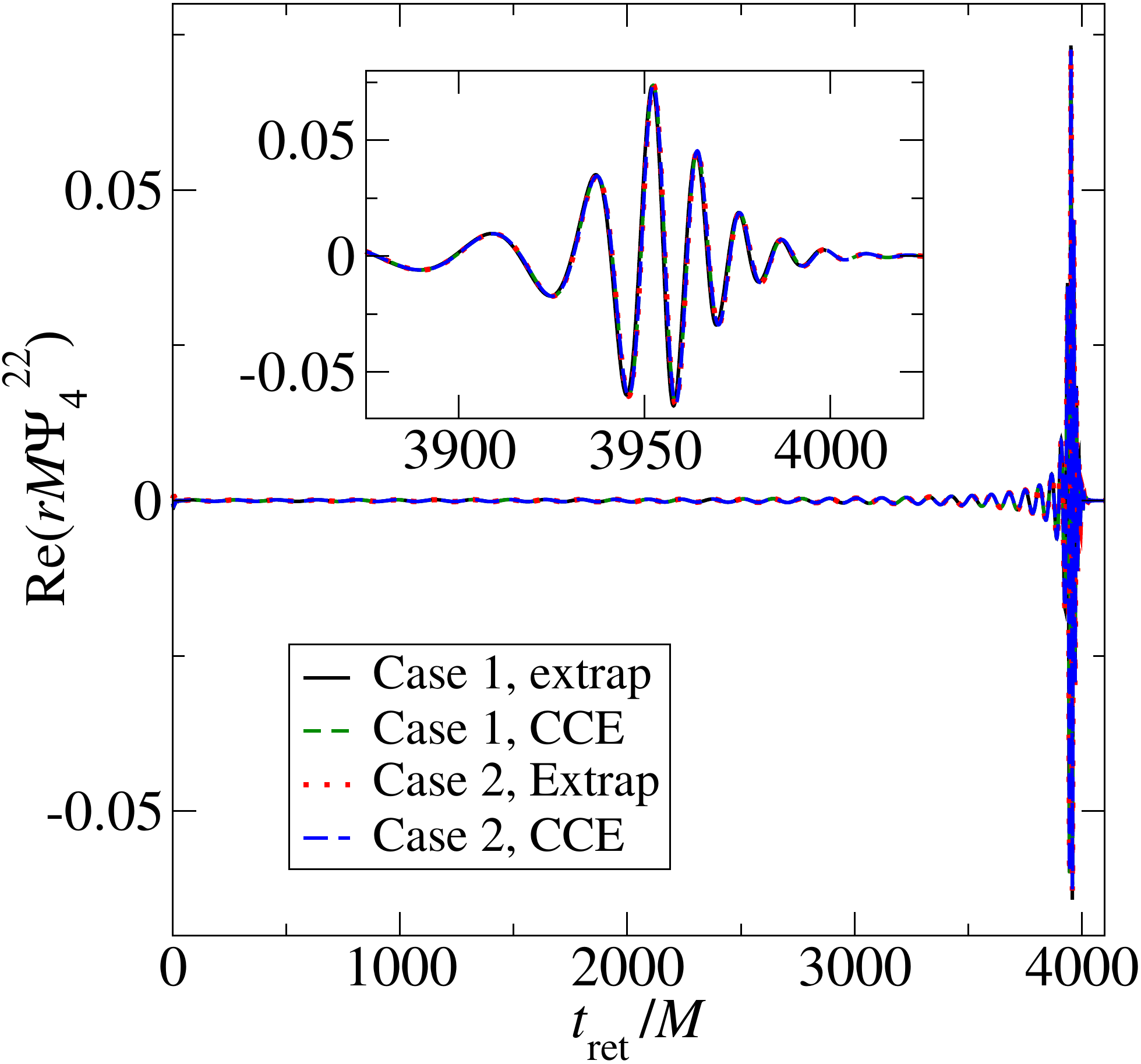}
  \caption{The real part of $rM\Psi_4^{2,2}$ 
    for both 
    extrapolated and CCE waveforms, for the first two simulations in
    Table~\ref{tab:models}.  Waveforms are aligned in the interval
    $[1000M,2000M]$.  The four curves agree very well.
    Time-averaged differences between these curves
    are shown in Figs.~\ref{fig:NPHistogramTwoGauges}
    through~\ref{fig:HistogramCCEVsNP} below.
  }
  \label{fig:Psi422TwoGauges}
\end{figure}

Figure~\ref{fig:Psi422TwoGauges} shows the dominant mode 
$\Psi_4^{2,2}$
as a function of time for both gauge choices, and for
both extrapolated and CCE waveforms.  All four plots in 
this figure
agree well, suggesting that both CCE and 
extrapolated waveforms for this
dominant mode are independent of gauge, 
at least on the scale
of the figure.  

\begin{figure}
  \includegraphics[width=1.\linewidth]{%
    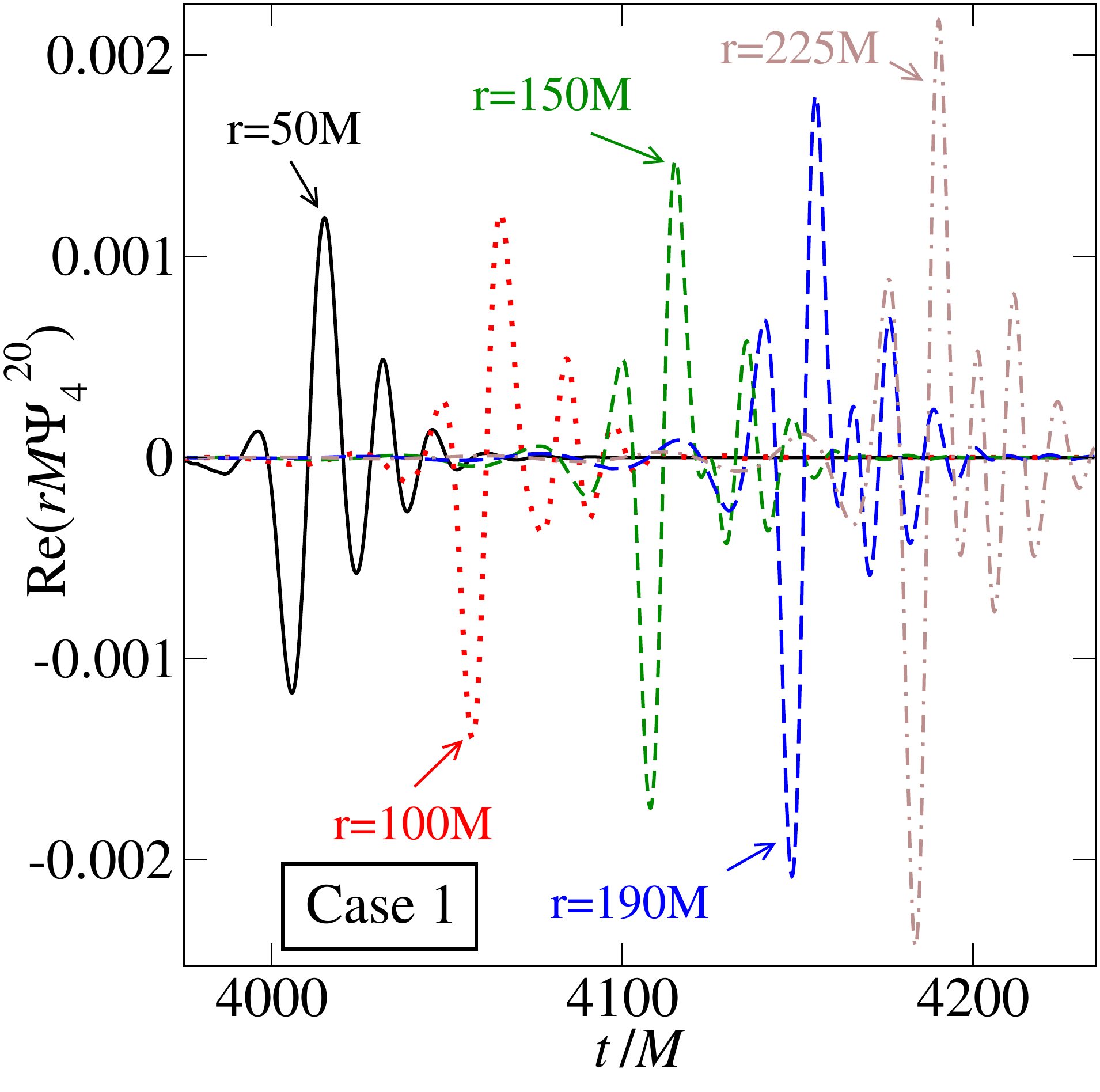}
  \caption{The real part of $rM\Psi_4^{2,0}$ extracted at multiple
    radii, before extrapolation, for the first simulation in
    Table~\ref{tab:models}.  
    Waveforms are shown only near peak amplitude because they are
    very small elsewhere.
    The waveform for each extraction radius $r$ is
    plotted versus time, rather than $\tr$, 
    so that waveforms extracted at larger
    $r$ reach their peak amplitude
    later.
    The increase in amplitude 
    with 
    extraction radius $r$
    indicates that 
    $\Psi_4^{2,0}$ falls off more slowly than $1/r$.
    We attribute this slow falloff to 
    the gauge condition used for simulation 1. The
    other simulations, which use a more robust 
    gauge condition, do not exhibit this 
    behavior.}
  \label{fig:Psi420ManyRadii}
\end{figure}

On the other hand, the extrapolated waveform for the subdominant 
mode $\Psi_4^{2,0}$ differs 
significantly between simulations 1 and 2.
In particular, for simulation 1, the gauge effects appear to be
so strong that it is difficult to even define the 
extrapolated $\Psi_4^{2,0}$ waveform.
To understand the difficulty, recall that the extrapolation procedure
assumes that $rM\Psi_4^{\ell,m}$
approaches a finite limit as $r\to\infty$.  However, if 
$rM\Psi_4^{2,0}$ from
simulation 1 is plotted at different extraction radii $r$, we find
that it appears to grow without limit as $r$ increases, 
as illustrated in
Fig.~\ref{fig:Psi420ManyRadii}.  
The assumption that the finite-radius
waveforms $rM\Psi_4^{\ell,m}(\tr,r)$ can be expanded in a convergent 
series in $1/r$ is thus violated in this case.
Note that this problem occurs only
for the gauge used in simulation 1; for the gauge used in the other
simulations, $rM\Psi_4^{2,0}$ approaches a 
finite limit as $r$ increases.

Although extrapolation fails to converge for the 
$\Psi_4^{2,0}$
waveform in simulation 1, we nevertheless compute the $N=5$
extrapolant for this mode for comparison purposes.
Based on Fig.~\ref{fig:Psi420ManyRadii}, we do not expect this
$N=5$ extrapolant to be very accurate.  It is worth noting, however, 
that this extrapolated waveform nevertheless agrees better with CCE than
the unextrapolated waveform measured at the outermost extraction radius.

\begin{figure}
  \includegraphics[width=1.\linewidth]{%
    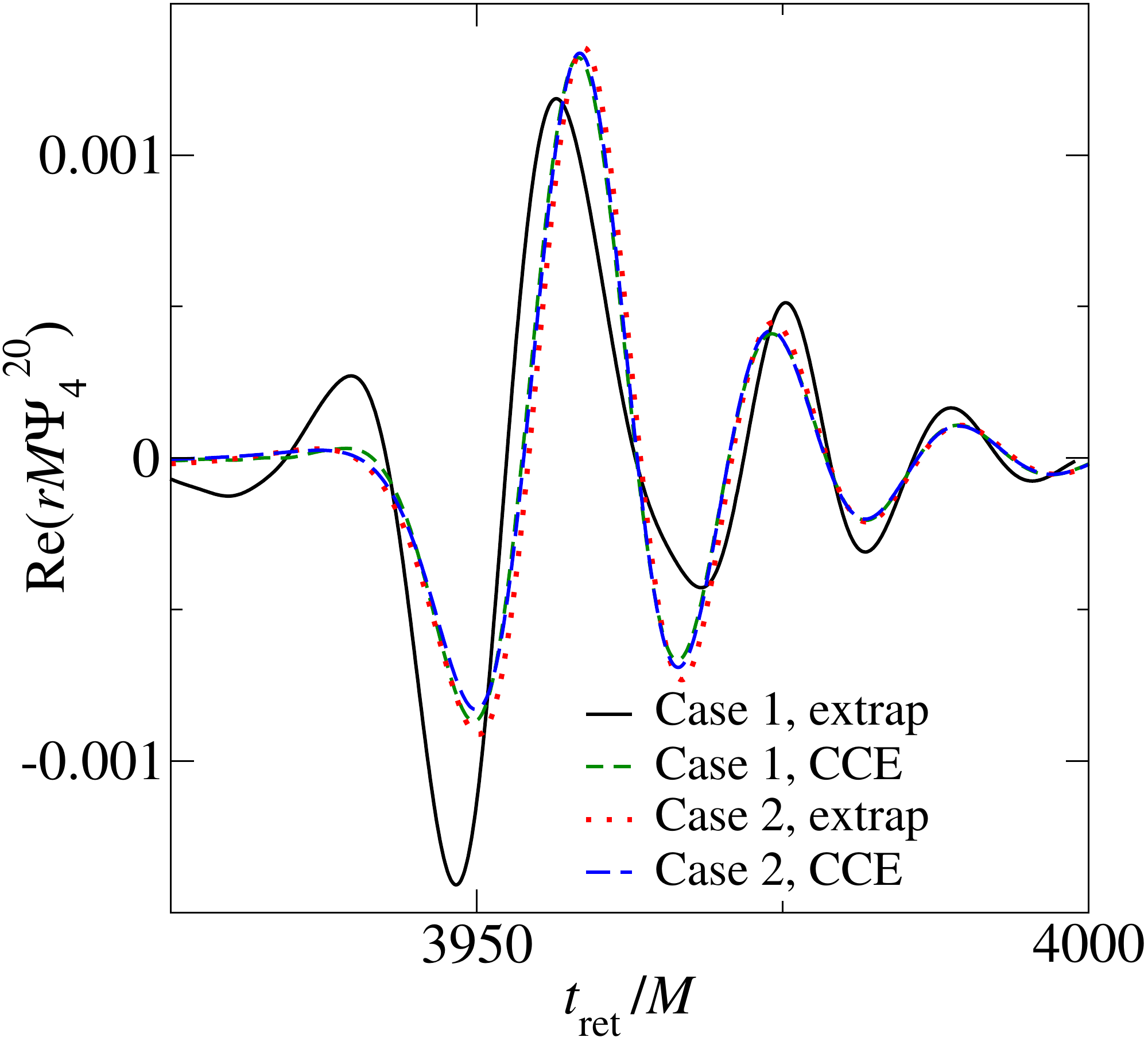}
  \caption{The real part of $rM\Psi_4^{2,0}$ 
    for both extrapolated and CCE waveforms, for the first two simulations in
    Table~\ref{tab:models}.  Waveforms are aligned in the
    interval $[1000M,2000M]$. We show only times near merger because
    the waveform is very small elsewhere.
    Although the difference between 
    CCE and extrapolated waveforms 
    for Case 2 is far smaller than for Case 1, even in Case 2 this
    difference is several times the numerical error
    Note that the time-averaged difference shown below in 
      Fig.~\ref{fig:HistogramCCEVsNP} for Case 2
      is dominated by the inspiral portion
      of the waveforms.
  }
  \label{fig:Psi420TwoGauges}
\end{figure}

Our expectations are confirmed by Fig.~\ref{fig:Psi420TwoGauges},
which shows $\Psi_4^{2,0}$ as a function of time for CCE and extrapolated
waveforms, and for simulations 1 and 2. 
This figure
is the same as Fig.~\ref{fig:Psi422TwoGauges}, except 
that it shows $\Psi_4^{2,0}$ instead of $\Psi_4^{2,2}$.  The extrapolated
waveforms in Fig.~\ref{fig:Psi420TwoGauges} are very different for
the two simulations, 
whereas the CCE waveforms
are almost 
indistinguishable.  
This provides
a demonstration of both the gauge-invariance of CCE, and of the
gauge-dependence of extrapolated waveforms.

\begin{figure}
  \includegraphics[width=1.\linewidth]{%
    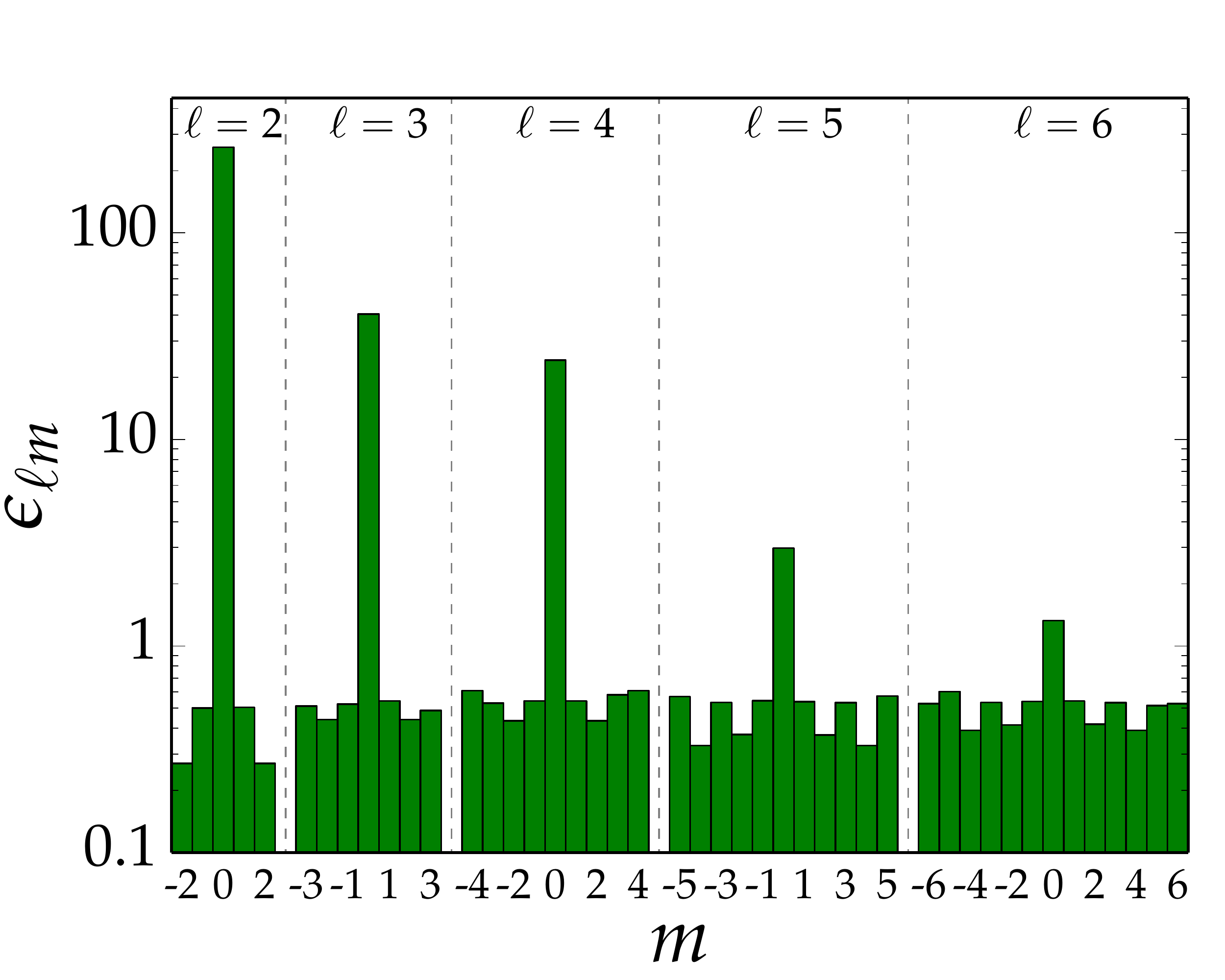}
  \caption{Fractional differences $\epsilon_{\ell m}$
    (cf. Eqs.~(\ref{eq:FractionalErrorMeasure}) 
    and~(\ref{eq:CombinedErrorBar}))
    between extrapolated $\Psi_4^{\ell,m}$
    from physically equivalent simulations with different gauge
    conditions (i.e., the first two simulations in
    Table~\ref{tab:models}), as a function of $(\ell,m)$.  The
    $(\ell,m)$ modes are labeled as in Fig.~\ref{fig:CauchyErrHist}.
    Waveforms are aligned in the interval $[1000M,2000M]$.  
  }
  \label{fig:NPHistogramTwoGauges}
\end{figure}

To make the above conclusions more precise, the differences between
these waveforms should be compared to the various sources of error
discussed in Section~\ref{sec:average-errors}.  
We construct a measure $\epsilon_{\ell m}$ 
of the fractional difference between the waveforms for each mode,
computed as the difference between
the extrapolated $\Psi_4^{\ell,m}$ 
from simulation 1 and the same
waveform from simulation 2, divided by a combined error bar for the
difference.  The combined error bar is defined as the
$L^1$ norm of the various sources of uncertainty that enter into 
the difference (cf. Section~\ref{sec:combination}).  In order to obtain a single
measure of the fractional agreement between the waveforms for
each mode, we also perform a time averaging of these fractional differences.
In other words, we define
\begin{equation}
  \label{eq:FractionalErrorMeasure}
  \epsilon_{\ell m} = \left\langle
  |\Psi_4^{l,m}{}^A - \Psi_4^{l,m}{}^B|/E \right\rangle,
\end{equation}
where $A$ and $B$ refer to the different simulations, and angle brackets
represent a time average.
The numerator is the error measure 
$\Delta_{\ell,m}$ of 
Eq.~\ref{eq:ComplexErrorMeasure}, and the
error bar in the denominator is computed here as 
\begin{equation}
  \label{eq:CombinedErrorBar}
  E = \frac{1}{2}\left(|C^E_A|+|C^E_B|\right) + |F_A| + |F_B|,
\end{equation}
where $C^E$ represents the Cauchy error computed using the 
extrapolated 
waveforms in simulation $A$ or $B$, and $F$ represents the extrapolation 
fit error.  Note that each of the these
error measures is computed separately for each $(l,m)$ mode and for each
time, and that the division in Eq.~(\ref{eq:FractionalErrorMeasure}) is done
before the time averaging. 

In Fig.~\ref{fig:NPHistogramTwoGauges}, we plot these time-averaged
fractional differences for all modes.
Values less than unity indicate
differences that are (on average) within the error bars.
Figure~\ref{fig:NPHistogramTwoGauges} shows
that for most $(\ell,m)$ modes, 
extrapolated waveforms
for the two different gauge choices
are essentially indistinguishable (i.e. within the error bars).
However, for the $m=0$ modes, extrapolated
waveforms are contaminated by significant gauge 
effects that are larger
than other sources of error.  As $\ell$ increases, 
the average fractional difference between $m=0$ modes 
$\epsilon_{\ell 0}$ 
decreases.  This is simply because the 
amplitude
of the modes decreases with increasing $\ell$, so eventually the differences
fall within the error bars.

\begin{figure}
  \includegraphics[width=1.\linewidth]{%
    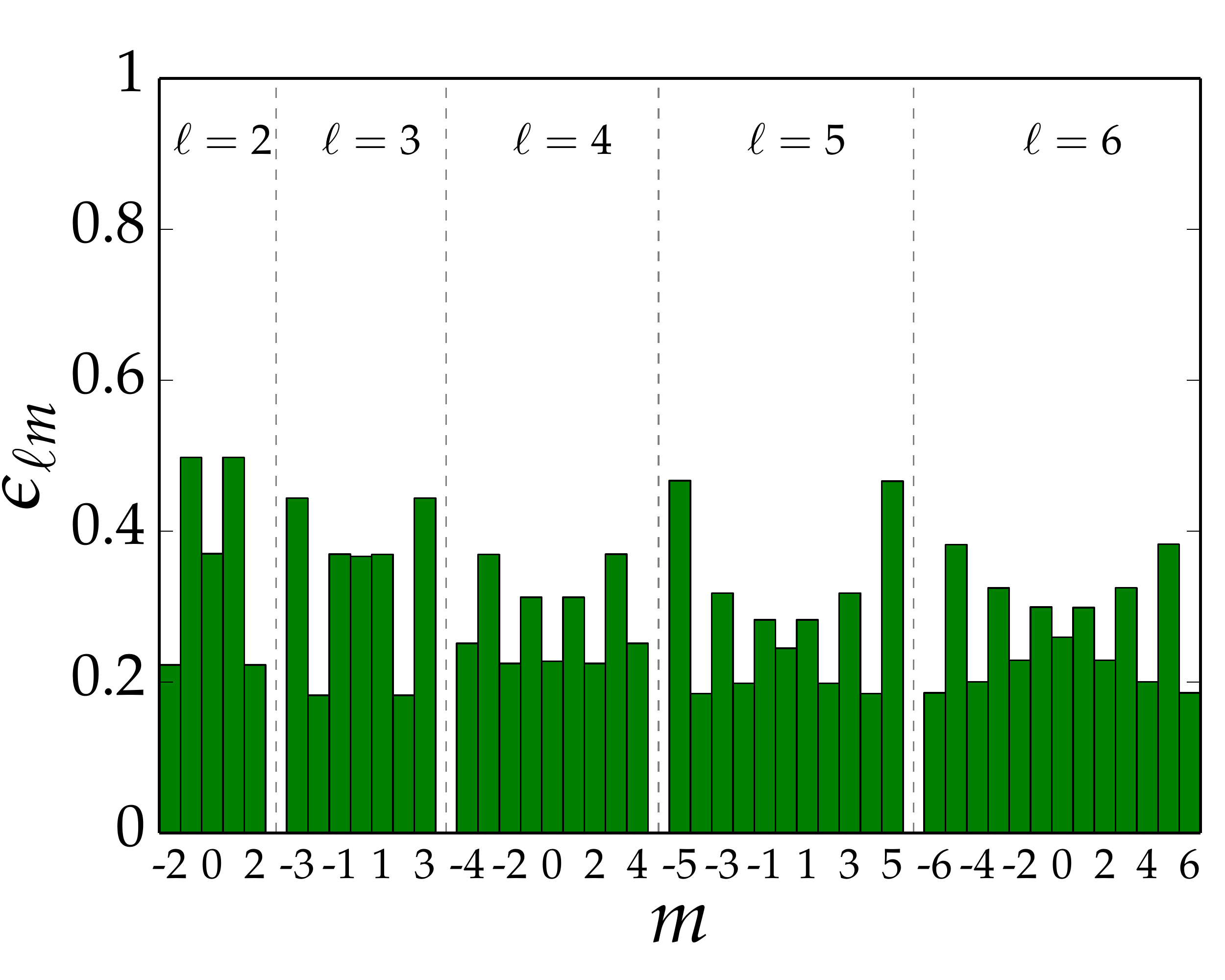}
  \caption{Fractional differences $\epsilon_{\ell m}$
    (cf. Eqs.~(\ref{eq:FractionalErrorMeasure}) 
    and~(\ref{eq:CombinedErrorBarCCE}))
    between CCE waveforms from the same simulations as shown in 
    Fig.~\ref{fig:NPHistogramTwoGauges}. Labels are the same as
    Fig.~\ref{fig:NPHistogramTwoGauges}, except here the differences
    are shown on a linear plot.
  }
  \label{fig:CCEHistogramTwoGauges}
\end{figure}

Figure~\ref{fig:CCEHistogramTwoGauges} 
shows fractional differences between waveforms from the same two
simulations as Figure~\ref{fig:NPHistogramTwoGauges}, but for CCE waveforms.  
Because CCE waveforms have different sources of error than extrapolated
waveforms, the denominator of Eq.~(\ref{eq:FractionalErrorMeasure})
is computed in this case as 
\begin{equation}
  \label{eq:CombinedErrorBarCCE}
  E = \frac{1}{2}\left(|C^C_A|+|C^C_B|\right) + |T_A| + |T_B| + |I_A| + |I_B|,
\end{equation}
where $C^C$ represents the Cauchy error computed using the CCE
waveforms in simulation $A$ or $B$, $T$ represents CCE truncation error,
and $I$ represents the CCE initial-data error.

The differences shown 
in Fig.~\ref{fig:CCEHistogramTwoGauges}
are smaller than unity,
verifying that CCE is indeed
gauge-invariant to the level of our numerical error, even for a gauge
(the gauge from simulation 1) that is sufficiently ill-behaved that
extrapolation fails to converge (cf. Fig.~\ref{fig:Psi420ManyRadii}).  
Moreover, comparing 
Fig.~\ref{fig:NPHistogramTwoGauges} with 
Fig.~\ref{fig:CCEHistogramTwoGauges} shows that the 
differences
between CCE waveforms from simulations 1 and 2 are on average smaller
than the differences between extrapolated 
waveforms from the same two
simulations.

\subsection{When is CCE necessary?}
\label{sec:comp-cce-extr}

\begin{figure*}
  \centering \subfigure[\ Simulation 1: q=1, nonspinning, gauge 1] {
    \includegraphics[width=\columnwidth]{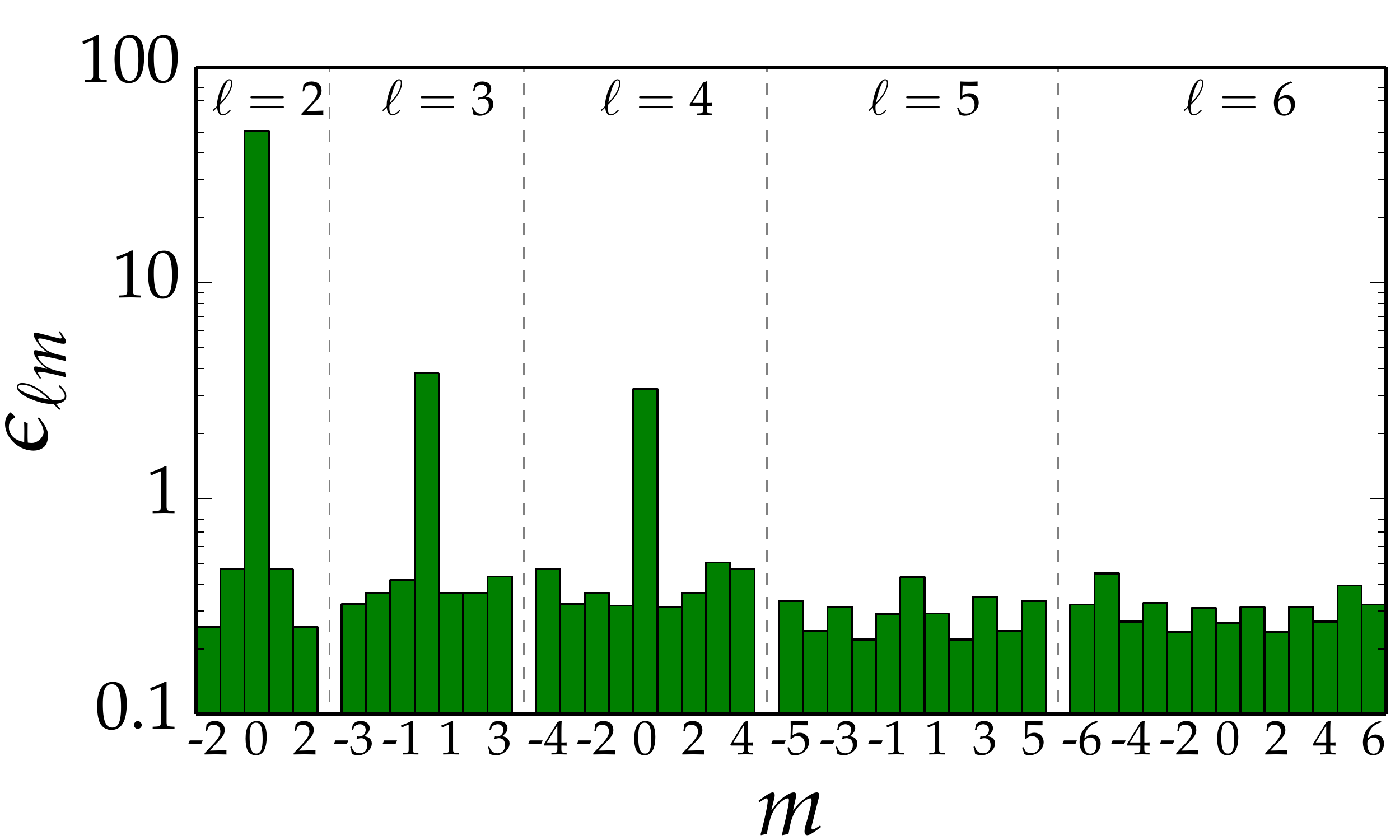}
  } \subfigure[\ Simulation 2: q=1, nonspinning, gauge 2] {
    \includegraphics[width=\columnwidth]{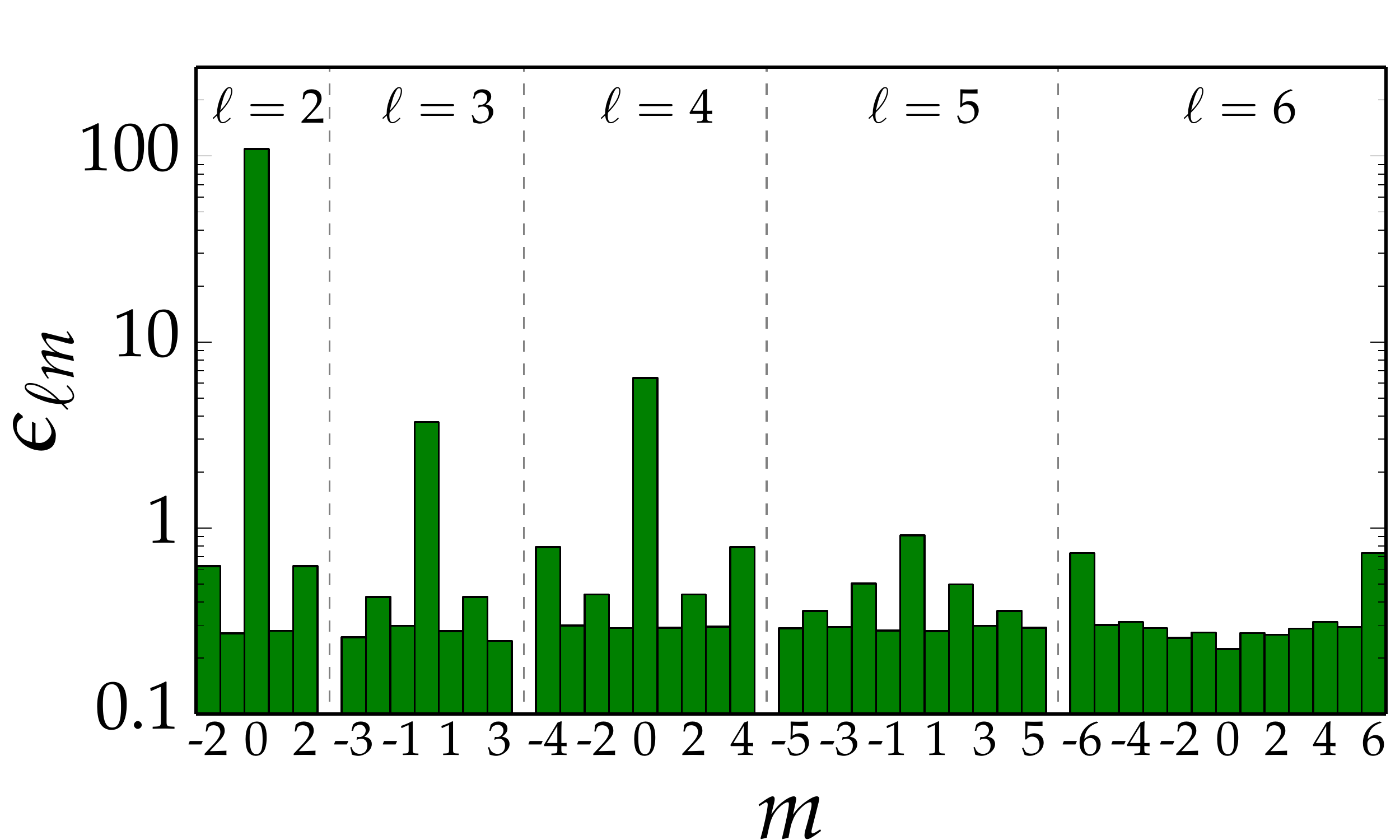}
  } \subfigure[\ Simulation 3: q=6, nonspinning] {
    \includegraphics[width=\columnwidth]{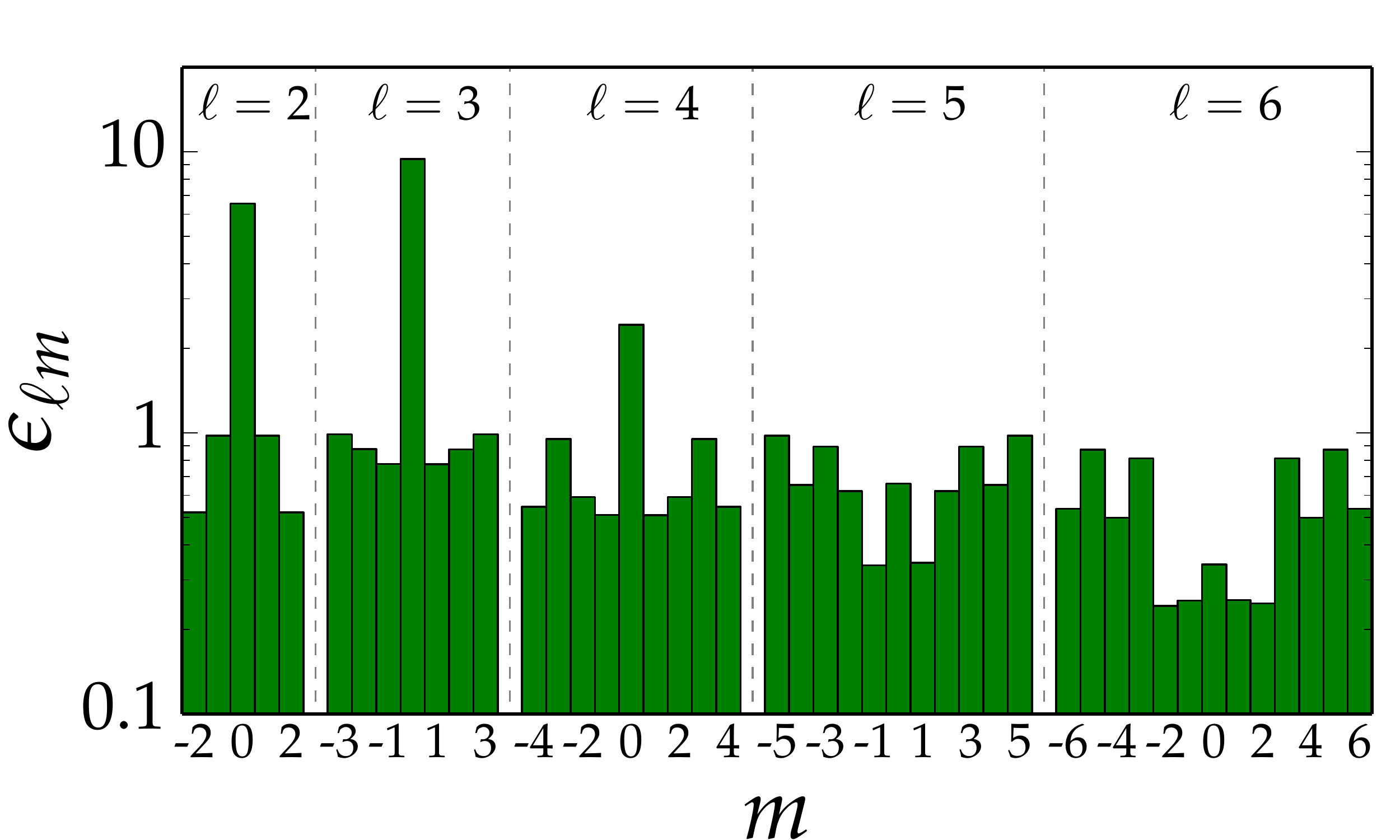}
  } \subfigure[\ Simulation 4: q=3, precessing] {
    \includegraphics[width=\columnwidth]{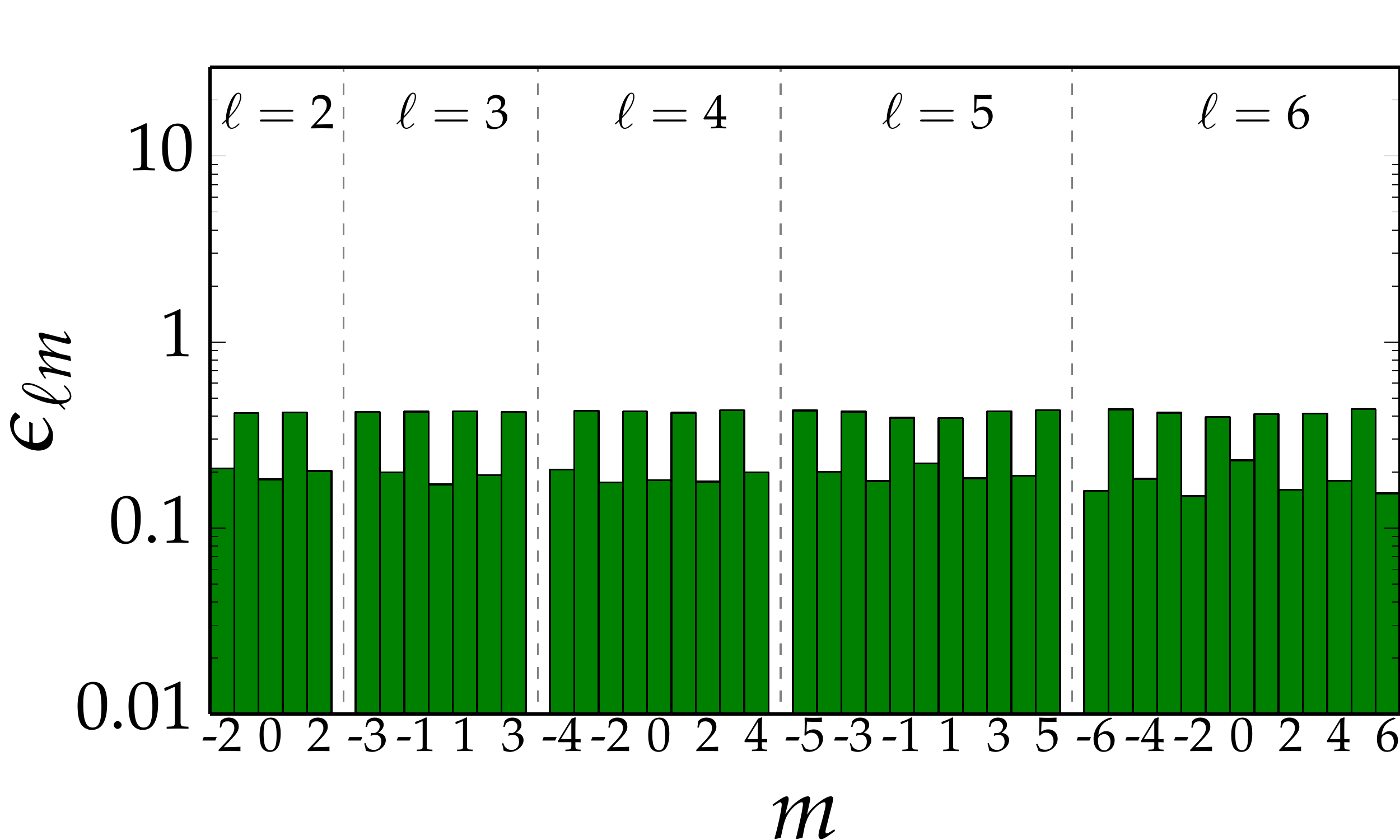}
  }
  
  \caption{Fractional differences between extrapolated and CCE
    $\Psi_4^{\ell,m}$ for all four cases in Table~\ref{tab:models}, as
    a function of $(\ell,m)$.  The $(\ell,m)$ modes are labeled as in
    Fig.~\ref{fig:CauchyErrHist}. Waveforms are aligned in the
    interval $[1000M,2000M]$.
  }
  \label{fig:HistogramCCEVsNP}
\end{figure*}

In the previous section, we found 
an example of extrapolated waveforms 
being significantly contaminated
by gauge effects.
In particular, the gauge used in simulation 1 of Table~\ref{tab:models}
results in 
waveforms for which some spherical harmonic modes (namely,
those with $m=0$) cannot be reliably extrapolated because they 
fall off more slowly than $1/r$.

This example raises the question of how reliable the extrapolation
method is in general.  It should be possible to 
find (or construct)
other examples in which extrapolation
yields the wrong waveform.  But will all of these examples exhibit
clear erroneous behavior such as the slow falloff
shown in Fig.~\ref{fig:Psi420ManyRadii}, or is it possible for
extrapolation to yield the incorrect result without any indication
of a problem?   In principle, the latter should be possible for
a sufficiently pathological gauge.
For instance, a gauge pulse traveling
outward and falling off exactly like $1/r$ would allow convergent
extrapolation, but would still contaminate the extrapolated waveform. 

Here we focus on a more specific question: for
simulations using the damped harmonic gauge 
condition~\cite{Lindblom2009c,Choptuik:2009ww,Szilagyi:2009qz} 
as currently implemented in \texttt{SpEC}, how 
reliable are extrapolated waveforms?  We
answer this question for the simulations in 
Table~\ref{tab:models},
by comparing extrapolated waveforms to 
gauge-invariant CCE waveforms.

This comparison is shown in 
Fig.~\ref{fig:HistogramCCEVsNP}, 
where we plot the average fractional differences 
between extrapolated and CCE waveforms
for all simulations in Table~\ref{tab:models}, and for 
all modes with $\ell\le 6$.  The quantity plotted is 
$\epsilon_{\ell m}$ as defined in Eq.~(\ref{eq:FractionalErrorMeasure}),
with the error bar defined by

\begin{equation}
  E = \frac{1}{2}\left(|C^C|+|C^E|\right) + |F| + |T| + |I|.
  \label{eq:CombinedErrorBarCCEVsNp}
\end{equation}
Here $C^C$ and $C^E$ are the respective Cauchy errors 
  computed using CCE and extrapolated waveforms, $T$ 
is the  
CCE truncation
error, $F$ is the extrapolation fit error, 
and $I$ is the CCE initial-data error.

If the magnitudes of the fractional errors plotted in
Fig.~\ref{fig:HistogramCCEVsNP} are 
less than unity, 
then the differences between CCE and extrapolated waveforms are smaller 
on average than 
the estimated error bars, 
and we can conclude that 
gauge errors in extrapolated waveforms
are unimportant.

We find that this is indeed the case for
almost all $(\ell,m)$ modes, 
including the dominant $(2,2)$ modes.
However, for the first few modes with $m=0$, we find that the
difference between CCE and extrapolated waveforms is 
larger than the estimated error, suggesting 
that for these modes the gauge
contamination in extrapolated waveforms is significant.  

Earlier in Fig.~\ref{fig:Psi420TwoGauges} we compared the $(2,0)$
mode between CCE and extrapolated waveforms, and we found that the
agreement was much better for
simulation 2
than for simulation 1.
But Fig.~\ref{fig:HistogramCCEVsNP} appears to support the opposite conclusion;
the fractional differences between CCE and extrapolated waveforms
in this figure are smaller for simulation 1 than simulation 2.  
This discrepancy can be explained by noting that the quantities in
Fig.~\ref{fig:HistogramCCEVsNP} are normalized by the 
error bar,
defined in Eq.~(\ref{eq:CombinedErrorBarCCEVsNp}), 
which is much larger for
simulation 1 than simulation 2.
Figure~\ref{fig:ComplexErrHist} 
shows that for most modes, 
the largest contribution to this error measure in simulation
1 is the extrapolation fit error.

For most modes with $m \ne 0$, the average fractional differences in 
Fig.~\ref{fig:HistogramCCEVsNP} are less than unity.  For the $q=6$ simulation 
(case 3 in Table~\ref{tab:models}), however, many of these modes have
average fractional differences that are very close to unity.  
Upon further examination, we find that for this case, 
using lower-order extrapolation seems to improve the
agreement with CCE for
most modes.  In fact, if we use order $N=2$
extrapolation, the average fractional differences between extrapolated
and CCE waveforms fall markedly below unity for every mode,
including $m=0$ modes.  This is potentially misleading, however, because
the primary reason for the improvement is that the estimated
extrapolation fit error is erroneously small when using higher-order
extrapolation.  The actual
difference (not normalized
by the error bar) between CCE and extrapolation is in fact
greater for lower-order extrapolation.  Evidently, the accuracy of the 
estimated extrapolation fit error decreases as order is increased.

\begin{figure}
  \includegraphics[width=1.\linewidth]{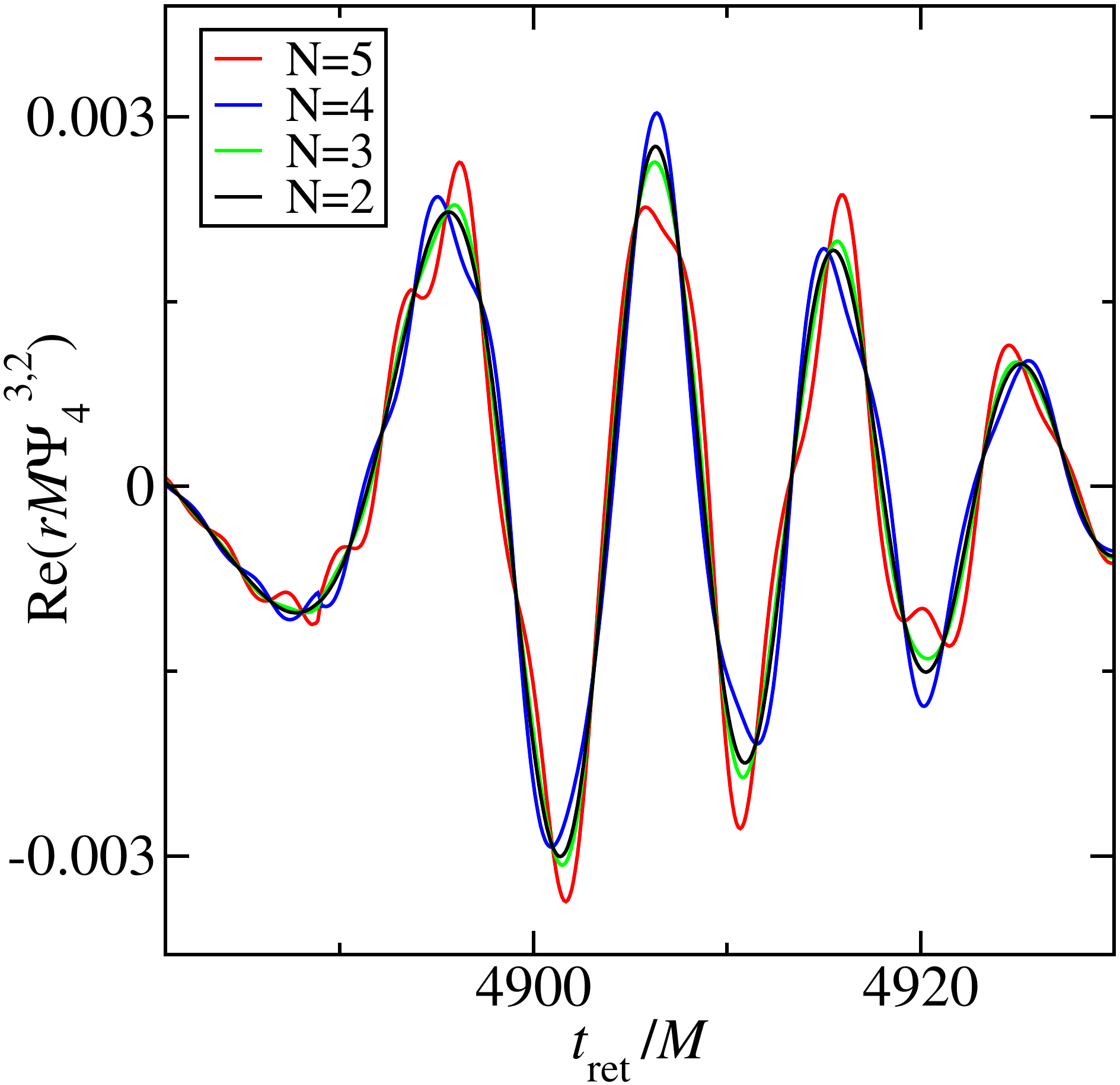}
  \caption{
      Merger portion of the real part of the extrapolated 
      $rM\Psi_4^{3,2}$
      mode for the $q=6$ case 
      (simulation 3 in Table~\ref{tab:models}).
      Divergence of the extrapolated waveform is evident
      as extrapolation order is increased.  Note that the order $N=2$  
      extrapolated waveform agrees well with CCE in this case.
      Maximum amplitude (of $rM\Psi_4^{2,2}$)
      is at $\tr\simeq 4901M$.
  }
  \label{fig:q06BadExtrap}
\end{figure}

This behavior could be at least somewhat anticipated
by inspecting the convergence of the extrapolated waveforms with 
extrapolation order.  We find that both the amplitude and phase of
many modes exhibit clear divergence as extrapolation order is 
increased, particularly for times near merger.  
Increasing extrapolation order produces increasing amounts of 
higher-frequency noise,
as shown in Fig.~\ref{fig:q06BadExtrap}.
This casts significant doubt on the reliability of any extrapolation
error estimate in this case.  
Note that the extrapolated $(2,2)$ mode in this simulation 
actually does 
converge for the first few 
extrapolation orders, and it also agrees
well with CCE.
Note also that for
the other BBH cases, 
there is no
clear lack of convergence in the extrapolated waveforms (for $m\ne0$ modes),
and lower-order
extrapolation does not improve the agreement with CCE.

\begin{figure}
  \includegraphics[width=1.\linewidth]{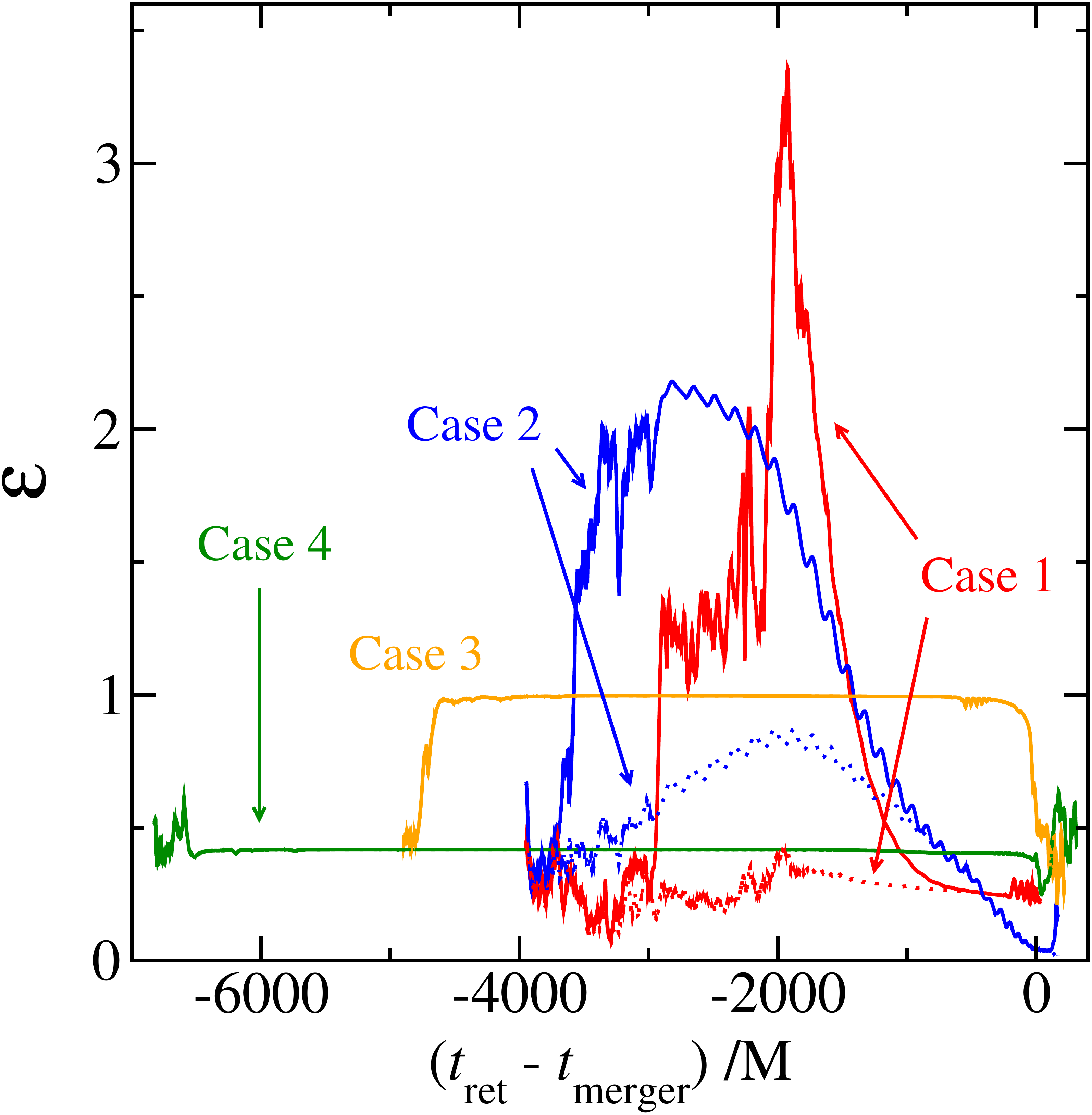}
  \caption{
    Differences between CCE and extrapolated waveforms integrated
      over the sky and normalized by error bars, computed according to 
    Eq.~(\ref{eq:errorsumovermodes}) for the four cases of 
    Table~\ref{tab:models}.  Curves have been shifted so that merger in 
    each case is at $t \simeq 0$.  For each case, solid lines are computed
    using all modes up to $L=8$, while dotted lines 
    are the same but with $m=0$ modes
    omitted.  For cases 3 and 4, 
    the dotted lines 
    are indistinguishable
    from the corresponding solid lines.
  }
  \label{fig:SumOverLFractionalDiffs}
\end{figure}

So far 
we have considered different $Y_{\ell m}$ modes separately.
Let us now briefly examine 
the difference between CCE and extrapolated
waveforms when summing over all modes, as is done when
computing the waveform in a particular sky direction.
In particular, we would like to investigate whether the large errors in 
extrapolated $m=0$ modes shown in Fig.~\ref{fig:HistogramCCEVsNP}
correspond to large errors after summing over all modes.
Instead
of choosing a single
direction on the sky, we integrate
the difference between CCE and extrapolated waveforms over all sky directions,
and use Eqs.~(\ref{eq:ComplexErrorDef}) and~(\ref{eq:ComplexErrorTotal})
to write this integral as a sum over modes.  We then normalize by the
quadrature sum of the errors in each mode.
Thus we compute the expression
\begin{equation}
  \epsilon = \frac{||\Psi_4^A - \Psi_4^B||}{\sum\limits_{\textrm{sources}} \big( \sum\limits_{\ell, m} E_{\ell,m}^2 \big)^{1/2}},  \label{eq:errorsumovermodes}
\end{equation}
where $A$ and $B$ in the numerator represent CCE and extrapolated waveforms,
and where the numerator is evaluated using 
Eq.~(\ref{eq:ComplexErrorTotal}). 
The sum in the denominator is over all sources of error, 
with the individual mode
contributions summed in quadrature, 
for each source of error.
The sources of error that enter into this calculation include the Cauchy error,
extrapolation fit error, CCE initial-data error, and CCE truncation error.
Figure~\ref{fig:SumOverLFractionalDiffs} shows 
the quantity $\epsilon$
for each of the four numerical simulations
we consider.  In the figure, curves have been 
shifted
in time so that the merger occurs at $t\simeq0$ for each case.  

To estimate the importance of $m=0$ modes in the sum over all modes, 
we compute
the sums in Eq.~(\ref{eq:errorsumovermodes}) twice---once with all modes
included (up to $L=8$), and again with $m=0$ modes omitted.
As shown in Fig.~\ref{fig:SumOverLFractionalDiffs}, 
including $m=0$ modes
substantially changes
the waveform agreement for the equal-mass, non-spinning configurations
(cases 1 and 2 in Table~\ref{tab:models}): 
in both cases $\epsilon<1$ when omitting the $m=0$ modes, and
$\epsilon > 1$ when including them.  
For the $q=6$ simulation (case 3), the difference
between CCE and extrapolated waveforms is the same size as the combined
error bar.
Including the $m=0$ modes
makes no noticeable difference
in this case,
even though $m=0$ modes were in disagreement
(albeit not by as much) in 
  Fig.~\ref{fig:HistogramCCEVsNP}.  
Including $m=0$ modes
makes no noticeable difference in the generic configuration 
(case 4) as well,
although this is to be expected because of the good agreement between
CCE and extrapolated waveforms for all modes
in this case.

  It may be somewhat surprising that the curves for cases 3 and 4 
  are largely constant in time.  This is because for many modes
  both the difference between CCE and extrapolated waveforms in the numerator
  of Eq.~(\ref{eq:errorsumovermodes}) and the estimated error bars in the 
  denominator
  are dominated by the CCE initial-data error, as shown in 
  Fig.~\ref{fig:ComplexErrHist}.
  This error manifests as a largely constant in time amplitude offset, 
  as illustrated in Fig.~\ref{fig:q01relamp22}.  
  This accounts both for the flatness of the
  case 3 and 4 curves in Fig.~\ref{fig:SumOverLFractionalDiffs} 
  as well as the negligible impact
  of $m=0$ modes for these cases.

The above considerations indicate 
that the question of whether
CCE is necessary to achieve accurate 
waveforms depends not only
on the various sources of error, but also on 
which $(\ell,m)$ modes are
of interest.  For general applications in which one is interested in
all $(\ell,m)$ modes, we find that without CCE, (presumed) gauge
errors can dominate the errors in our waveforms.


\section{Discussion}
\label{sec:discussion}

Comparisons between different methods of waveform extraction are meaningful
only when considering the various sources of uncertainty that affect the
final waveform.  We have estimated the key error 
contributions for a handful of simulations.
In all of the cases we considered, the CCE null-grid truncation error
was by far the smallest source of uncertainty.  
This suggests that the relative expense of 
CCE could be reduced by running
at lower CCE resolution, without significant impact on the results.
The extrapolation fit error was the most
significant source of error in the first equal-mass simulation 
(case 1 in Table~\ref{tab:models}), presumably because of the
gauge condition used (described in Sec.~\ref{sec:binary-black-hole}).
In the other cases, which used
harmonic or damped harmonic gauge, the extrapolation
fit, CCE initial-data, and Cauchy errors were 
generally comparable.

A potential improvement to extrapolation
would be to use a time-varying extrapolation order, with higher
order in the early inspiral and lower
order near merger, so
that the order decreases with decreasing wavelength.  This could be
achieved smoothly by combining 
extrapolants of different
orders, each weighted by the (inverse) variances of the
polynomial fit, and suitably normalized.  Such a procedure would not only
provide for more accurate extrapolation, but would
also reduce the magnitude of the
estimated extrapolation fit error.

We were somewhat surprised to find that the CCE initial-data error
was often quite significant, sometimes dominating the other
source of error.
Reducing the magnitude
of this error could be achieved by using 
a larger Cauchy computational domain 
(so that the worldtube
radius could be larger), which would increase the 
computational cost of the simulations.  The
extra cost
would be modest for codes (like \texttt{SpEC}) that use spherical outer domains
rather than Cartesian-aligned grids, except for the extra evolution
time necessary for the gravitational waves to reach the more distant
worldtube.  
The CCE initial-data error could also be reduced, in principle,
by using improved initial data in the 
characteristic code~\cite{Bishop:2011iu}.

By explicitly comparing two simulations with identical physical parameters,
differing only in the gauge condition used for
the Cauchy simulation, we showed
in Sec.~\ref{sec:errors-from-diff} that CCE 
waveforms are gauge-independent to within
uncertainties.  
We found that extrapolated waveforms, on the other hand, had
significant gauge dependence for $m=0$ modes.
It was clear from Fig.~\ref{fig:Psi420ManyRadii} that extrapolation would 
fail for $m=0$ for
the simulation with the gauge
condition of case 1 from Table~\ref{tab:models},
and that therefore another method such as CCE 
was required.
In the $q=6$ simulation, 
the poor convergence of extrapolation made it clear that an alternate
extraction method was required.
However, for $m=0$ modes in
case 2, there was no {\it a priori\/}
indication that extrapolated waveforms would be inaccurate.  

We find that large-amplitude modes (such as $\Psi_4^{22}$)
generally agree well between CCE and extrapolated waveforms.
However, the $m=0$ ``memory'' modes
disagree significantly in almost every case.
This disagreement is not necessarily a result of 
gauge effects alone.  
The long wavelength of the $m=0$ modes may lead to inherent difficulties 
in the polynomial fit, resulting in poor extrapolation, as discussed
at the end of Sec.~\ref{section:extrapolation}.
Indeed, we find that most
of the difference in the $(2,0)$ mode in the upper right
panel of Fig.~\ref{fig:HistogramCCEVsNP}, for example,
comes from the inspiral, where the wavelength is longer.
The fractional difference between the extrapolated and CCE 
waveforms is greater than unity during the merger in this case as well,
but it is orders of magnitude
less there than is it during the inspiral.

Unlike in the other cases, extrapolated and CCE waveforms were found
to agree quite well for all (including $m=0$) modes 
in the precessing configuration (case 4 in Table~\ref{tab:models}).
One reason for this is that the uncertainties are
larger in this case than in the others, 
as shown in Fig.~\ref{fig:ComplexErrHist}.  Even with larger error bars, 
however, it is somewhat surprising that $m=0$ modes do not stand out
in Fig.~\ref{fig:HistogramCCEVsNP}, as they do for the other cases.
We do not know the reason for this, 
but we note that this is the only 
simulation we consider that utilized a 
damped harmonic gauge condition
for the majority of the inspiral, as described in 
Sec.~\ref{sec:binary-black-hole}.

Because of the potential disagreement in
$m=0$ modes, we recommend using
CCE in applications for which all modes are important.
Additionally, Even though we found above in the $q=6$
simulation that extrapolated waveforms agreed with CCE
for $m\ne0$ modes, we do not consider this a confirmation of the reliability
of extrapolated waveforms.
When no convergence at any order is evident in the extrapolation
procedure, the waveforms and error estimates simply cannot be trusted.
For this reason, we also recommend CCE in cases where extrapolation
fails to show convergence for at least the first few orders.
We caution that each mode of interest must be individually 
checked for convergence.  For instance, as 
discussed above for the $q=6$ simulation,
the extrapolated $(2,2)$ mode was convergent, while the other modes were not.

When extrapolation does show reasonable 
convergence, however,
the uncertainties in the two waveform extraction methods are 
comparable.
In this case, because of the simplicity and 
reduced computational expense, 
extrapolated waveforms are preferred
for $m \ne 0$ modes.
Nevertheless, even if the extrapolation is convergent, 
we recommend doublechecking with
CCE waveforms for simulations that use 
new gauge conditions
or for new regions of parameter space.


\acknowledgments

We thank Nigel Bishop, Ian Hinder, Lee Lindblom,
Harald Pfeiffer, and Jeffrey Winicour for helpful discussions.
We thank Christian Ott for help in initiating and completing this 
project.
We gratefully acknowledge support from the Sherman Fairchild Foundation;
from NSF grants PHY-1068881, PHY-1005655, and 
DMS-1065438 at Caltech; 
and from NSF grants PHY-0969111 and PHY-1005426,
and NASA Grant NNX09AF96G at Cornell.
CR acknowledges support by NASA through
Einstein Postdoctoral Fellowship grant number PF2-130099 awarded by
the Chandra X-ray center, which is operated by the Smithsonian
Astrophysical Observatory for NASA under contract NAS8-03060.
Simulations used in this work
were computed with \texttt{SpEC}~\cite{SpECwebsite}.  Computations
were performed on the Zwicky cluster at Caltech, which is supported by
the Sherman Fairchild Foundation and by NSF award PHY-0960291;
on the NSF XSEDE network under grant TG-PHY990007N; 
and on the GPC supercomputer at the SciNet HPC
Consortium~\cite{scinet}. SciNet is funded by the Canada Foundation
for Innovation under the auspices of Compute Canada, the Government
of Ontario, Ontario Research Fund--Research Excellence, and the
University of Toronto.


\bibliographystyle{apsrev}


\end{document}

%% file: Macros.tex
\DeclareSymbolFontAlphabet{\mathrsfs}{rsfs}
\DeclareMathAlphabet{\mathcal}{OMS}{cmsy}{m}{n}
\newcommand{\scri}{\mathrsfs{I}}

\newcommand{\mTwoYlm}[1]{\ensuremath{\scripts{_{-2}}{Y}{_{#1}}}}

\newcommand{\define}{\coloneqq}

\newcommand{\OfOrder}{\mathcal{O}}

\newcommand{\e}{\ensuremath{\mathrm{e}}}

\def\beq{\begin{equation}}
\def\eeq{\end{equation}}

\newcommand{\Eadm}{\ensuremath{M_{\text{ADM}}}}

\newcommand{\ADMMass}{\Eadm}

\newcommand{\tr}{\ensuremath{t_{\text{ret}}}}
\newcommand{\tri}{\ensuremath{t_{\text{ret},i}}}

\newcommand{\tcorr}{\ensuremath{t_{\text{corr}}}}

\newcommand{\ra}{\ensuremath{r_{\text{ar}}}}

\newcommand{\rt}{\ensuremath{r_{\ast}}}

\makeatletter
\newcommand{\foreign}[1]{{#1}}
\newcommand{\etal}{\textit{et~al}\@ifnextchar{\relax}{.\relax}{\ifx\@let@token.\else\ifx\@let@token~.\else.\@\xspace\fi\fi}}
\newcommand{\etc}{\foreign{etc}\@ifnextchar{\relax}{.\relax}{\ifx\@let@token.\else\ifx\@let@token~.\else.\@\xspace\fi\fi}}
\newcommand{\eg}{\foreign{e.g}\@ifnextchar{\relax}{.\relax}{\ifx\@let@token.\else\ifx\@let@token~.\else.\@\xspace\fi\fi}}
\newcommand{\ie}{\foreign{i.e}\@ifnextchar{\relax}{.\relax}{\ifx\@let@token.\else\ifx\@let@token~.\else.\@\xspace\fi\fi}}
\newcommand{\scripts}[3]{%
  \@mathmeasure\z@\displaystyle{#2}%
  \global\setbox\@ne\vbox to\ht\z@{}\dp\@ne\dp\z@
  \setbox\tw@\box\@ne
  \@mathmeasure4\displaystyle{\copy\tw@#1}%
  \@mathmeasure6\displaystyle{#2#3}%
  \dimen@-\wd6 \advance\dimen@\wd4 \advance\dimen@\wd\z@
  \hbox to\dimen@{}{\kern-\dimen@\box4\box6}%
}
\makeatother
